\documentclass[review]{elsarticle}

\usepackage{hyperref}
\usepackage{amsmath, graphicx,amssymb,epsfig,psfrag,subfigure,bm,bbm}
\journal{Journal of Nuclear Materials}

\begin{document}

\begin{frontmatter}

\title{Surrogate representation of sink strengths and the long-term role of crystalline interfaces in the development of irradiation-induced bubbles}


\author[DUT1]{Jing Luo}

\author[NPI]{Yong Xin}

\author[DUT1]{Zhengcheng Zhou}

\author[DUT1,DUT2,DUT3]{Yichao Zhu\corref{mycorrespondingauthor}}
\cortext[mycorrespondingauthor]{Corresponding authors}
\ead{yichaozhu@dlut.edu.cn}

\author[DUT1,DUT2,DUT3]{Xu Guo\corref{mycorrespondingauthor}}
\ead{guoxu@dlut.edu.cn}

\address[DUT1]{Department of Engineering Mechanics, Dalian University of Technology, Dalian, 116023, P. R. China}
\address[NPI]{Science and Technology on Reactor System Design Technology Laboratory, Nuclear Power Institute of China, Chengdu, 610213, P. R. China}
\address[DUT2]{State Key Laboratory of Structural Analysis for Industrial Equipment, Dalian University of Technology}
\address[DUT3]{International Research Center for Computational Mechanics, Dalian University of Technology}

\begin{abstract}
  The present article addresses an early-stage attempt on replacing the analyticity-based sink strength terms in rate equations by surrogate models of machine learning representation. Here we emphasise, in the context of multiscale modelling, a combinative use of machine learning with scale analysis, through which a set of fine-resolution problems of partial differential equations describing the (quasi-steady) short-range individual sink behaviour can be asymptotically sorted out from the mean-field kinetics. Hence the training of machine learning is restrictively oriented, that is, to express the local and already identified, but analytically unavailable nonlinear functional relationships between the sink strengths and other local continuum field quantities. With the trained models, one is enabled to quantitatively investigate the biased effect shown by a void/bubble being a point defect sink, and the results are compared with existing ones over well-studied scenarios. Moreover, the faster diffusive mechanisms on crystalline interfaces are distinguishingly modelled by locally planar rate equations, and their linkages with rate equations for bulk diffusion are formulated through derivative jumps of point defect concentrations across the interfaces. Thus the distinctive role of crystalline interfaces as partial sinks and quick diffusive channels can be investigated. Methodologicalwise, the present treatment is also applicable for studying more complicated situation of long-term sink behaviour observed in irradiated materials.
\end{abstract}


\end{frontmatter}


\section{Introduction}
Irradiation can induce deleterious changes in mechanical properties of crystalline materials  \cite{Lucas_JoNM_1993,Dai_JNM_2003,Materna-Morris_JNM_2009,was_fundamentals_2017}, which are microscopically carried out by the mutually intertwined development of various species of crystalline defects. Hence theoretical studies about the long-term behaviour of irradiated materials entail proper integrations of models built on multiple time and length scales. On one hand, it is necessary to conduct (sub) nano- or mesoscopic studies \cite[e.g.][]{Holmstrom_PRB_2008, Fu_PRL_2004, Liu_JMS_2012, ChenY_NatCom2015, DiazdelaRubia_Nature2000, Arsenlis_PhilosMag2004, CuiYN_PRL2018, Nordlund_JNucMater2018}, so as to identify the distinct underlying features of irradiated materials. On the other hand, we also need kinetic models incorporating the irradiation-induced microstructural features on a continuum background, so as to enable predictions made on a time scale reasonable for engineering applications.

Rate theories, or theories under other names which are conceptually formulated through the so-called rate equations (REs) \cite[e.g. as reviewed by][]{Nordlund_JNucMater2019}, are probably the most widely-used category of kinetic models for predicting long-term irradiation-induced behaviour. In rate theories, microstructural species are all described in an averaged sense through their density distributions. Hence their predictive accuracies naturally rely on how consistently the involved continuum terms summarise the underlying dynamics. In particular, terms quantifying the loss of point defects (PDs) to sinks, that is, the sink strengths, are of great importance in reflecting how PDs interact with other types of defects (or defect clusters) \cite{Nordlund_JNucMater2019}.

The present article is aimed to use machine learning models to express the sink strength terms in REs, whose evaluations normally rely on analytically solutions to (often idealised) problems resolving the underlying behaviour of individual sinks. Besides, we will not only consider PD evolution in the bulk, but also that on crystalline interfaces, so as to investigate the distinctive role of interfaces in the development of radiation-induced bubble. The novelties demonstrated by the present article can be summarised in the following three aspects.

Firstly, cross-scale formulation of the sink strength terms is traditionally derived through the conceptualisation of an effective medium, in which a set of locally high-resolution (LHR) problems of partial differential equations (PDEs) are established to describe the underlying defect sink behaviour \cite[e.g.][]{Yoo_JNucMater1976, Mansur_JNucMater1978, Mansur_NucTech1978, Brailsford_PhilosMag1981, Ahlgren_JNucMater2012}. But to effectively transit high-resolution results to mean-field formulation, explicit analytical solutions to the LHR problems are often required \cite{Stoller_JNucMater2008}. For upscaling PD-void interactions, for example, a widely adopted concession is to assume spherical symmetry over an infinitely large effective medium \cite{Brailsford_PhilosMag1981}. Such restrictions on solution analyticity naturally pose challenges in modelling complicated situations, where multiple mutually-intertwined irradiation-induced mechanisms or finite-size image effects should be considered. Over this issue, machine learning (ML) tools \cite[e.g.,][]{Rumelhart_Nature1986, Rumelhart_1987} are turned to. Here we emphasise, in a context of multiscale modelling, a combinative use of machine learning tools with scale analysis, or more rigorous asymptotic analysis. Through scale analysis, one manages to sort the (quasi-steady) short-range PD-sink interactions from the (mean-field) kinetics of (long-range) PD transition, and a set of locally high-resolution (LHR) PDE cell problems get identified, so as to resolve the local individual sink behaviour. Upon such treatments, the system kinetics is still formulated through REs, while the local, analytically unavailable, nonlinear inter-relationships between the sink strengths and other onsite continuum field quantities should thus be fully unlocked via well-trained machine learning models. In this viewpoint, although the present work is simply focused on PD-bubble interactions (partly for demonstrating the method), its underlying methodology should be applicable for far more complicated situations involving multiple sink mechanisms taking place in fission or future planned fusion materials.

The second point of novelty shown by the present work lies in studying the biased behaviour of voids/bubbles as PD sinks. Note that the local elastic strain exerted by an individual void/bubble introduces differences in the moving tendency of an self-interstitial atom (or an interstitial gas atom) against a vacancy site \cite{Wolfer_JAP1975, Balluffi_book2005}. For nanometer-sized cavities, such stress induced preferred absorption has recently reassured by object kinetic Monte-Carlo (OKMC) calculations incorporating coefficients obtained on a density functional basis \cite{Carpentier_ActaMater2017}. For more general cases, such as large-sized voids and bubbles whose internal gas pressure progressively builds up, using full continuum calculations seems to be more realistic. But the corresponding local PDE problems are even more complicated for bubbles, because an elasticity problem defined in a finite region should also be included, and various factors, such as bubble spacings, bubble sizes, inner gas pressure, as well as the applied stress, should all be taken into account. This issue can now be properly investigated using the present treatment, which is independent of the complexities exhibited by the underlying LHR cell problems. The sink bias measurement \cite{Heald_ActaMetal1975} is thus expressed by a well-trained machine learning model, and the results are compared with existing ones over properly-studied cases \cite{Carpentier_ActaMater2017}.

Thirdly, it has been widely recognised that crystalline interfaces, such as grain boundaries, are natural (partial) sinks to mobile defects \cite[e.g., see][]{Beyerlein_ProgMaterSci2015, ChenY_NatCom2015,Vattre_NatCom2016,GuYJ_JMPS2017}, i.e., a number of point defects get trapped by an interface while there are still a certain portion penetrating through. Nowadays, a dominating view in modelling the partial sink behaviour of interfaces is to include a normal sink term to rate equations. But this sees its limitation in two aspects. First, a mathematically bounded normal sink term in REs is set under the presumption that the sinks occupy a volume in space, while an interface is of zero volume. Second, the diffusive process on interfaces at a higher speed than that in the bulk is not properly resolved. For modelling crystalline interfaces as partial planar sinks (of zero volume), we extend the results by \cite{Zhu_PRL_2018} to impose a jump condition in the normal derivatives of point defect concentration across the interface, instead of using a normal sink term. This jump condition is actually equivalent to a normal sink term, which contains a Dirac-$\delta$ function identifying the manifolds represented by crystalline interfaces. Besides, a set of (locally planar) REs are also presented in conjunction with that in the bulk, so as to capture the distinctive diffusive processes on crystalline interfaces.

With the derived sink strength terms of bubbles to point defects, the long-term bubble growth behaviour in irradiated materials can be simulated. In particular, the distinctive role played by crystalline interfaces is examined with the present model under various conditions controlled by factors such as irradiation dose, external load, etc.. Studies along this direction are expected to provide a quantitative rationale to crack initiation (likely from grain boundaries) as observed in nuclear fuels \cite{Gandhi_ActaMetal1979, Kapoor_JNucMater2007}, and this issue will be discussed after numerical results are presented.

The article is arranged as follows. In Sec.~\ref{Sec_rate_eqn}, rate equations with crystalline interfaces modelled as partial defect sinks are presented. In Sec.~\ref{Sec_homogenisation}, scale analysis is conducted to sort out, from the mean-field REs, a set of LHR cell problems formulating the underlying PD-bubble interactions, based on which a curriculum for machine learning is devised in Sec.~\ref{Sec_ML}, and the rational quadratic Gaussian procession regression scheme is selected among a comparative test over nineteen popular machine learning algorithms. Then simulations of bubble growth in the presence of crystalline interfaces are presented in Sec.~\ref{Sec_numerics}, and the roles of various involved factors are analysed. The article concludes with a further discussion in Sec.~\ref{Sec_conclusion}. Throughout the article, a subscript $\alpha=$i, v or g is affiliated with a quantity indicating that its association with self-interstitial atoms, vacancies and noble gas atoms, respectively.

\section{Rate equations\label{Sec_rate_eqn}}
\subsection{Microstructural mechanisms in consideration\label{Sec_micro_mechanism}}
For better illustration of the present upscaling strategy, we are focused on formulating the sink strength of bubbles to three PD species: self-interstitial atoms (SIAs), vacancy sites and irradiation-induced noble gas atoms (NGAs), both in the bulk and on crystalline interfaces. The following microstructural behaviour will be taken into account.
\begin{enumerate}
    \item Generation of mobile PDs as a result of a series of (unspecified) primary radiation damage events.
    \item PD diffusion in the bulk.
    \item Partial absorption of PDs by a crystalline interface.
    \item Faster diffusion of absorbed PDs on the crystalline interface.
    \item Further recombination of SIAs with vacancy sites, both in the bulk and on the interface.
    \item PDs sunk to bubbles, both in the bulk and on the interface.
    \item Evolution of bubble properties, such its size and interior pressure, upon absorption of PDs.
\end{enumerate}
Several issues are noted. First, for simplicity, mobile defects are restricted of point defect type. For instance, a di-vacancy is simply envisaged as two neighbouring mono-vacancies. Second, voids are simply treated as special cases of bubbles. Third, the mechanisms of void nucleation are not considered here, and a distribution of void embryos is assumed in advance. Fourth, (immobile) interstitial clusters are not considered for the moment, but an extension of the present treatment to cover their role seems relatively straight forward. This point will be discussed further in the conclusion session. Finally, the underlying dislocation substructures are not explicitly considered, but their biased sink behaviour to self interstitial atoms over vacancies are implicitly formulated in a (net) source term for PD generation.

\subsection{Rate equations with crystalline interfaces being partial sinks}
\subsubsection{Definition of mean-field variables}
In rate theories or their derivatives, PDs are represented by their concentration distributions in space, conventionally denoted by $C_{\alpha}$ ($\sim$ number per volume), where $\alpha=$v, i, or g, corresponding to the species of vacancies, SIAs, or NGAs, respectively. Alternatively, one may also define (non-dimensional) field quantities of fractional concentration, denoted by $c_{\alpha}$, which equals the number of the species of interest within a representative volume, divided by the total number of atoms in it. The two sets of quantities differ correspondingly by a factor of $v_0$, which is the volume occupied by a single atom from the hosting materials, i.e., $c_{\alpha} = C_{\alpha}v_0$.

In this article, we adopt the fractional concentration $c_{\alpha}$ for further modelling. This is because (faster) PD evolutions on interfaces are modelled individually, and the definition of $C_{\alpha}$ is not consistent as moving from the bulk ($\sim$m$^{-3}$) to an interface ($\sim$m$^{-2}$). Here a symbol of ``~$\tilde{\,}~$'' is affiliated with a quantity implying its association with crystalline interfaces. On an interface, we define
\begin{equation} \label{c_transition_2d}
    \tilde{c}_{\alpha} = \tilde{C}_{\alpha}v_0^{\frac2{3}},
\end{equation}
where the interface thickness is assumed to be $v_0^{\frac1{3}}$, roughly an atomic spacing. Given the continuity in $c_{\alpha}$ across an interface, the symbol of ``~$\tilde{\,}~$'' can actually be dropped from $\tilde{c}_{\alpha}$.

\subsubsection{Rate equations in the bulk}
With regards to the microstructural dynamics listed above, rate equations in the bulk read
\begin{equation}
\label{sec2-c-evlution-eq}
\frac{\partial c_{\alpha}}{\partial t} = - \nabla\cdot\mathbf{j}_{\alpha} + K_{\alpha} - \delta_{\alpha} k_{\text{iv}} c_{\text i}c_{\text v} - k_{\alpha \text{B}}^2 D_{\alpha} c_{\alpha}, \quad \text{in }\Omega \setminus \Gamma,
\end{equation}
where $\Omega$ denotes the overall domain of interest; $\Gamma$ represents a collection of manifolds in $\Omega$ identifying the crystalline interfaces in consideration; $\mathbf{j}_{\alpha}$ ($\sim$ m$\cdot$s$^{-1}$) is a vector measuring the fractional PD flux in space; $K_{\alpha}$ formulates the PD generation (less those absorbed by background dislocations); $k_{\text{iv}}$ is a coefficient of vacancy-interstitial recombination; $\delta_{\alpha}=1$ for $\alpha=$i or v and $\delta_{\alpha}=0$ for $\alpha$=g; $k_{\alpha \text{B}}^2$ ($\sim$m$^{-2}$) measures the sink strength of bubbles to species $\alpha$.

The PD fractional flux $\mathbf{j}_{\alpha}$ in Eq.~\eqref{sec2-c-evlution-eq} is formulated by \cite{Sutton_book1995}
\begin{equation}\label{sec2-flux}
\mathbf{j}_{\alpha} = - D_{\alpha} \nabla c_{\alpha} + D_{\alpha} \cdot \frac{\lambda_{\alpha} \Delta v_{\alpha}}{k_{\text B} T} c_{\alpha} \nabla p,
\end{equation}
where $D_{\alpha}$ is known as the diffusivity; ``$\nabla$'' denotes the ``spatial gradient of''; $\lambda_{\alpha}=\pm1$; $p$ is the hydrostatic pressure field; $\Delta v_{\alpha}$ measures the relaxition volume when a PD is removed; $k_{\text B}$ is the Boltzmann constant; $T$ is temperature. The second term on the right side of Eq.~\eqref{sec2-flux} arises due to the fact that the PD movement brings about changes in local free volume. Here we require $\lambda_{\alpha}=1$ for $\alpha=$v, and $\lambda_{\alpha}=-1$ for $\alpha=$i, g in Eq.~\eqref{sec2-flux}.

Alternatively, a quantity of PD flux $\mathbf{J}$ ($\sim$ number$\cdot$m$^{-2}\cdot$s$^{-1}$) can also be introduced, and it is linked with the corresponding fractional quantity by $\mathbf{j}_{\alpha} = v_0\mathbf{J}_{\alpha}$.

The source term $K_{\alpha}$ in Eq.~\eqref{sec2-c-evlution-eq} is often quantified by \cite{was_fundamentals_2017}
\begin{equation} \label{source_term}
    K_{\alpha} = k_{\alpha}^{\text{s}} \dot{s}_{\text{dpa}},
\end{equation}
where $\dot{s}_{\text{dpa}}$ measures the rate of displacement per atom (d.p.a.), and $k_{\alpha}^{\text{s}}$ is an empirical coefficient. To take the effect of dislocations as biased sinks to PDs, one may (temporarily) assume different values for $k_{\text{i}}^{\text{s}}$ against $k_{\text{v}}^{\text{s}}$.

\subsubsection{Rate equations on crystalline interfaces\label{Sec_interface}}
A dominant way of modelling interfaces as PD sinks is to include an extra sink term in REs \cite{Beyerlein_ProgMaterSci2015}, which sees its limitation in two aspects. First, faster diffusive processes, such as channelling diffusion, usually take place on interfaces, which can not be resolved by a mathematically bounded normal sink term. Second, an interface is of zero volume. Thus a Dirac-$\delta$ function should appear in the corresponding sink term, which effectively leads to discontinuity in PD flux as formulated by Eq.~\eqref{PD_absorption}.

Given the fact that $\tilde{D}_{\alpha} \gg D_{\alpha}$ with $\tilde{D}$ the diffusivity on an interface, the diffusive mechanisms near an interface should be modelled as a locally two-dimensional (in-plane) process. Thus the PD evolution on an interface $\Gamma$ is formulated by
\begin{equation}\label{sec2-c_gb-evlution-eq0}
\frac{\partial c_{\alpha}}{\partial t} = - \tilde{\nabla}\cdot\tilde{\mathbf{j}}_{\alpha} + \tilde{K}_{\alpha} - \delta_{\alpha} k_{\text{iv}} c_{\text i}c_{\text v} - \tilde{k}_{\alpha}^2 \tilde{D}_{\alpha} c_{\alpha},\quad \text{on }\Gamma,
\end{equation}
where ``$\tilde{\nabla}$'' denotes the local surface gradient, that is, the projection of the full gradient vector in three-dimensional space onto the local tangent plane to the interface; $\tilde{\mathbf{j}}_{\alpha}$ is the interface flux given by
\begin{equation}\label{sec2-flux_gb}
\tilde{\mathbf{j}}_{\alpha} = -\tilde{D}_{\alpha} \left[\nabla c_{\alpha} - \lambda_{\alpha} \cdot \frac{\Delta v_{\alpha}c_{\alpha}}{k_{\text B} T} \nabla p \right];
\end{equation}
$\tilde{K}_{\alpha}$ measures the PD source on the interface whose formulation will be given next; $\tilde{k}_{\alpha}^2$ measures the sink strength of bubbles on the interface.

\begin{figure}[!ht]
  \centering
  \includegraphics[width=.3\textwidth]{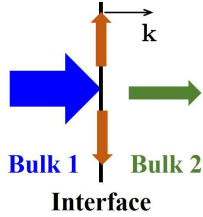}
  \caption{A diagram on the collective behaviour of PDs near a crystalline interface. A crystalline interface serves as a partial sink, i.e., certain portion of PDs are absorbed by the interface, and get quickly transited elsewhere on the interface, while there are still some PDs penetrating to the neighbouring bulk.} \label{Illu_pipe_diffusion}
\end{figure}

Besides, a crystalline interface serves as a partial sink to PDs, as schematically shown in Fig.~\ref{Illu_pipe_diffusion}.
A certain portion of PDs are absorbed by the interface, and get quickly transited elsewhere on the interface governed by Eq.~\eqref{sec2-c_gb-evlution-eq0}. But there are still some PDs penetrating to the neighbouring bulk. As proposed by \cite{Zhu_PRL_2018}, the PD loss (per area) to a planar partial sink can be modelled by means of the difference in PD flux across it, that is,
\begin{equation} \label{PD_absorption}
    \left.\mathbf{J}_{\alpha}\right|_{\text{bulk 1}} - \left.\mathbf{J}_{\alpha}\right|_{\text{bulk 2}} = - \frac{D_{\alpha}}{v_0}\left[\frac{\partial c_{\alpha}}{\partial k}\right]_-^+,
\end{equation}
where ``$\left[\cdot\right]_-^+$'' denotes the difference in the values across an interface along its normal $\mathbf{k}$; ``$\frac{\partial}{\partial k}$'' denotes the normal derivative along $\mathbf{k}$. Note that the definition of $\left[\cdot\right]_-^+$ should be associated with a directional vector, e.g. $\mathbf{k}$. For the scenario shown in Fig.~\ref{Illu_pipe_diffusion}, $\left[\cdot\right]_-^+$ is defined such that the value on the bulk 2 side less that on the bulk 1 side. Eq.~\eqref{PD_absorption} is derived with the usage of Eqs.~\eqref{sec2-flux} and the identity $\mathbf{j}_{\alpha} = v_0\mathbf{J}_{\alpha}$. Since the fractional concentration $c_{\alpha}$ and the hydrostatic pressure gradient $\nabla p$ are both continuous across the interface, the jump in PD fluxes across the interface actually arises from the difference in the normal derivative of $c_{\alpha}$.

Meanwhile, the (partial) PD absorption from the bulk serves as a sourcing mechanism to the diffusive process on the interface. To derive the source term of Eq.~\eqref{sec2-c_gb-evlution-eq0} for fractional PD concentrations, a factor of $v_0^{\frac2{3}}$ which measures the area occupied by an atom on the interface, should be multiplied to the right side of Eq.~\eqref{PD_absorption} (with a negative sign added), yielding
\begin{equation} \label{source_interface}
\tilde{K}_{\alpha} = \frac{D_{\alpha}}{v_0^{\frac1{3}}}\left[\frac{\partial c_{\alpha}}{\partial k}\right]_-^+.
\end{equation}

Inserting Eq.~\eqref{source_interface} into \eqref{sec2-c_gb-evlution-eq0}, we finally derive the PD evolution on $\Gamma$ to be
\begin{equation}\label{sec2-c_gb-evlution-eq}
\frac{\partial c_{\alpha}}{\partial t} = - \tilde{\nabla}\cdot\tilde{\mathbf{j}}_{\alpha} + \frac{D_{\alpha}}{v_0^{\frac1{3}}}\left[\frac{\partial c_{\alpha}}{\partial k}\right]_-^+ - \delta_{\alpha} k_{\text{iv}} c_{\text i}c_{\text v} - \tilde{k}_{\alpha}^2 \tilde{D}_{\alpha} c_{\alpha},\quad \text{on }\Gamma.
\end{equation}
Note again that through the term containing a jump in the derivative of $c_{\alpha}$ in Eq.~\eqref{sec2-c_gb-evlution-eq}, a crystalline interface is modelled as a partial sink to PDs.

\subsection{Challenges in quantifying the sink strength}
The effectiveness of rate equations is highly affected by
how their sink strength terms get evaluated. But evaluations of $k_{\alpha}^2$ of Eq.~\eqref{sec2-c-evlution-eq} and $\tilde{k}_{\alpha}^2$ of Eq.~\eqref{sec2-c_gb-evlution-eq} entail resolving the local sink behaviour of individual bubbles, whose resolutions should be higher than that of the mean-field rate theories. And for effective transition between scales, analyticity in solutions to the locally high-resolution models is generally necessitated \cite{Stoller_JNucMater2008}.

In the upcoming two sections, we try to demonstrate that a combinative use of scale analysis and machine learning may help remove the restriction on solution explicitness of the LHR problems for upscaling. First we will use scale analysis to show that the (short-range) PD - bubble interactions can be asymptotically treated as quasi-steady processes at the mean-field level, and a set of LHR cell problems can thus be sorted out. Although explicit solutions to the LHR cell problems are unavailable, one may solve the LHR cell problems for several sampled scenarios. With the generated data, machine learning models are then trained to express the sink strengths in REs \eqref{sec2-c-evlution-eq} and \eqref{sec2-c_gb-evlution-eq}.

\section{Identification of LHR problems describing individual sink behaviour\label{Sec_homogenisation}}
In this section, we identify a set of locally high-resolution problems describing the local PD-bubble interactions underlying the rate equations in the bulk and on the interfaces. First a family of necessary field variables are introduced, so as to represent the spatial distribution of bubbles. Then a set of LHR problems are presented by sorting the short-range PD-bubble interactions from the long-range PD migration. Hence definite but implicit functional relationships linking the sink terms with other local continuum variables are given in the end of the section. Here our analysis always starts with the cases in the bulk, then with the cases on the interface.

\subsection{Continuum representations of spatially-varying bubbles\label{Sec_continuum_representation}}
On a length scale where rate equations are set up, microstructural bubbles should also be represented in a continuum sense. Here we describe the spatial distribution of bubbles with a bubble density distribution denoted by $\varrho$, a size distribution of bubble denoted by $r$, and a number distribution of interior NGAs per bubble denoted by $n$. For their quantifications, we refer to a statistically averaging strategy described as follows. Centred at a spatial point $\mathbf{x}$, a representative volume $\Omega_{\text{r}}$ containing several bubbles is taken. Then the bubble density $\varrho(\mathbf{x},t)$ ($\sim$ m$^{-3}$) is defined to be the number of bubbles within $\Omega_{\text{r}}$ divided by its volume; the size distribution $r(\mathbf{x},t)$ ($\sim$ nm) equals the averaged radius defined under the condition that the overall volume occupied by bubbles is conserved in $\Omega_{\text{r}}$; the distribution of NGA number per bubble $n(\mathbf{x},t)$ equals the averaged NGA number per bubble defined under the condition that the overall number of NGAs is conserved in $\Omega_{\text{r}}$.

Here the spatial distribution of bubbles is summarised by the first statistical moments (or the mean values) of the corresponding physical quantities. It lies on a presumption that the underlying bubbles should take a locally ordered configuration, although there may still be long-range variations. Such a presumption is supported by certain experimental and theoretical evidence \cite{Johnson_Nature1978, Evans_bubblelattice1990, Ghoniem_JCompAidMaterDes2002}, but whether other statistical moments (such as the variance) should be involved for upscaling is also an interesting issue worth further investigations.

In a similar sense, another set of field quantities $\{\tilde{\rho}, \tilde{r}, \tilde{n}\}$ are also introduced for bubble representation on crystalline interfaces $\Gamma$. Note that the bubble density $\tilde{\rho}$ then is of unit m$^{-2}$. Here we require the average bubble spacing should be continuous as transiting from the bulk to the interface, and this gives
\begin{equation} \label{void_density_transition}
    \tilde{\rho}^{\frac1{2}} = \lim_{\mathbf{x}\rightarrow\Gamma} \varrho^{\frac1{3}}.
\end{equation}

\subsection{LHR cell problems resolving local PD-bubble interactions}
Note that $c_{\alpha}$ of Eq.~\eqref{sec2-c-evlution-eq} effectively defines fractional concentration in a mean-field sense. If viewed on a fine scale where individual bubbles are resolved, the corresponding profiles of the fractional concentration, should be highly oscillatory. This is because on bubble surfaces, the fractional concentration should equal the thermally equilibrated value $c_{\alpha}^{\text{e}}=0$. But it may quickly surge to a considerably different value, so as to accommodate the overall PD transition. As a consequence, the time scale associated with the short-range PD-bubble interactions (driven by the sharp spatial gradients of fractional concentration) should be far shorter than that of the long-range PD transition (driven by the mean-field gradient). Hence the short-range dynamics can simply be treated as quasi-steady states embedded into the mean-field kinetics. As for upscaling, the corresponding (steady-state) LHR cell problems should be parameterised by the onsite continuum quantities.

Such scale-separable features are also displayed by the fine-scale stress field, which has to accommodate both the gas pressure on bubble surfaces and external loads.

It is noted that a more self-consistent way to sort out the underlying LHR problems is to carry out asymptotic analysis for homogenisation, while the present treatment is conducted in a reasonably approximating manner.

\subsubsection{Domain of definition}
\begin{figure}[!ht]
  \centering
  \subfigure[Bulk cell]{
  \includegraphics[width=0.4\textwidth]{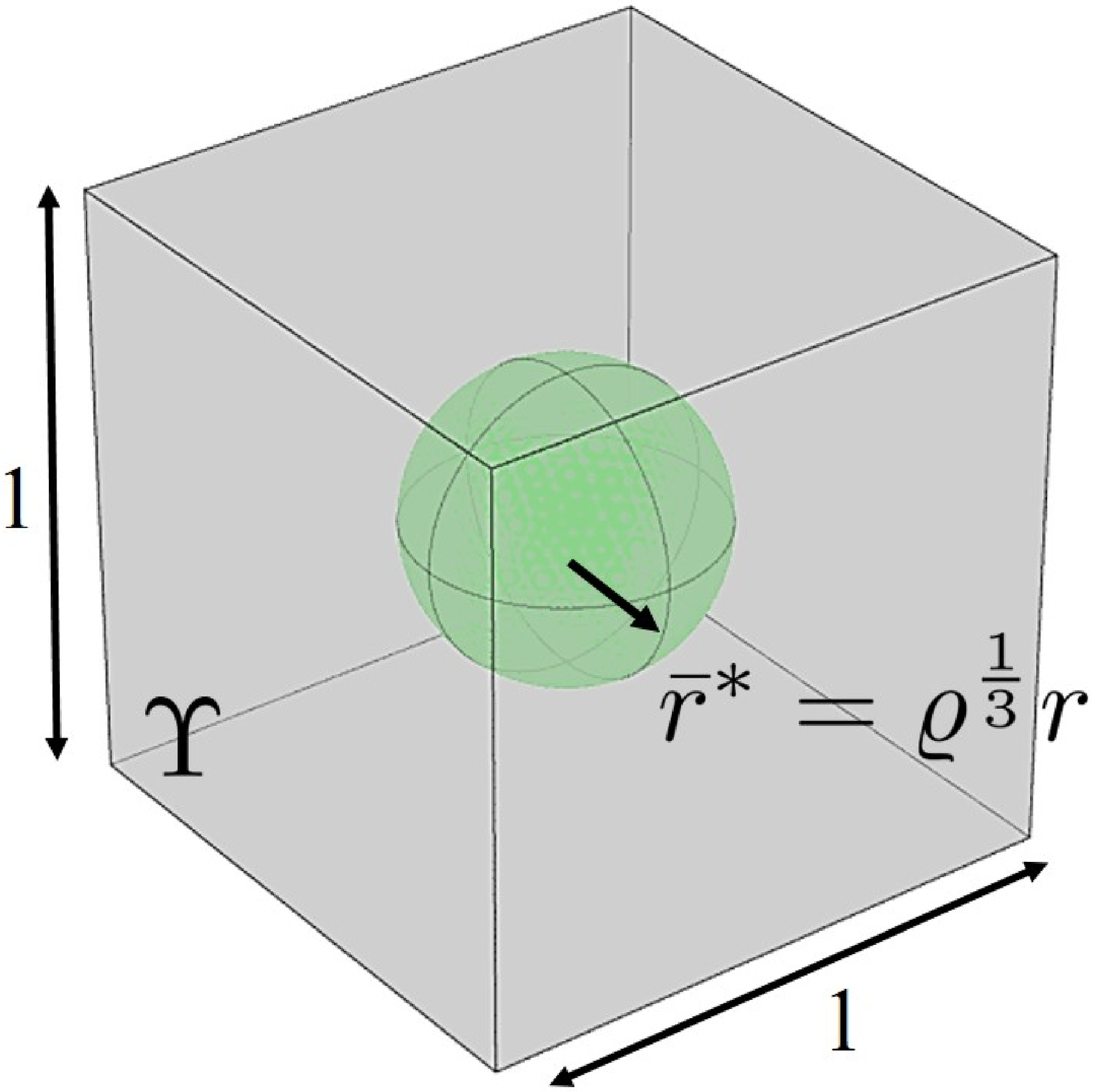}}
  \quad
  \subfigure[Interface cell]{
  \includegraphics[width=0.36\textwidth]{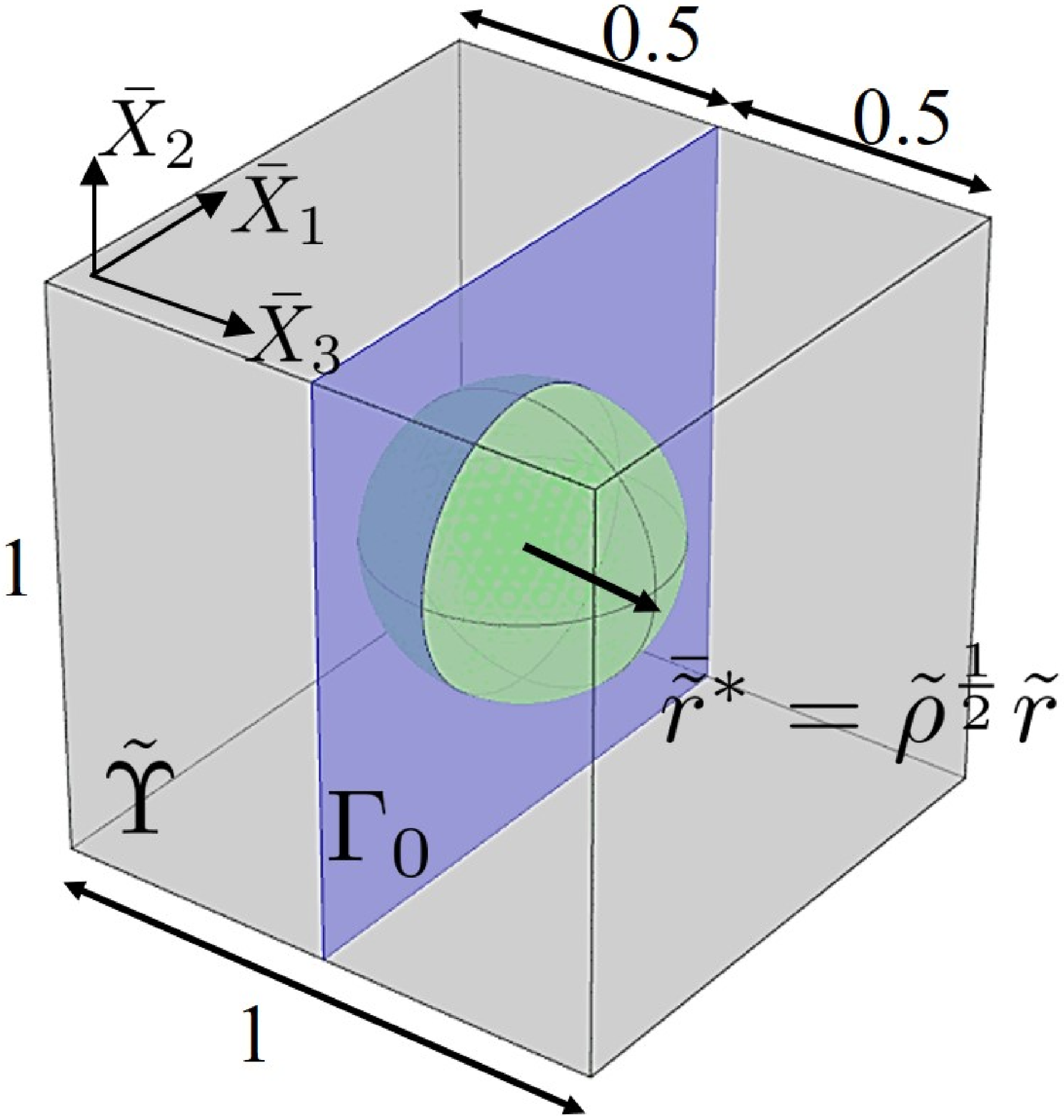}}
  \caption{Computational domain for setting up the cell problems (a) in a bulk cell and (b) in an interface cell.}\label{sec2-voids_on_GB}
\end{figure}

The mentioned LHR problems for PD-bubble interactions are defined in microscopic ``cells'', as schematically shown in Fig.~\ref{sec2-voids_on_GB}.
In the centre of a ``bulk cell'' (parameterised by its position $\mathbf{x}$ in the bulk), as depicted by Fig.~\ref{sec2-voids_on_GB}(a), a spherical hole is located in its middle to denote a bubble sampled near $\mathbf{x}$.
When measured on a mean-field scale, the cell size should be as diminishingly short as $\varrho^{-\frac1{3}}$. This defines a short length scale characterised by another (non-dimensional) spatial variable
\begin{equation} \label{X_star}
   \bar{\mathbf{X}} = \varrho^{\frac1{3}} \mathbf{x}.
\end{equation}
Here a bar is affiliated with a variable always indicating that it is defined in a non-dimensional sense.

In the (non-dimensional) space measure in $\bar{\mathbf{X}}$, the (rescaled) cell, denoted by $\Upsilon$, becomes a unit cube containing a spherical cavity, as shown in Fig.~\ref{sec2-voids_on_GB}(a), and we set
\begin{equation} \label{bulk_cell}
    \Upsilon = \left[-\frac1{2},\frac1{2}\right]^3 \setminus \mathcal{O}(\bar{r}^*),
\end{equation}
where $\mathcal{O}(\bar{r}^*)$ denotes a sphere centred at the origin and of radius $\bar{r}^*$ and
\begin{equation} \label{r_rescaled}
\bar{r}^* = \varrho^{\frac1{3}}r
\end{equation}
measures the non-dimensional radius of the cavity.

Now we can introduce $\hat{c}_{\alpha}(\mathbf{x};\bar{\mathbf{X}})$ to denote the locally high-resolution fractional concentration of species $\alpha$ in the bulk cell. Here a hat is affiliated with a variable indicating that it is defined in a LHR sense. Note that $\bar{\mathbf{X}}$ are the (non-dimensional) spatial variables of $\hat{c}_{\alpha}$, while $\mathbf{x}$ are simply parameters indicating which cell is referred to. In a similar sense, we can also introduce a LHR stress field denoted by $\hat{\boldsymbol{\sigma}}(\mathbf{x};\bar{\mathbf{X}})$ and a LHR hydrostatic pressure field denoted $\hat{p}(\mathbf{x};\bar{\mathbf{X}})$ to a bulk cell.

Similarly, an ``interface cell'', denoted by $\tilde{\Upsilon}$, is also introduced for modelling the short-range PD-bubble interactions taking place on crystalline interfaces, as shown in Fig.~\ref{sec2-voids_on_GB}(b). We can also introduce corresponding LHR quantities of $\hat{\tilde{c}}_{\alpha}$, $\hat{\tilde{\boldsymbol{\sigma}}}$ and $\hat{\tilde{p}}$ to denote the fractional concentration, the stress field and the hydrostatic pressure field in an interface cell.

Several of its distinguishable features are noted regarding the definition of an interface cell. Firstly, the space is normalised with $\rho^{-\frac1{2}}$, the average bubble spacing on the interface. Secondly, an interface section, denoted by $\Gamma_0$, is located in the middle of $\hat{\Upsilon}$ perpendicular to the $\bar{X}_3$ direction, and we have
\begin{equation} \label{Gamma0}
\Gamma_0 = \left[-\frac1{2},\frac1{2}\right]^2 \setminus \mathcal{O}_{\text{2d}} \left(\tilde{\rho}^{\frac1{2}}\tilde{r}\right),
\end{equation}
where $\mathcal{O}_{\text{2d}} \left(\bar{\tilde{r}}\right)$ denotes a circle centred at the origin and of radius $\bar{\tilde{r}}$. In an interface cell, we are mainly interested in the PD evolution on $\Gamma_0$, while the bulk-interface exchange in the value of $\hat{\tilde{c}}_{\alpha}$ has been taken into account by the derivative jump in Eq.~\eqref{sec2-c-evlution-eq}.

\subsubsection{Governing equations}
To formulate the behaviour of individual sink at its steady state, we write a differential equation for the LHR fractional concentration given by
\begin{equation} \label{eqn_flux_sr}
    \nabla_{\bar{\mathbf{X}}} \cdot \left(\nabla_{\bar{\mathbf{X}}} \hat{c}_{\alpha} - \frac{\lambda_{\alpha} \Delta v_{\alpha} \hat{c}_{\alpha}}{k_0T} \nabla_{\bar{\mathbf{X}}} \hat{p}\right) = 0, \quad \text{in }\Upsilon,
\end{equation}
where $\nabla_{\bar{\mathbf{X}}}$ denotes taking derivative with respect to $\bar{\mathbf{X}}$. Note that the term in the bracket of Eq.~\eqref{eqn_flux_sr} measures the (localised) PD flux due to the (high-frequency) oscillation of $\hat{c}_{\alpha}$ in the bulk cell.

The expression for the LHR hydrostatic pressure field $\hat{p}(\mathbf{x};\bar{\mathbf{X}})$ is also needed for expressing Eq.~\eqref{eqn_flux_sr}. This means one should solve a linear elasticity problem in the bulk cell as well. Hence we reach a (localised) force equilibrium equation given by
\begin{equation} \label{eqn_fb_sr}
    \nabla_{\bar{\mathbf{X}}}\cdot \hat{\boldsymbol{\sigma}} = \boldsymbol{0}, \quad \text{in }\Upsilon,
\end{equation}
which is coupled with a (localised) Hookean law described by
\begin{equation} \label{eqn_Hookean_sr}
    \hat{\boldsymbol{\sigma}} = \lambda \varrho^{\frac1{3}} \text{tr}(\nabla_{\bar{\mathbf{X}}} \hat{\mathbf{u}})\mathbf{I} + \mu \varrho^{\frac1{3}} \left(\nabla_{\bar{\mathbf{X}}}\hat{\mathbf{u}} + (\nabla_{\bar{\mathbf{X}}}\hat{\mathbf{u}})^{\mathrm{T}}\right),
\end{equation}
where $\lambda$ and $\mu$ are two material constants known as the Lam\'{e} constants; $\hat{\mathbf{u}}$ is the localised high-resolution displacement field.

Once $\hat{\boldsymbol{\sigma}}$ is obtained, the high-resolution hydrostatic pressure field $\hat{p}$ is calculated by
\begin{equation} \label{p_sr}
    \hat{p} = -\frac1{3}\text{tr}\left(\hat{\boldsymbol{\sigma}}\right) = -\frac1{3} \left(\hat{\sigma}_{11} + \hat{\sigma}_{22} + \hat{\sigma}_{33}\right).
\end{equation}

As for the case associated with in an interface cell $\tilde{\Upsilon}$ as shown in Fig.~\ref{sec2-voids_on_GB}(b), the hydrostatic pressure can be computed similarly as in the case of bulk cells. But the PD - bubble interaction is only formulated on the interface section $\Gamma_0$, and this formualtes a steady-state equation for $\hat{\tilde{c}}_{\alpha}$, which is given by
\begin{equation} \label{eqn_flux_gb_sr}
    \tilde{\nabla}_{\bar{\mathbf{X}}} \cdot \left(\tilde{\nabla}_{\bar{\mathbf{X}}} \hat{\tilde{c}}_{\alpha} - \frac{\lambda_{\alpha} \Delta v_{\alpha} \hat{\tilde{c}}_{\alpha}}{k_0T} \tilde{\nabla}_{\bar{\mathbf{X}}} \hat{p}\right) = 0, \quad \text{on }\Gamma_0,
\end{equation}
where $\tilde{\nabla}_{\bar{\mathbf{X}}} = \left(\frac{\partial}{\partial \bar{X}_1}, \frac{\partial}{\partial \bar{X}_2}\right)^{\mathrm{T}}$.

\subsubsection{Linkage with mean-field quantities through boundary conditions}
First we consider the boundary conditions for the elasticity problem. On the interior spherical surface of a bulk cell, the gas pressure due to the NGAs within should be accommodated \cite{was_fundamentals_2017}, and we thus have
\begin{equation} \label{BC_stress_inner_sr}
    \left.\hat{\boldsymbol{\sigma}}\right|_{\partial\mathcal{O}(\bar{r}^*)}\cdot \mathbf{m} = -\left( \frac{3n k_0T}{4\pi r^3} - \frac{2\gamma}{r}\right) \mathbf{m},
\end{equation}
where $\mathbf{m}$ is the unit vector normal to the spherical surface and pointing towards the centre; $\gamma$ ($\sim$ N/m) is a surface tension coefficient. The first term on the right side of Eq.~\eqref{BC_stress_inner_sr} measures the gas pressure to the NGAs within the (rescaled) bubble of interest, and the second term stems from the surface tension effect. On the outer surface of a cell, the LHR stress field should be matched with the mean-field stress.

As for the boundary conditions of the PDE problem about the LHR concentration $\hat{c}_{\alpha}$, it should equal the thermally equilibrated value on the interior surface, that is,
\begin{equation} \label{BC_c_inner_sr}
    \left.\hat{c}_{\alpha}\right|_{\partial \mathcal{O}(\bar{r}^*)} = 0.
\end{equation}
The other boundary condition comes from the fact that the mean-field fractional concentration $c_{\alpha}$ should equal the mean-value of the LHR concentration over the bulk cell, i.e.,
\begin{equation} \label{c_macro_to_micro}
    c_{\alpha} = \int_{\Upsilon} \hat{c}_{\alpha}(\mathbf{x};\bar{\mathbf{X}})\, \mathrm{d}\bar{\mathbf{X}}.
\end{equation}
Besides, on the outer surface of the bulk cell, we let the fractional PD concentration stay uniform, which is formulated by
\begin{equation} \label{BC_c_outer_sr}
    \left.\hat{c}_{\alpha}\right|_{\bar{X}_i=\pm\frac1{2}} = c_{\alpha}^0, \text{ for }i=1,2,3,
\end{equation}
where the parameter $c_{\alpha}^0$ should be evaluated such that Eq.~\eqref{c_macro_to_micro} holds.

The boundary conditions for completing the LHR problem defined in an interface cell can be set up likewise.

Therefore, a set of LHR cell problems describing the local PD-bubble interactions are established, with the onsite mean-field quantities, such as $c_{\alpha}$ and the mean-field stress field $\boldsymbol{\sigma}$, being their controlling parameters.

\subsubsection{Definite but implicit expressions for the sink terms}
Once the LHR quantities of $\hat{c}_{\alpha}$ are computed, the sink term $A_{\alpha}$ of Eq.~\eqref{sec2-c_gb-evlution-eq} should equal the rate in the (fractional) number of PDs that are absorbed by the (rescaled) sphere located at the bulk cell centre. Mathematically, this is described by
\begin{equation} \label{A_macro_to_micro}
\begin{aligned}
    k_{\alpha\text{B}}^2 D_{\alpha} c_{\alpha} = - \varrho^{\frac2{3}} D_{\alpha}\int_{\partial\mathcal{O}(\bar{r}^*)} \left(\frac{\partial \hat{c}_{\alpha}}{\partial m} - \frac{\lambda_{\alpha}\Delta v_{\alpha} \hat{c}_{\alpha}}{k_0T} \cdot \frac{\partial \hat{p}}{\partial m}\right) \, \mathrm{d} S_{\bar{\mathbf{X}}},
\end{aligned}
\end{equation}
where $\frac{\partial}{\partial m} = \mathbf{m}\cdot\nabla_{\bar{\mathbf{X}}}$ denotes the spatial derivative along the spherical inward normal direction $\mathbf{m}$; $\mathrm{d} S_{\bar{\mathbf{X}}}$ represents a non-dimensional infinitesimal area on the spherical surface.

Upon integration with respect to $\bar{\mathbf{X}}$, $k_{\alpha\text{B}}^2$ in Eq.~\eqref{A_macro_to_micro} should solely depend on the macroscopic position $\mathbf{x}$. To be more precise, the sink terms depend on the geometrical feature of the bulk cell $\Upsilon$ (determined by the mean-field quantity $\varrho^{\frac1{3}} r$), as well as the boundary conditions defining the cell problems (formulated in terms of the mean-field quantities of $n$, $c_{\alpha}$ and $\boldsymbol{\sigma}$). This theoretically implies a functional relationship of
\begin{equation} \label{A_implicit_relation}
    k_{\alpha\text{B}}^2  = \chi_{\alpha}\left(c_{\alpha}, \boldsymbol{\sigma} , \varrho , r , n ; \Lambda\right),
\end{equation}
where the vector $\Lambda$ contains the temperature field $T$ (which is set as a constant here) and all associated material constants. To evaluate the functional relationship of $\chi_{\alpha}$, we turn to machine learning tools.

Meanwhile, the sink strength term $\tilde{k}_{\alpha\text{B}}^2$ appearing in Eq.~\eqref{sec2-c_gb-evlution-eq} on a crystalline interface can be calculated similarly by
\begin{equation} \label{A_macro_to_micro_gb}
\begin{aligned}
    \tilde{k}_{\alpha\text{B}}^2 \tilde{D}_{\alpha} c_{\alpha} = - \tilde{\rho} \tilde{D}_{\alpha} \int_{\partial\mathcal{O}_{\text{2d}}(\bar{\tilde{r}}^*)} \left(\frac{\partial \hat{\tilde{c}}_{\alpha}}{\partial m} - \frac{\lambda_{\alpha}\Delta v_{\alpha} \hat{\tilde{c}}_{\alpha}}{k_0T} \cdot\frac{\partial \hat{p}}{\partial m}\right)\, \mathrm{d} s_{\bar{\mathbf{X}}},
\end{aligned}
\end{equation}
where $\frac{\partial}{\partial m} = m_1\frac{\partial}{\partial \bar{X}_1} + m_2\frac{\partial}{\partial \bar{X}_2}$; $\mathrm{d} s_{\bar{\mathbf{X}}}$ is a non-dimensional infinitesimal arclength on $\partial \mathcal{O}_{\text{2d}}(\bar{\tilde{r}}^*)$. Another implicit relationship implied by
\begin{equation} \label{A_implicit_relation_gb}
    \tilde{k}_{\alpha\text{B}}^2 = \tilde{\chi}_{\alpha}\left(\tilde{c}_{\alpha}, \boldsymbol{\sigma}, \tilde{\rho} , \tilde{r} , \tilde{n} ; \Lambda\right)
\end{equation}
should also exist.

With the sink strength calculated, one can also evolve the corresponding bubble distribution, which may be formulated by the so-called master equations. However, as bubbles are treated as immobile sinks here, the evolution of their spatial distribution can be simply described by a set of ordinary differential equations in $t$ (with $\mathbf{x}$ just parameters). For instance, the net income of vacancies less self-interstitial atoms results in size growth of bubbles. Thus we have
\begin{equation} \label{eqn_r}
    \frac{4\pi}{3}\frac{\partial \left(\varrho r^3\right)}{\partial t} = k_{\text{vB}}^2 D_{\text{v}} c_{\text{v}} - k_{\text{iB}}^2 D_{\text{i}} c_{\text{i}}.
\end{equation}
Moreover, the NGA number distribution also evolves as a bubble absorbs surrounding NGAs, and we have
\begin{equation} \label{eqn_n}
    \frac{\partial \left(\varrho n\right)}{\partial t} = \frac{k_{\text{gB}}^2 D_{\text{g}} c_{\text{g}}}{v_0},
\end{equation}
where $v_0$ is recalled to be the volume occupied by a single atom in the hosting materials. The ODEs describing the bubble evolution on interfaces can be derived likewise.

\section{Machine learning\label{Sec_ML}}
In this section, machine learning models are adopted to further express the implicit functional relations of $A_{\alpha}$ and $\tilde{A}_{\alpha}$ formulated by Eqs.~\eqref{A_implicit_relation} and \eqref{A_implicit_relation_gb}, respectively. We will first show that a rescaling of the cell problems outlined in Sec.~\ref{Sec_homogenisation} helps in 1) minimising the number of input arguments for machine learning; 2) devising expressions in consistency with the format of the sink strengths appearing in the rate equations. Nineteen machine learning algorithms are tested, and the rational quadratic Gaussian procession regression (GPR) scheme is selected for evaluating the sink strengths.

\subsection{Rescaling the cell problems\label{Sec_reduction}}
In general, the more input arguments are there for regression analysis, the more data are needed. Thus a reduction in the number of input arguments for machine learning, whenever necessary, is often meaningful. Again we start with the case of bulk cells. In fact, certain correlations exist among the continuum quantities serving as input arguments of Eq.~\eqref{A_implicit_relation}. To sort out such inter-relationships, the cell problems outlined in Sec.~\ref{Sec_homogenisation} are rescaled further. With the cumbersome details given in Appendix, we find that upon rescaling, the number of input arguments for machine learning is decreased to five, and the implicit relation given by Eq.~\eqref{A_implicit_relation} can be simplified as
\begin{equation} \label{sink_strength}
k_{\alpha \text{B}}^2 = \varrho^{\frac2{3}} \cdot \left(1-\frac{4\pi \varrho r^3}{3}\right) \cdot F\left(\beta_1, \beta_2, \beta_3, \beta_4, \beta_5\right),
\end{equation}
where
\begin{subequations} \label{beta_appendix}
\begin{equation} \label{beta1-2_appendix}
    \beta_1 = \varrho^{\frac1{3}}r , \, \beta_2 = \lambda_{\alpha} \left(\frac{3n \Delta v_{\alpha}}{4\pi r^3} - \frac{2\gamma \Delta v_{\alpha}}{k_0T r}\right),
\end{equation}
\begin{equation} \label{beta3-5_appendix}
    \beta_3 = \frac{\sigma_1 \lambda_{\alpha}\Delta v_{\alpha}}{k_0T}, \, \beta_4 = \frac{\sigma_2 \lambda_{\alpha}\Delta v_{\alpha}}{k_0T}, \, \beta_5 = \frac{\sigma_3 \lambda_{\alpha}\Delta v_{\alpha}}{k_0T}
\end{equation}
\end{subequations}
with $\sigma_1 \ge \sigma_2 \ge \sigma_3$ denoting the principle components of the local mean-field stress. Note that the expression for the sink strength given by Eq.~\eqref{sink_strength} no longer depends on the local fractional concentration $c_{\alpha}$, but only on the local bubble configurations. This coincides with the normal settings for sink strength terms in REs.

Now a curriculum for machine learning is devised, that is, to regress for a five-input function $F(\boldsymbol{\beta})$ given by
\begin{equation} \label{F_star_def}
    F(\boldsymbol{\beta}) = F(\beta_1,\beta_2,\beta_3,\beta_4,\beta_5).
\end{equation}

As for the cases in association with interface cells, we can similarly derive
\begin{equation} \label{sink_strength_gb}
    \tilde{k}_{\alpha \text{B}}^2 = \tilde{\rho} \cdot \left(1- \tilde{\rho} \tilde{r}^2\right) \cdot \tilde{F}\left(\tilde{\beta}_1, \tilde{\beta}_2, \tilde{\beta}_3, \tilde{\beta}_4, \tilde{\beta}_5\right).
\end{equation}
where $\tilde{\beta}_1 = \tilde{\rho}^{\frac1{2}}\tilde{r}$ and $\tilde{\beta_2}$ to $\tilde{\beta}_5$ are expressed the same as the corresponding $\beta_s$ given by Eqs.~\eqref{beta_appendix}.

\subsection{Data collection}
\subsubsection{Curriculum for data collection\label{Sec_curriculum}}
A curriculum guiding the data generation process is thus summarised by the flow chart shown by Fig.~\ref{sec6-MLflowchat}. The derivations underpinning this curriculum can be found in Appendix, and here we simply list the five key steps for data collection.
\begin{figure}[!ht]
  \centering
  \includegraphics[width=0.85\textwidth]{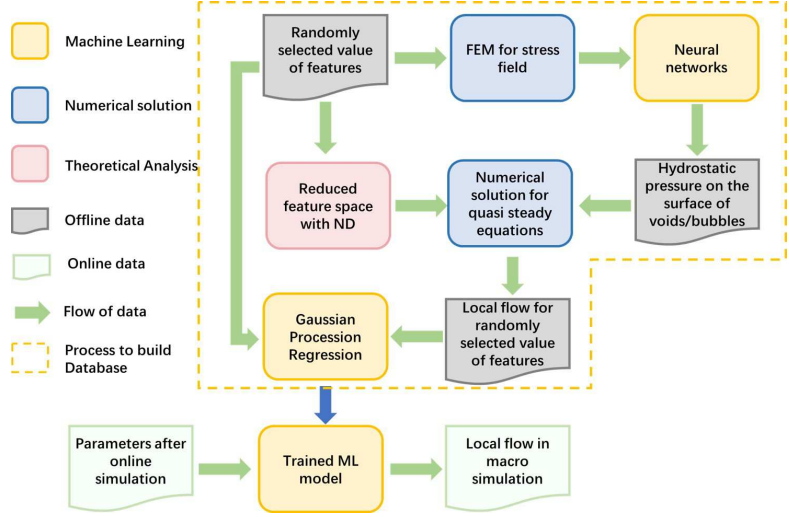}
  \caption{A curriculum of flow chart guiding the implementation of machine learning tools}\label{sec6-MLflowchat}
\end{figure}

In step 1, we generate five random numbers and assign them to the entries of the (five-dimensional) vector $\boldsymbol{\beta}$. The corresponding domain for solving the (rescaled) LHR cell problems is then determined as
\begin{equation}
    \Upsilon = \left[-\frac1{2}, \frac1{2}\right]^3 \setminus \mathcal{O}(\beta_1).
\end{equation}

In step 2, we solve a rescaled elasticity problem governed by
\begin{equation}\label{problem_mechanical_rescale}
    \left\{\begin{aligned}
    & \nabla_{\bar{\mathbf{X}}} \cdot \boldsymbol{\sigma}^* = \boldsymbol{0}, \quad \text{in }\Upsilon;\\
    & \boldsymbol{\sigma}^* = \frac{\lambda v_0}{k_0T}  \text{tr}(\nabla_{\bar{\mathbf{X}}}\mathbf{u}^*) + \frac{\mu v_0}{k_0T} \left(\nabla_{\bar{\mathbf{X}}}\mathbf{u}^* + (\nabla_{\bar{\mathbf{X}}} \mathbf{u}^*)^{\mathrm{T}}\right) , \quad \text{in }\Upsilon;\\
    & \boldsymbol{\sigma}^* \cdot \mathbf{m} = -\beta_2, \quad \text{on }\partial\mathcal{O}(\beta_1);\\
    & \left.\boldsymbol{\sigma}^* \mathbf{m}\right|_{\bar{X}_1 = \pm\frac1{2}} = \pm\beta_3\mathbf{e}^1,\, \left.\boldsymbol{\sigma}^* \mathbf{m}\right|_{\bar{X}_2 = \pm\frac1{2}} = \pm\beta_4 \mathbf{e}^2, \, \left.\boldsymbol{\sigma}^* \mathbf{m}\right|_{\bar{X}_3 = \pm\frac1{2}} = \pm \beta_5 \mathbf{e}^3,
    \end{aligned}\right.
\end{equation}
where $\{\mathbf{e}^i\}_{i=1}^3$ form an orthogonal triad; $\mathbf{u}^*$ represents a rescaled LHR displacement field.

In step 3, a Laplacian equation is solved for the rescaled hydrostatic pressure $p^*$, given by
\begin{equation}\label{problem_pressure_rescale}
    \left\{\begin{aligned}
    & \nabla_{\bar{\mathbf{X}}}^2 p^* = 0, \quad \text{in }\Upsilon;\\
    & \left.p^*\right|_{\partial\mathcal{O}(\beta_1)} = - \left. \frac{\text{tr}(\boldsymbol{\sigma}^*)}{3}\right|_{\partial\mathcal{O}(\beta_1)};\\
    & \left.p^*\right|_{\bar{X}_i = \pm \frac1{2}} = - \left. \frac{\text{tr}(\boldsymbol{\sigma}^*)}{3}\right|_{\bar{X}_i = \pm \frac1{2}},
    \end{aligned}\right.
\end{equation}
for $i=1$, $2$ and $3$.

In step 4, we consider solving an equilibrium-state problem for the rescaled LHR fractional concentration $c^*$ governed by
\begin{equation}\label{problem_c_rescale}
    \left\{\begin{aligned}
    & \nabla_{\bar{\mathbf{X}}} \cdot \left(\nabla_{\bar{\mathbf{X}}} c^* - c^* \nabla_{\bar{\mathbf{X}}} p^*\right) = 0, \quad \text{in }\Upsilon;\\
    & c^* = 0, \quad \text{on }\partial\mathcal{O}(\beta_1);\\
    & \left.c^*\right|_{\bar{X}_i = \pm \frac1{2}} = 1, \quad \text{for }i=1,2,3.
    \end{aligned}\right.
\end{equation}

In step 5, we determine the corresponding value of $F_{\alpha}$ by
\begin{equation} \label{F_star}
    F(\boldsymbol{\beta}) = - \frac1{\int_{\Upsilon} c^* \,\mathrm{d} \bar{\mathbf{X}}} \int_{\partial \mathcal{O}(\beta_1)} \mathbf{m}\cdot \nabla_{\bar{\mathbf{X}}}c^* \, \mathrm{d} S_{\bar{\mathbf{X}}}.
\end{equation}

Several issues regarding the curriculum for data collection are noted. First, for simplicity, we assume that a bulk cell is always in alignment with the local principle stress components. The present method also works for more complicated stress situations, where the number of input arguments should be more. Second, the (rescaled) hydrostatic pressure field $p^*$ is not computed by the stress gradient, as suggested by \eqref{p_sr}. This is because computation of stress gradients requires the resulting displacement field to be differentiable twice, which is not favoured by general finite element algorithms using piecewisely linear basis functions. Rather, $p^*$ is calculated through a Laplacian equation \cite{ZhuYC_JMPS2017a}. Third, the values of $k_{\alpha \text{B}}^2$ at different (mean-field) locations should be identical, provided that all the continuum variables involved stay the same. Thus the curriculum is position-independent. Finally, no data exchange takes place between data preparation for different scenarios. Thus computational parallelism may be employed to speed up the data preparation process.

As for the cases of interface cells, the corresponding curriculum is the same until step 3. In step 4, one needs to solve a concentration equilibrium problem on a hollow square of $\Gamma_0$ by Eq.~\eqref{Gamma0}, i.e.
\begin{equation}\label{problem_c_rescale_gb}
    \left\{\begin{aligned}
    & \tilde{\nabla}_{\bar{\mathbf{X}}} \cdot \left(\tilde{\nabla}_{\bar{\mathbf{X}}} \tilde{c}^* - \tilde{c}^* \nabla_{\bar{\mathbf{X}}} p^*\right) = 0, \quad \text{in }\Upsilon;\\
    & \tilde{c}^* = 0, \quad \text{on }\partial\mathcal{O}_{\text{2d}}(\beta_1);\\
    & \left.\tilde{c}^*\right|_{\bar{X}_i = \pm \frac1{2}} = 1,
    \end{aligned}\right.
\end{equation}
for $i=1$ and $2$. With $\tilde{c}^*$ determined, we can evaluate $\tilde{F}(\boldsymbol{\beta})$ by
\begin{equation} \label{F_star_gb}
    \tilde{F}(\boldsymbol{\beta}) = - \frac1{\int_{\Gamma_0} \tilde{c}^* \,\mathrm{d} \bar{X}_1 \mathrm{d} \bar{X}_2} \int_{\partial \mathcal{O}_{\text{2d}}(\beta_1)} \tilde{\mathbf{m}}\cdot \tilde{\nabla}_{\bar{\mathbf{X}}}\tilde{c}^* \, \mathrm{d} s_{\bar{\mathbf{X}}}.
\end{equation}

\subsubsection{Results}
Following the key steps listed above, 1000 data points are generated for the cases of bulk cells, and 1241 points are generated for the cases of interface cells. For obtaining them, the values of inputs $\boldsymbol{\beta}$ are randomly generated following certain uniform distributions given by
\begin{equation} \label{input_generation}
    \beta_1 \thicksim \mathcal{U}\left(0.1,0.45\right), \qquad\beta_s \thicksim \mathcal{U}  \left(-2,2\right),
\end{equation}
for $s=2$, $\cdots$, $5$.

Note that $\beta_1 = \varrho^{\frac{1}{3}}r $ or $\tilde{\rho}^{\frac{1}{2}}\tilde{r}$ has a definite range, that is, $\beta_1\in(0,0.5)$. But we did not consider the interval $(0.45,0.5)$, since the two neighbouring bubbles are so closely spaced that cracks may have already initiated then. Moreover, the situations where $\beta_1\in(0,0.1)$ are also not considered, as the finite element meshes become quite dense then. But it is known that $k_{\alpha \text{B}}^2\rightarrow0$, as $\beta_1\rightarrow0$. Thus linear interpolation is used to patch the case of $\beta_1\in(0,0.1)$ with the outcomes from machine learning models.

Note that for most machine learning tools, it is preferential to normalise the individual inputs within the interval of unity $[0,1]$. Here we use the symbol $\bar{\beta}_s$ to denote the normalised quantities of $\beta_s$, for $s=1$, $\cdots$, $5$.

\begin{figure}[!ht]
\centering
\subfigure[]{
\includegraphics[width=0.3\textwidth]{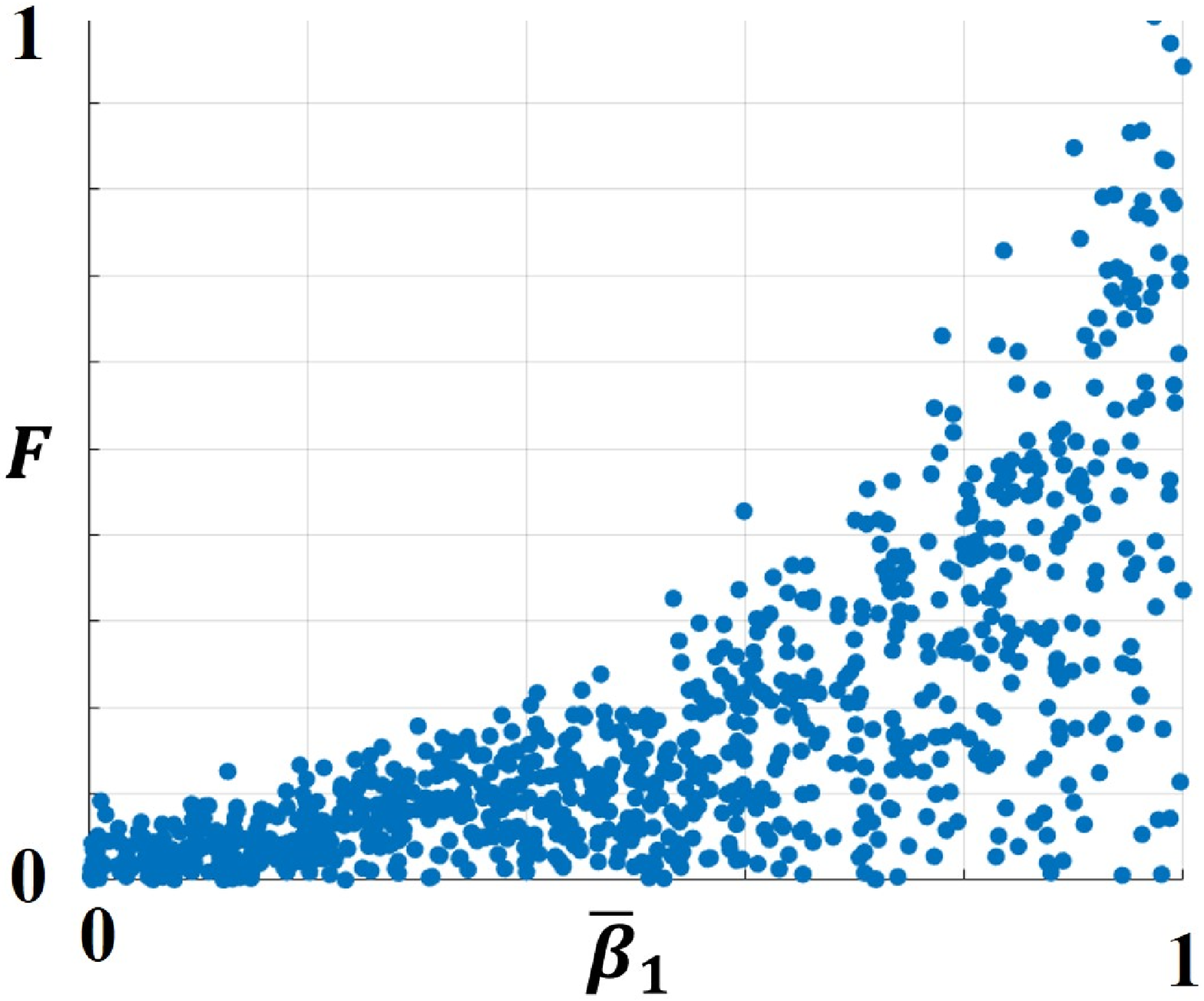}
}
\subfigure[]{
\includegraphics[width=0.3\textwidth]{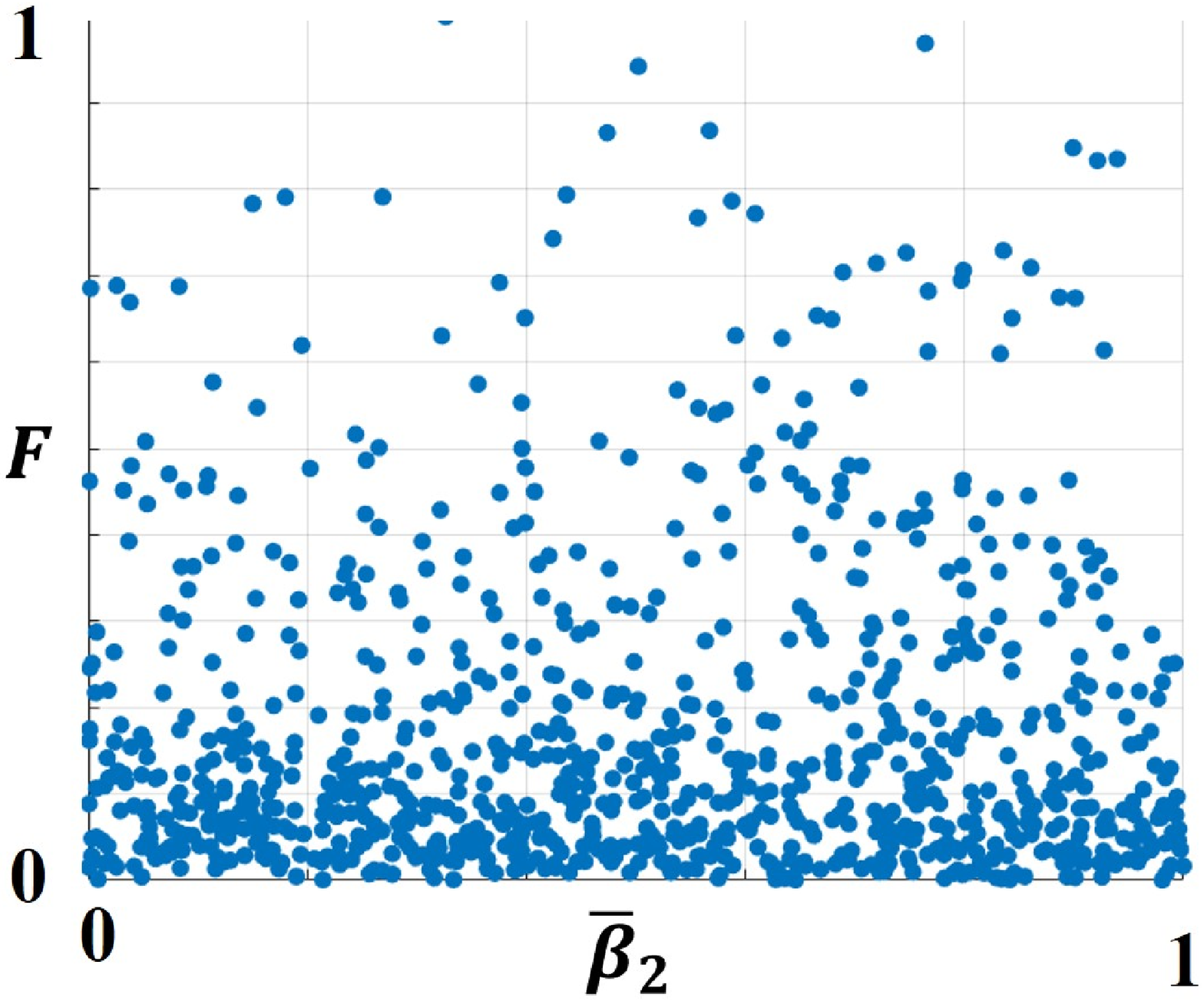}
}
\subfigure[]{
\includegraphics[width=0.3\textwidth]{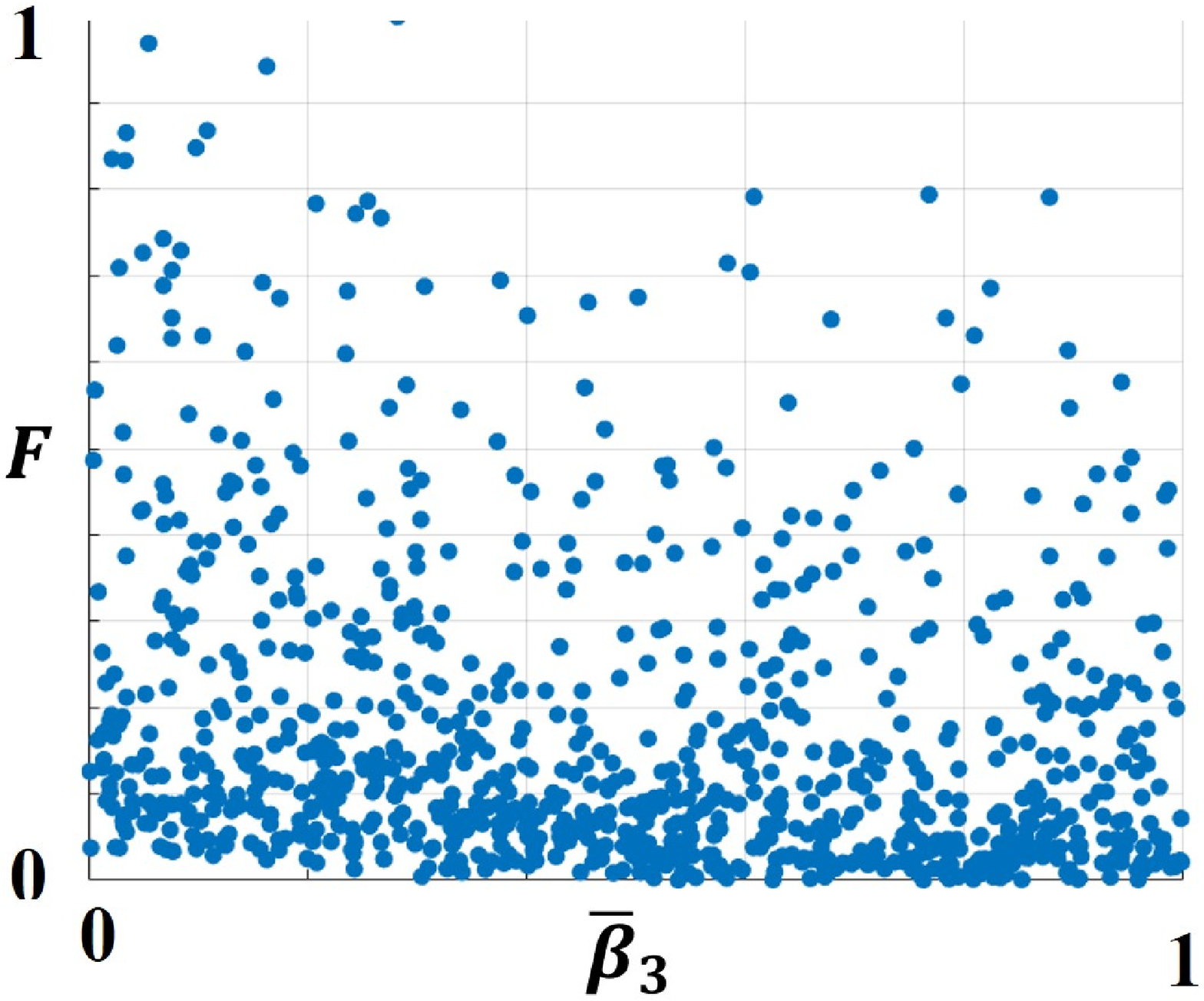}
}
\subfigure[]{
\includegraphics[width=0.3\textwidth]{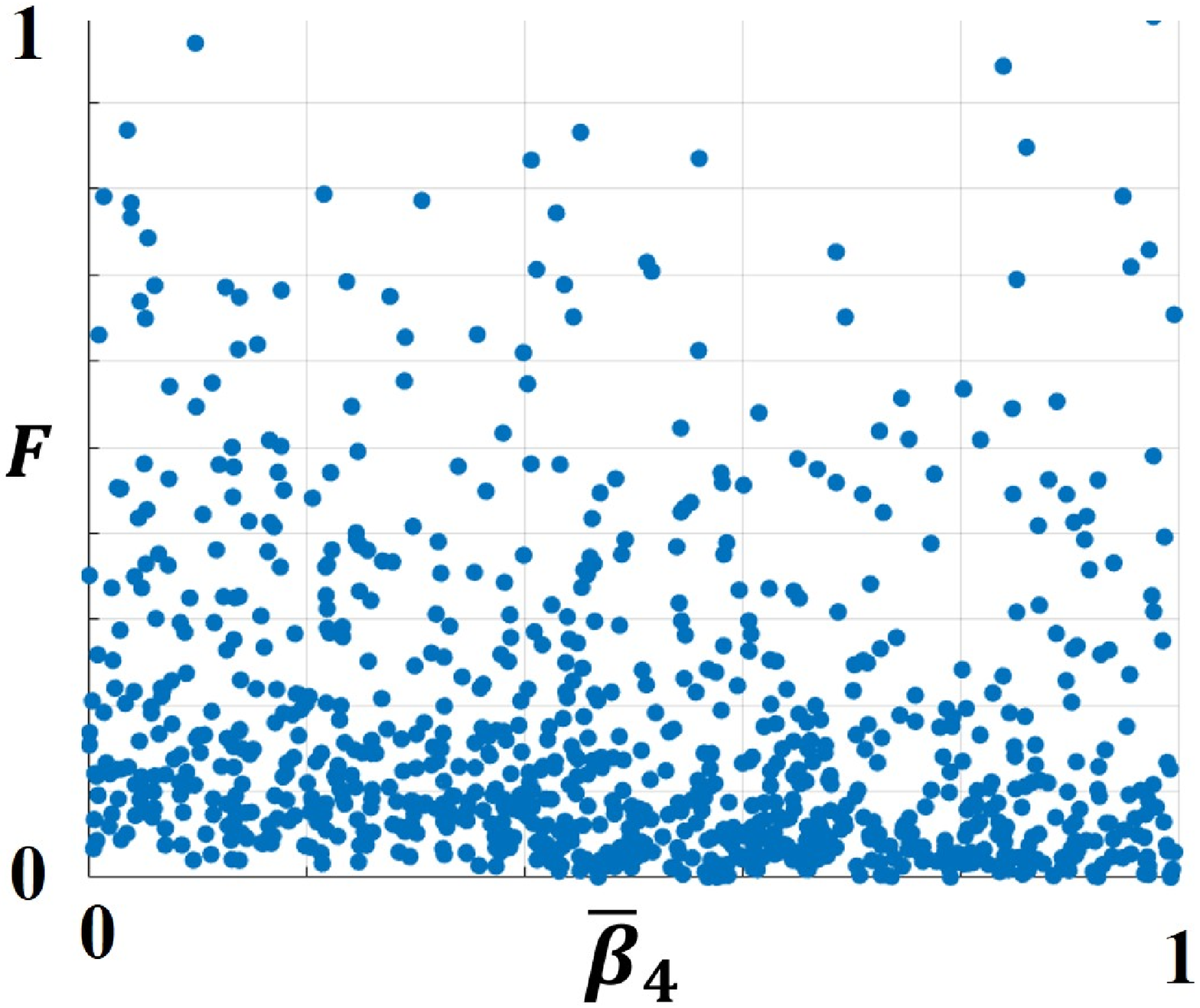}
}
\subfigure[]{
\includegraphics[width=0.3\textwidth]{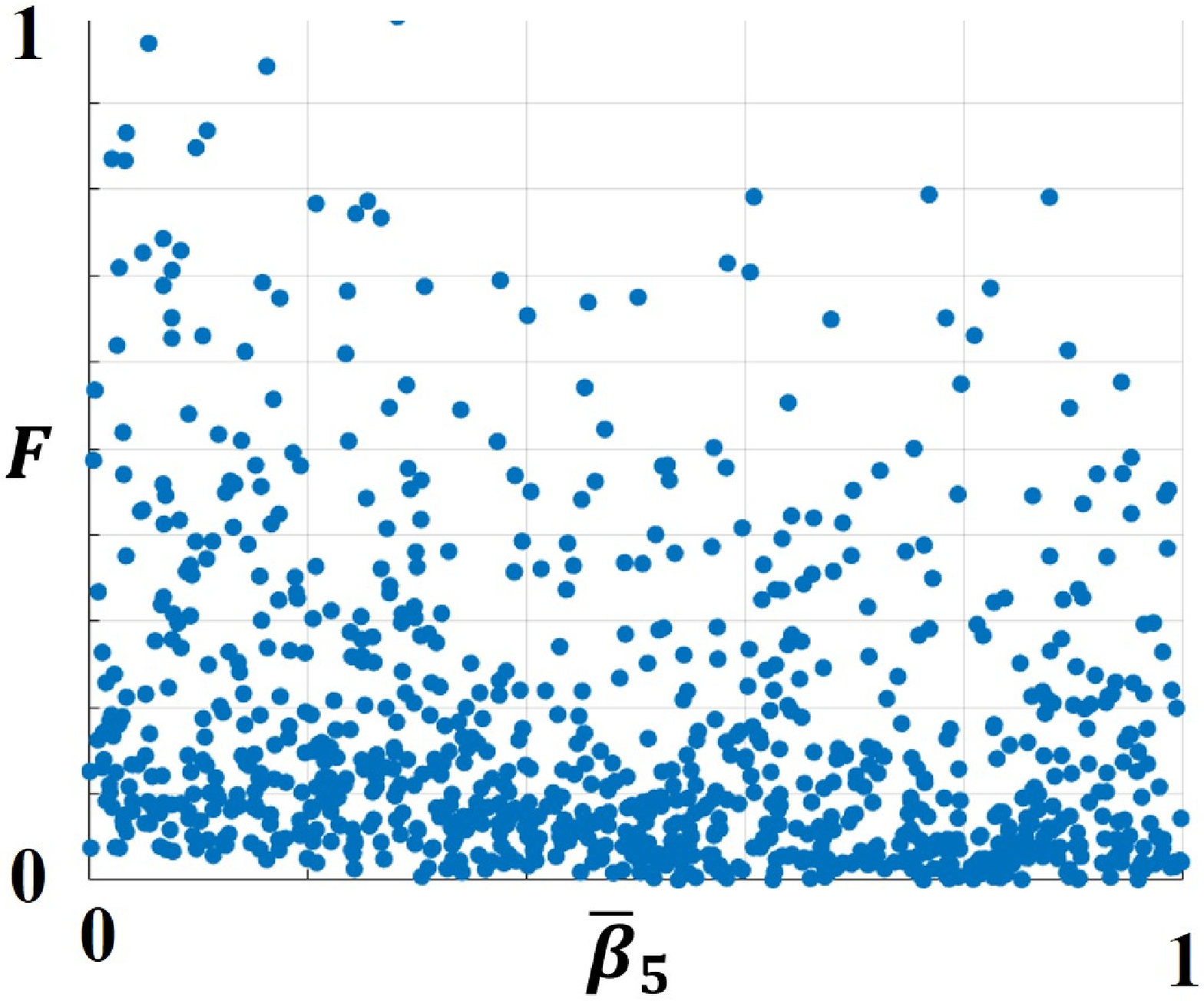}
}
\caption{Projection of data points onto the $F-\bar{\beta}_s$ for the cases of bulk cells: $F$ denotes the output from the functional relationship defined by Eq.~\eqref{F_star_def}, and $\bar{\beta}_s$ denotes the normoalised quantities of $\beta_s$ with $s=1$, $\cdots$, $5$.}
\label{sec6-bulk-visualisation}
\end{figure}

For visualisation, the data points collected for the cases of bulk cells are projected onto the output-input planes of $F-\bar{\beta}_s$, for $s=1$, $\cdots$, $5$, where $F$ denotes the outcome of the functional relationship given by Eq.~\eqref{F_star_def}, and the results are shown in Fig.~\ref{sec6-bulk-visualisation}.
Note that the $F-\bar{\beta}_1$ trend is relatively clear. This is because a larger $\bar{\beta}_1$, which is the normalised radius of the hollow region in $\Upsilon$, gives rise to a larger surface area for PD absorption.

\begin{figure}[!ht]
\centering
\subfigure[]{
\includegraphics[width=0.3\textwidth]{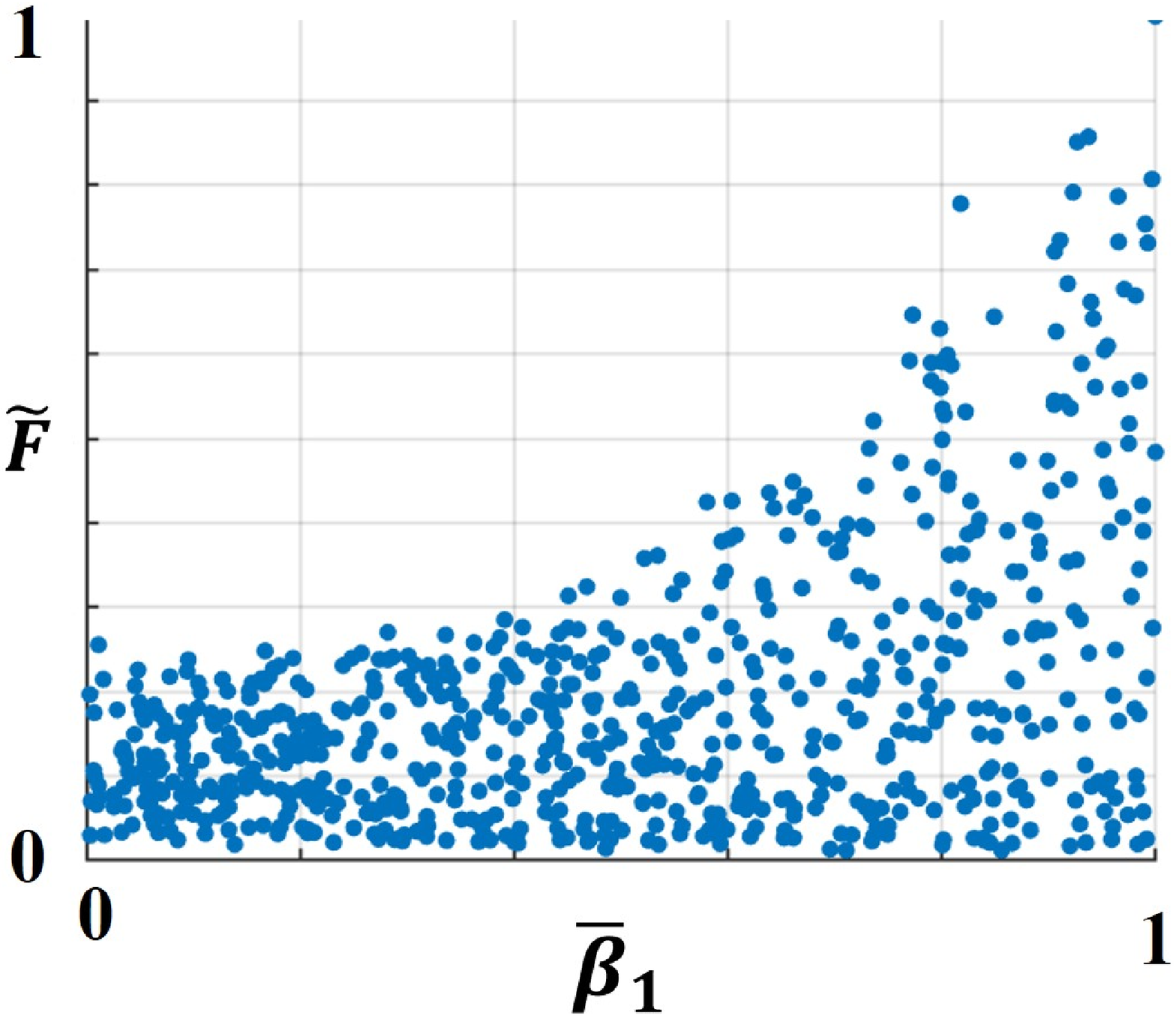}
}
\subfigure[]{
\includegraphics[width=0.3\textwidth]{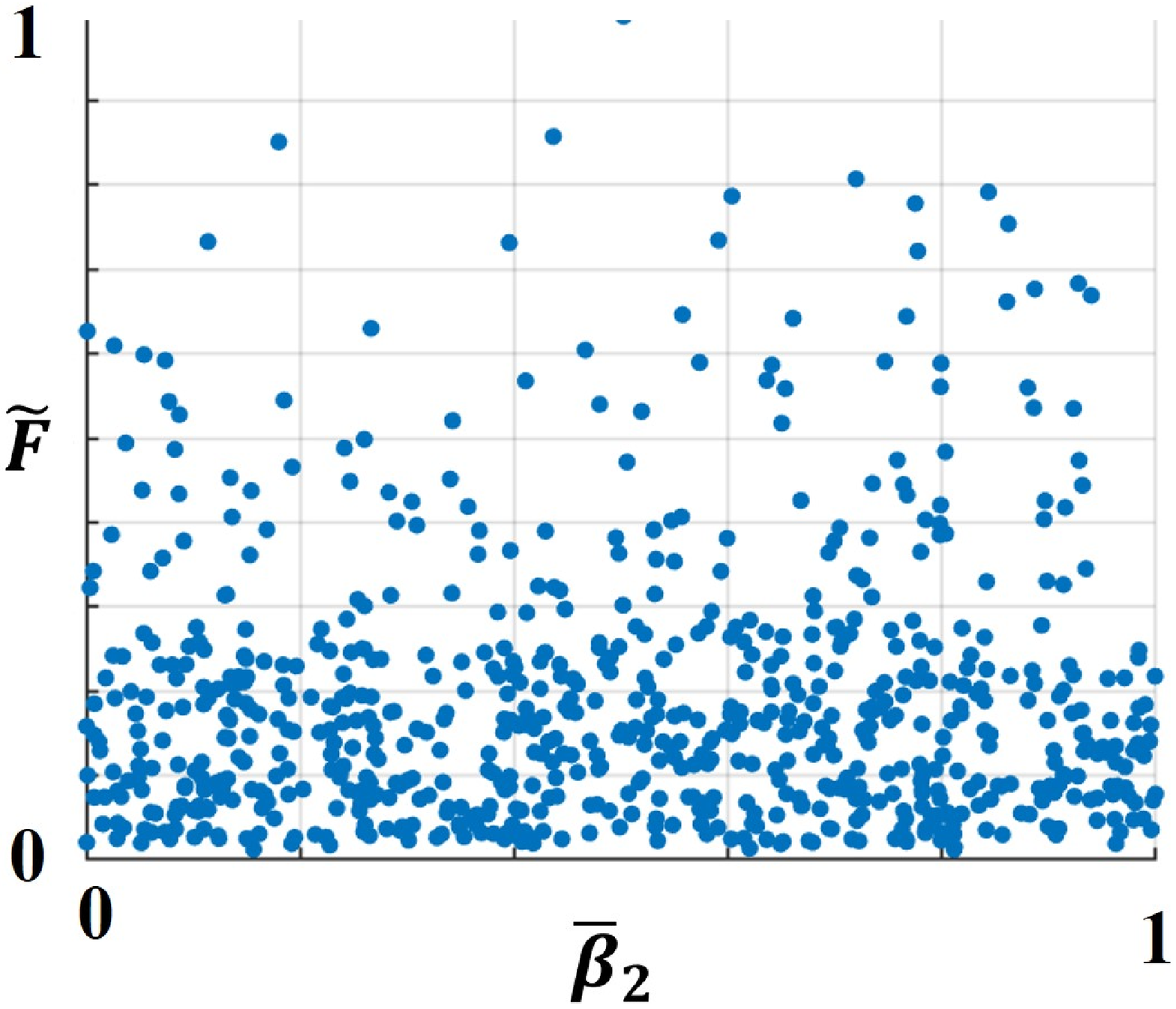}
}
\subfigure[]{
\includegraphics[width=0.3\textwidth]{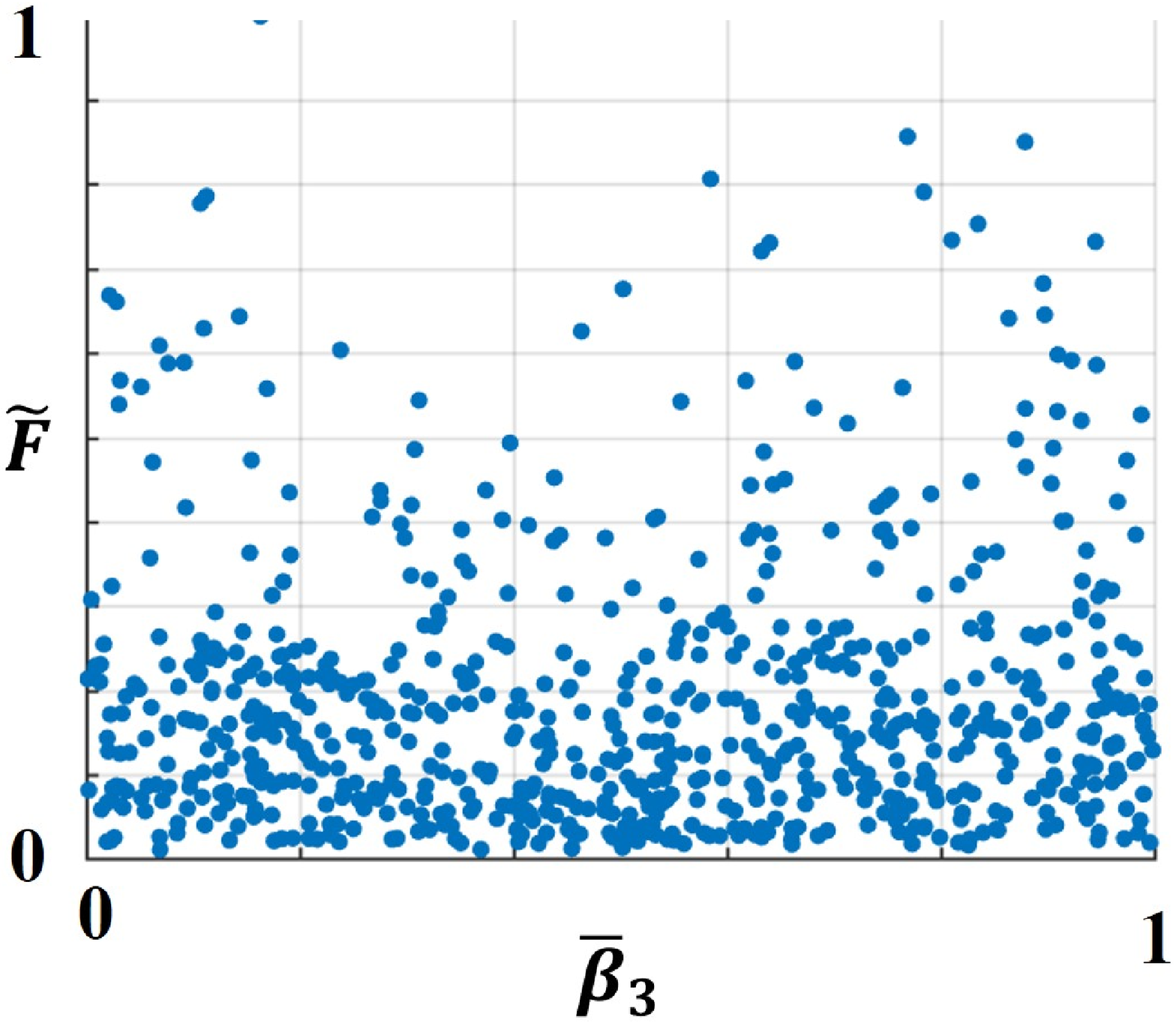}
}
\subfigure[]{
\includegraphics[width=0.3\textwidth]{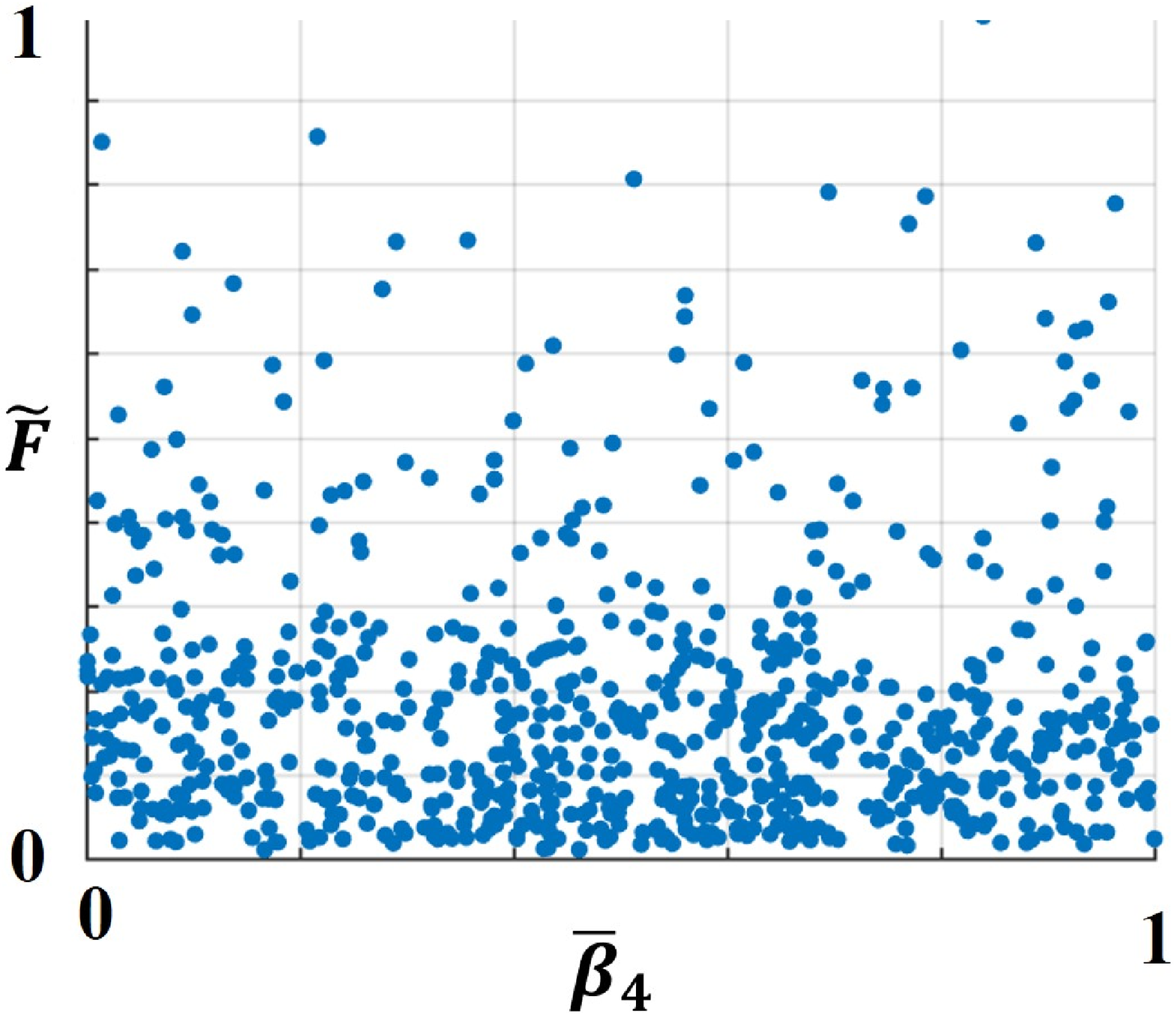}
}
\subfigure[]{
\includegraphics[width=0.3\textwidth]{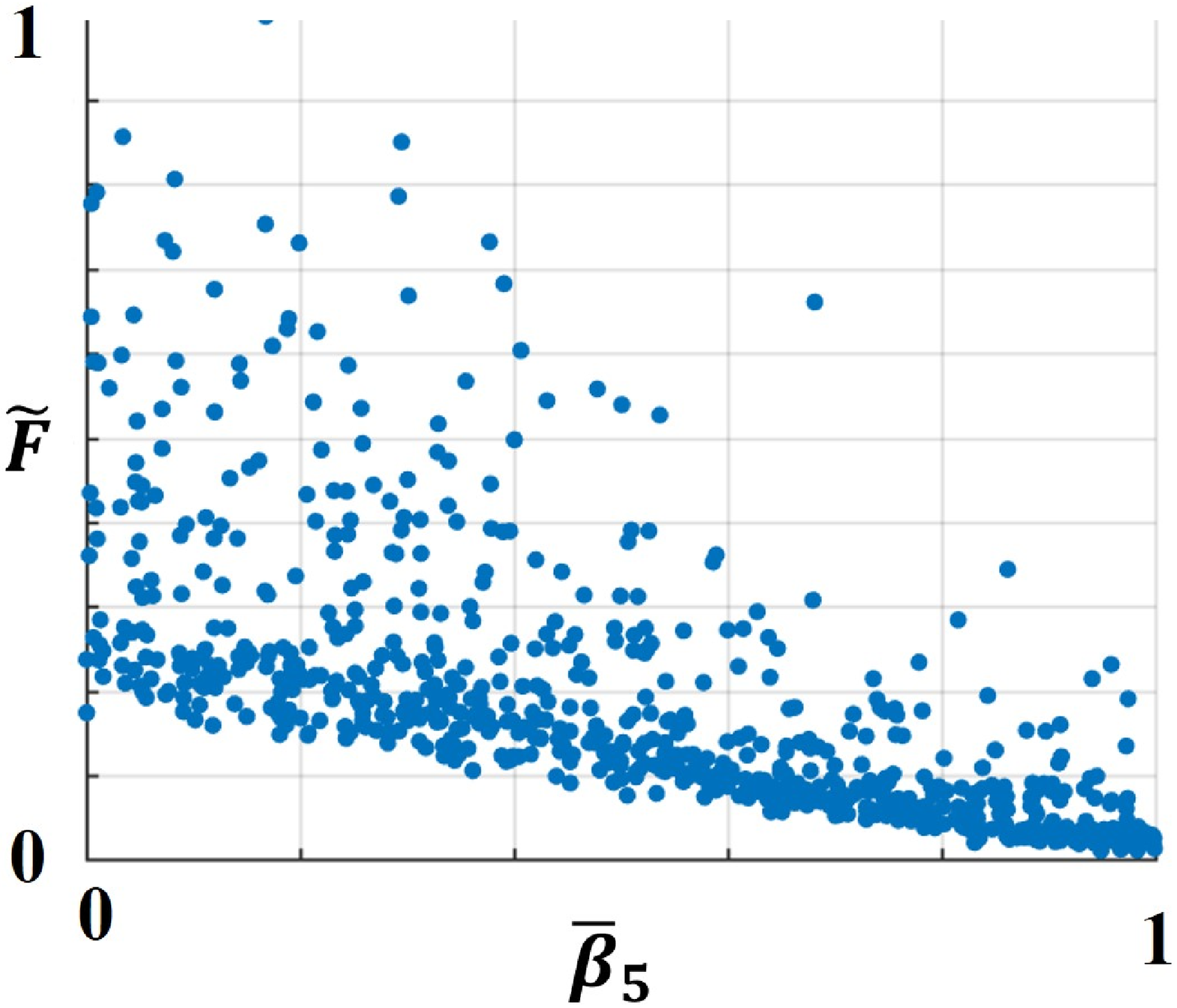}
}
\caption{Projection of data points onto the $\tilde{F}-\bar{\beta}_s$ for the cases of bulk cells: $F$ denotes the output from the functional relationship defined by Eq.~\eqref{F_star_gb}, and $\bar{\beta}_s$ denotes the normoalised quantities of $\beta_s$ with $s=1$, $\cdots$, $5$.}
\label{sec6-GB-visualisation}
\end{figure}

In a similar sense, the data points collected for the cases of interface cells are visualised in Fig.~\ref{sec6-GB-visualisation}.
Note that the difference lying between the distributional profiles of the two data sets should result from the effect of bulk diffusion against diffusion simply on the interface.


\subsection{Selection of machine learning methods}\label{sec-ML_selection}
A number of machine learning models can be trained based on the obtained data set. However, it remains unclear at this stage, which machine learning models are more suitable for analysing the present problem. Note that the functional relation given by Eq.~\eqref{F_star} should possess certain degree of continuity over its input $\mathbf{\beta}$. Hence a machine learning model that maintains functional continuities may be preferred for the present studies.

\begin{figure}[!ht]
  \centering
  \includegraphics[width=\textwidth]{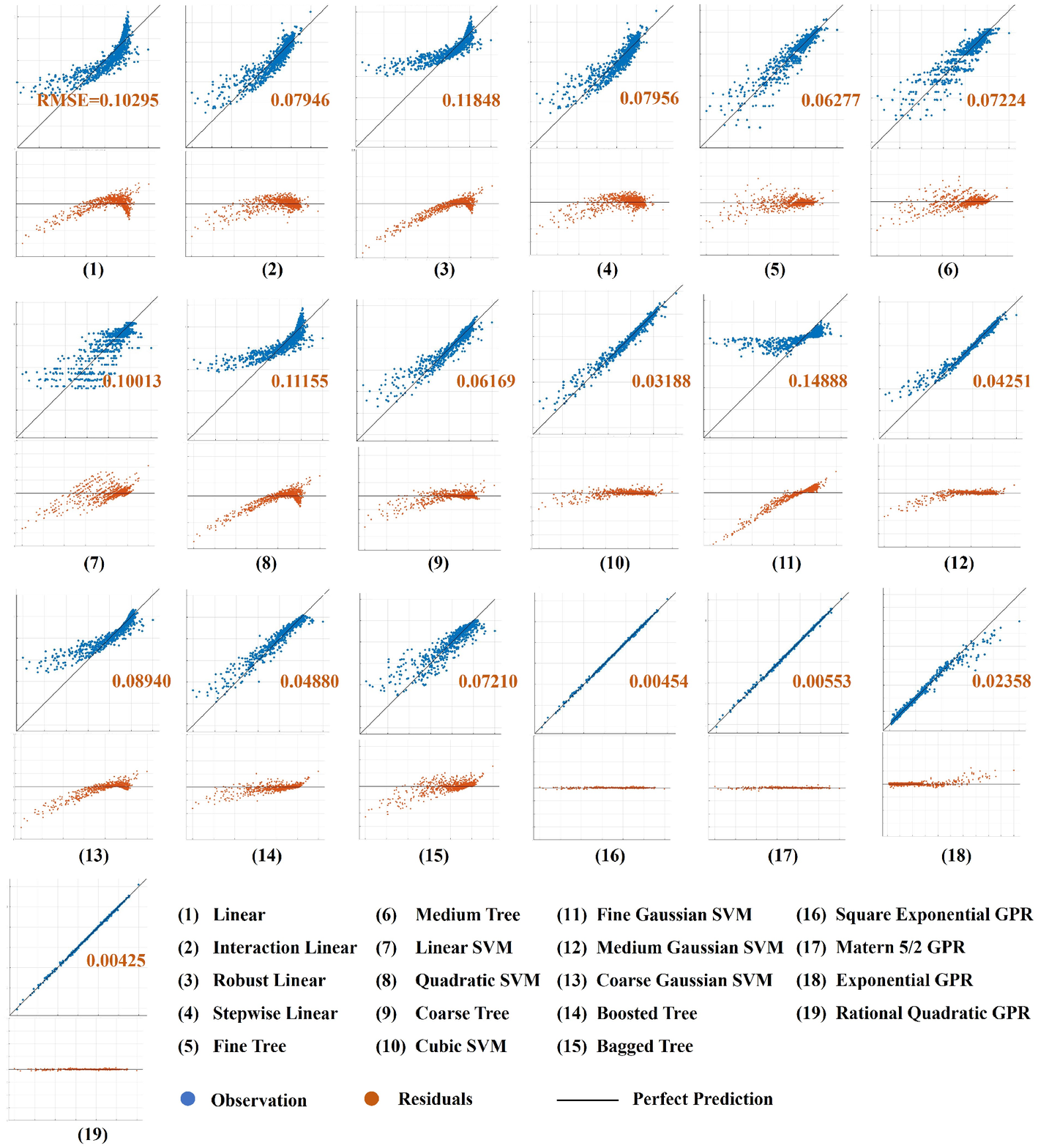}
  \caption{Performance shown by different types of machine learning models for the case of bulk cells. For each case, the predicted values of $F$ given by Eq.~\eqref{F_star} are drawn against their target values in the upper figure. The inclined line  of $y=x$ corresponds to the case of perfect predictions. The figure at the bottom for each case illustrates the deviation of the predicted data away from their corresponding targets, while the horizontal line stands for a perfect prediction. The RSME value is also marked correspondingly in each case.}\label{sec6-bulk}
\end{figure}

\begin{figure}[!ht]
  \centering
  \includegraphics[width=\textwidth]{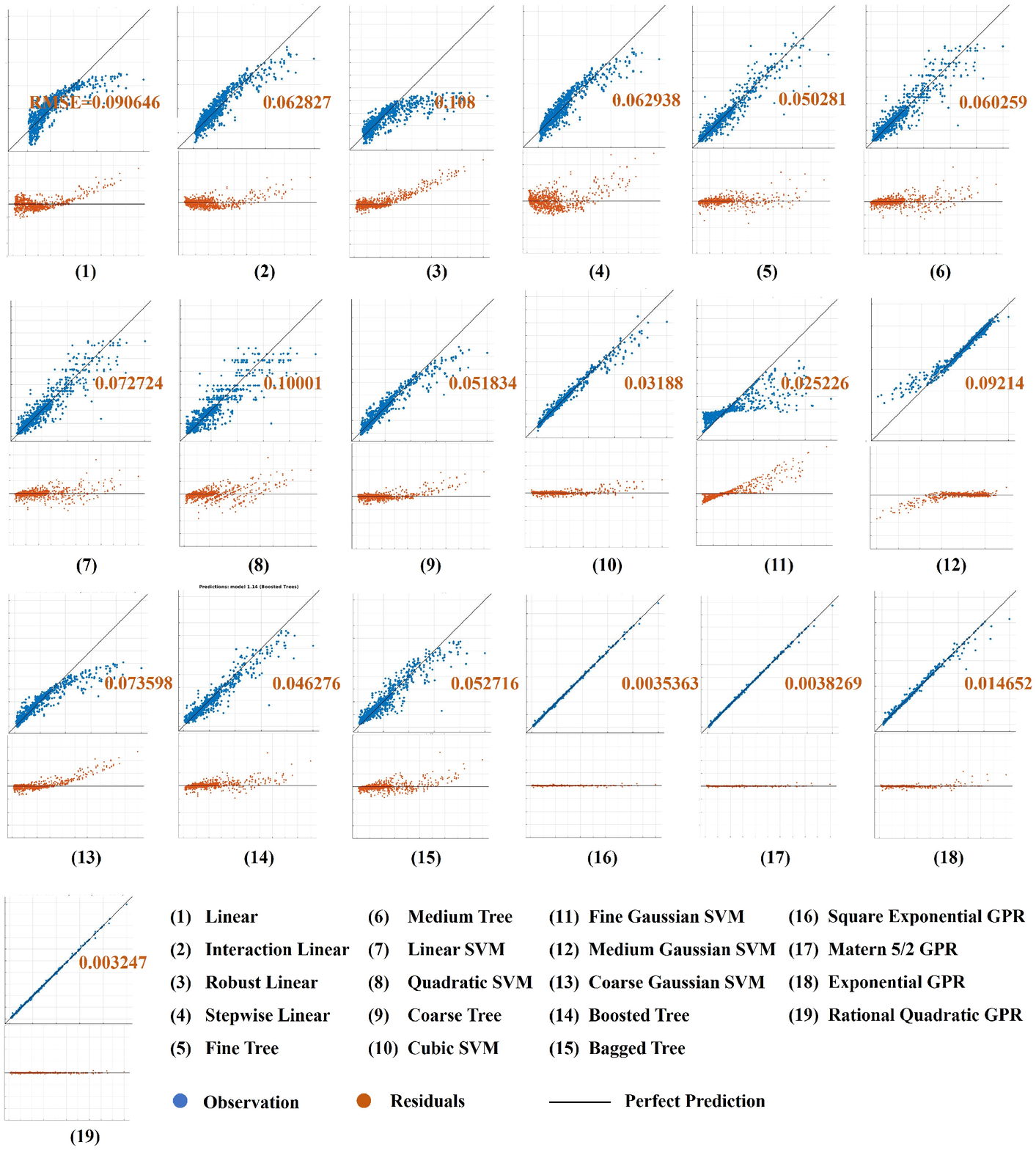}
  \caption{Performance shown by different types of machine learning models for the case of interface cells.}\label{sec6-GB}
\end{figure}

Up to nineteen algorithms implemented in the machine learning toolbox of Matlab 2019a have been tested in this work. The results are collectively shown in Fig.~\ref{sec6-bulk} and Fig.~\ref{sec6-GB}, for the cases of bulk cells and interface cells, respectively.
Take Fig.~\ref{sec6-bulk} as an example, for each type of machine learning method, two diagrams are produced for accessing its performance over the present problem guided by Eq.~\eqref{F_star}. In the upper diagram for each case, the predicted values of $F$ given by Eq.~\eqref{F_star} are drawn against their target values. The inclined line  of $y=x$ corresponds to the case of perfect predictions. The more compact of the data distribution is surrounding it, the better performance we should have. The diagram at the bottom for each case illustrates the deviation of the predicted data away from their corresponding targets, while the horizontal line stands for a perfect prediction. Besides, the value of the corresponding rooted square mean error (RSME), an important quantity for evaluating a machine learning model, is also marked correspondingly for each case. It is observed that the algorithm of rational quadratic Gaussian procession regression delivers the best performance among the nineteen tested models for the present problem. Same observation can also be from the situations of interface cells, as shown in Fig.~\ref{sec6-GB}.

\subsection{An application - measuring the sink bias of bubbles}
\begin{figure}[!ht]
  \centering
  \includegraphics[width=.6\textwidth]{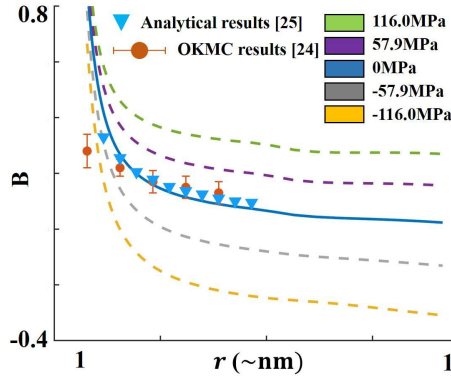}
  \caption{Measuring sink bias of voids against radius $r$ under various applied stress, where a positive value means a compressive load. The dots are from OKMC calculations \cite{Carpentier_ActaMater2017}; the triangles are from theoretical calculations \cite{Heald_ActaMetal1975}. The parameters are chosen \cite{Carpentier_ActaMater2017} as follows: $\gamma=1$J$\cdot$m$^{-2}$; $\varrho=3.74\times10^{21}$m$^{-3}$; $T=300$K; $\Delta v_{\text{i}}=\Delta v_{\text{v}}=0.015$nm$^3$.}\label{Fig_sink_bias}
\end{figure}

Certain reports suggest that voids/bubbles are actually biased sinks of SIAs against vacancies. The reason is that the two PD species exhibit opposite moving tendencies in a same elastic field caused by a void/bubble. One measurement of the sink bias of bubbles is given by \cite{Heald_ActaMetal1975}
\begin{equation} \label{sink_bias}
B = \frac{k_{\text{iB}}^2 - k_{\text{vB}}^2}{k_{\text{iB}}^2}.
\end{equation}
Theoretical and OKMC calculations have been conducted \cite{Borodin_JNucMater1993, Carpentier_ActaMater2017} to measure such sink bias of voids. But the present results enable one to examine more general cases of bubbles. Some results are shown in Fig.~\ref{Fig_sink_bias}.
Here we consider the case of voids in the bulk under varying (normal) stress components applied to the LHR cells. The stress-free case is compared with OKMC results with $r\le5$nm \cite{Carpentier_ActaMater2017, Heald_ActaMetal1975}. It appears that the machine learning models deliver a high predictive accuracy for large-size voids, which complement the OKMC calculations whose efficiency is high for small size voids. Another interesting observation is that $B$ may turn negative when the cell experiences strong traction. This is because the cell boundary becomes the high hydrostatic pressure region where SIAs are more favoured. Note that the machine learning models can also be used for making predictions for bubbles, both in the bulk and on crystalline interfaces.

\subsection{Summarising remarks}
Up to now, the sink strength terms $k_{\alpha \text{B}}^2$ and $\tilde{k}_{\alpha \text{B}}^2$, as appearing in REs \eqref{sec2-c-evlution-eq} (for bulk diffusion) and \eqref{sec2-c_gb-evlution-eq} (for interface diffusion), respectively, can be fully evaluated through the mentioned machine learning models, whose training is performed in advance in an offline stage.

\section{Simulation results and discussion\label{Sec_numerics}}
In this section, REs underpinned by trained machine learning models are evolved. One of our goals here is to capture the role played by crystalline interfaces as partial sinks in irradiation-induced bubble growth. Attentions will also be paid to the (localised) stress effects on the biased sink behaviour of bubbles. For this purpose, certain idealised settings are adopted for simulations.

\subsection{Simulation setting}
\subsubsection{Domain and parameters}
\begin{figure}[!ht]
  \centering
  \includegraphics[width=.95\textwidth]{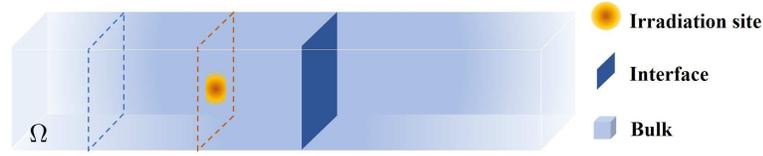}
  \caption{Simulation settings}\label{Results-model}
\end{figure}

The computational domain $\Omega$ on which simulations are carried out is set up as shown in Fig.~\ref{Results-model}.
It takes a cuboid shape of size $2\mu\text{m}\times0.5\mu\text{m}\times0.5\mu\text{m}$. A crystalline interface is located roughly in coincidence with its middle plane. Periodic boundary conditions are imposed on the four side surfaces of the domain $\Omega$, while flux-free boundary conditions are imposed on its two ends. The PD source terms $K_{\alpha}$ of Eq.~\eqref{sec2-c-evlution-eq} are confined within a small region in $\Omega$ as shown in Fig.~\ref{Results-model}. Such a setting albeit slightly non-physical, is aimed to capture the distinction in diffusivity in the bulk against on the interface.

The involved parameters are evaluated as follows. The PD source terms are chosen in proportion to the d.p.a. rate \cite{was_fundamentals_2017}, and they are quantified by $K_{\text{v}} = 10^{-2}\dot{s}_{\text{dpa}}$, $K_{\text{i}}= 10^{-6} \dot{s}_{\text{dpa}}$ and $K_{\text{g}}=10^{-3} \dot{s}_{\text{dpa}}$. The reason that we let $K_{\text{v}}\gg K_{\text{i}}$ is twofold. Certain practical mechanisms of interstitial consumption, such as dislocations as biased sinks, are implicit taken into account in the (net) source term. Moreover, higher concentration in vacancy helps us to capture the process of bubble growth more quickly. Nonetheless, the present model also works for other parameter values, which may be closer to the actual situations. The recombination coefficient is evaluated by $k_{\text{iv}}=2.5\times10^{-4}$s$^{-1}$. The system is also pre-assigned with a regular distribution of void embryos, with a density of $\varrho=10\mu\text{m}^{-3}$ and an initial size of $2.3$nm.

For material constants, we set the bulk diffusivity by $D_{\text{v}}= 10^{-15}$ m$^2\cdot$s$^{-1}$, $D_{\text{i}} = 20 D_{\text{v}}$ and $D_{\text{g}}= 0.2 D_{\text{v}}$, respectively. The diffusivity on the interface is correspondingly evaluated by $\tilde{D}_{\alpha} = 20 D_{\alpha}$. The volume occupied by a single matrix (aluminium) atom $v_0= 0.0167$nm$^3$, and the relaxation volumes are selected to be $\Delta v_{\text{v}}=0.7v_0$, $\Delta v_{\text{i}} =0.7v_0$ and $\Delta v_{\text{g}}=0.8v_0$, which could be relatively larger than the normal choices, but this helps identify the stress effect on bubble development. The surface tension is given by $\gamma=0.35$J$\cdot$m$^{-2}$.

The present model is also able to capture the role of external mechanical loads (through the hydrostatic pressure gradient field). Here we assume the system is loaded with a longitudinal stress component $\sigma_{\text{a}}$, i.e. $\sigma_{33}=\sigma_{\text{a}}$ and $\sigma_{ij}=0$ otherwise.

\subsubsection{Numerical Schemes}
Concerning the numerical schemes for evolving the derived system, the central finite difference scheme is used for discretising the rate equations \eqref{sec2-c-evlution-eq} and \eqref{sec2-c_gb-evlution-eq}. Noted that the time step associated with the rate equation \eqref{sec2-c_gb-evlution-eq} on an interface should be far less than that of the rate equation in the bulk. This is because the diffusivity on the interface is roughly 20 times larger than their counterpart in the bulk. But to maintain computational efficiency, the general time step $\Delta t$ is still selected with respect to the rate equation~\eqref{sec2-c-evlution-eq} in the bulk, while much finer time step $\Delta \tilde{t}=\frac{1}{20}\Delta t$ is only adopted when equation~\eqref{sec2-c_gb-evlution-eq} is evolved on the interface. Therefore, in each time step, we first evolve one (larger) time step for the bulk concentration $c_{\alpha}$ based on Eq.~\eqref{sec2-c-evlution-eq}. Then we evolve 20 steps for the concentration $\tilde{c}_{\alpha}$ on the interface, while the ``$[\cdot]$'' term in Eq.~\eqref{sec2-c_gb-evlution-eq} are still treated to be quasi-steady constants.

\subsection{A demonstrative example}
A demonstrative example is firstly presented to show the model effectiveness. For this case, we set the d.p.a. rate to be $\dot{s}_{\text{dpa}}= 1.25\times 10^{-2}$s$^{-1}$ (which is considerably faster than that in an actual reactor) for a time period of $t=1.4 L^2/D_{\text{v}}$, that is, for roughly 1.5 hours (which is considerably short than that in an actual reactor). It is shown in \ref{Sec_role_dose} that a change in d.p.a. rate does not induce too big difference in the microstructural trend against the irradiation dose. Hence it is reasonable to mimic the actual situations using the present model with a relatively high d.p.a. rate (for the present case). For this demonstrative example, the specimen is free of external load, i.e., $\sigma_\text a=0$.

\begin{figure}[!ht]
\centering
\subfigure[Illustration of data points]{
\includegraphics[width=0.45\textwidth]{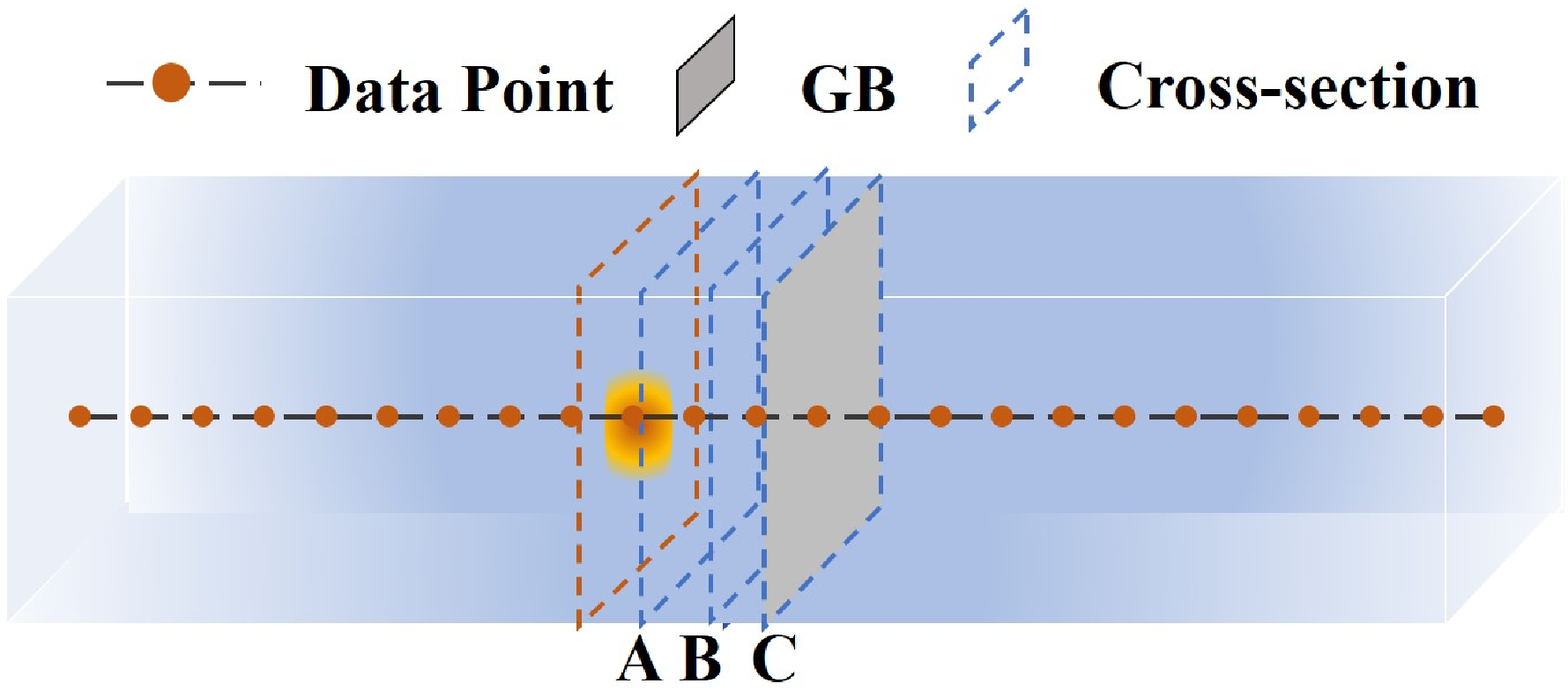}
}
\quad
\subfigure[Vacancies]{
\includegraphics[width=0.45\textwidth]{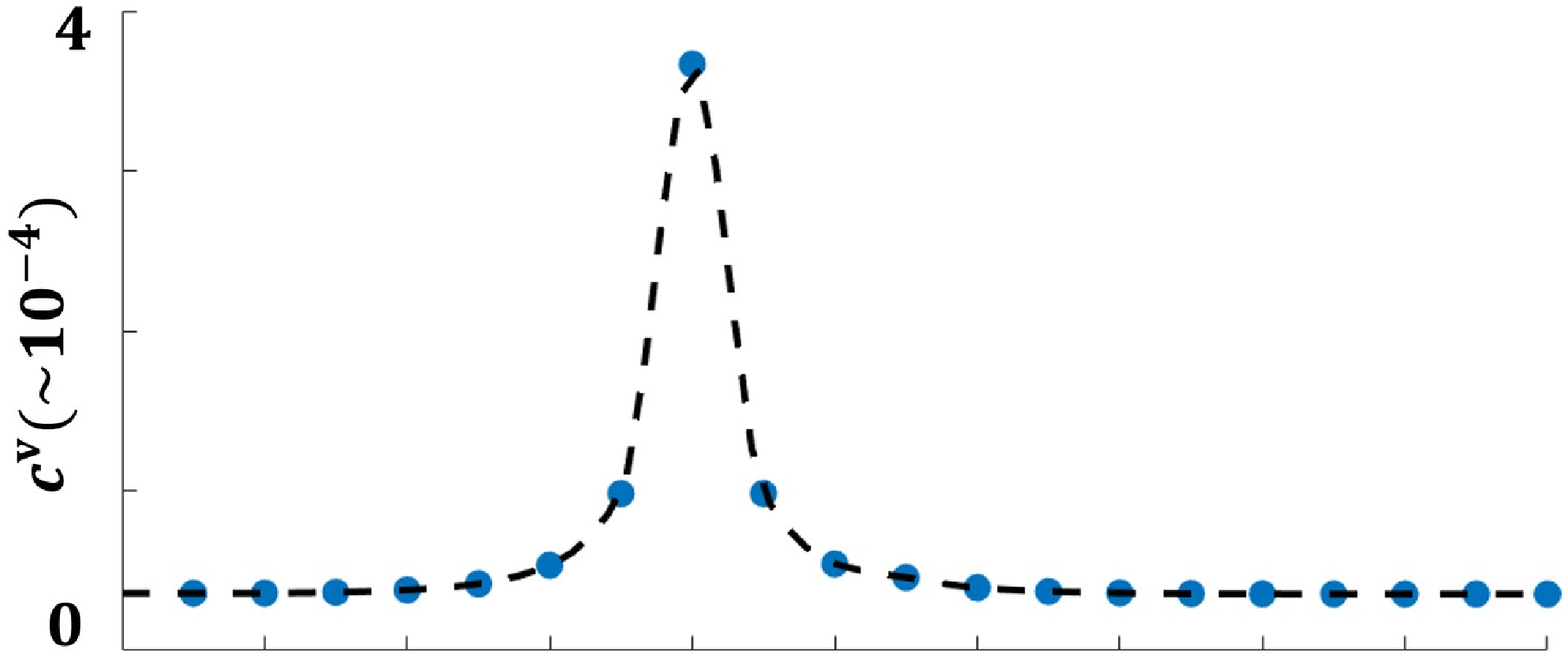}
}
\subfigure[Interstitial atoms]{
\includegraphics[width=0.45\textwidth]{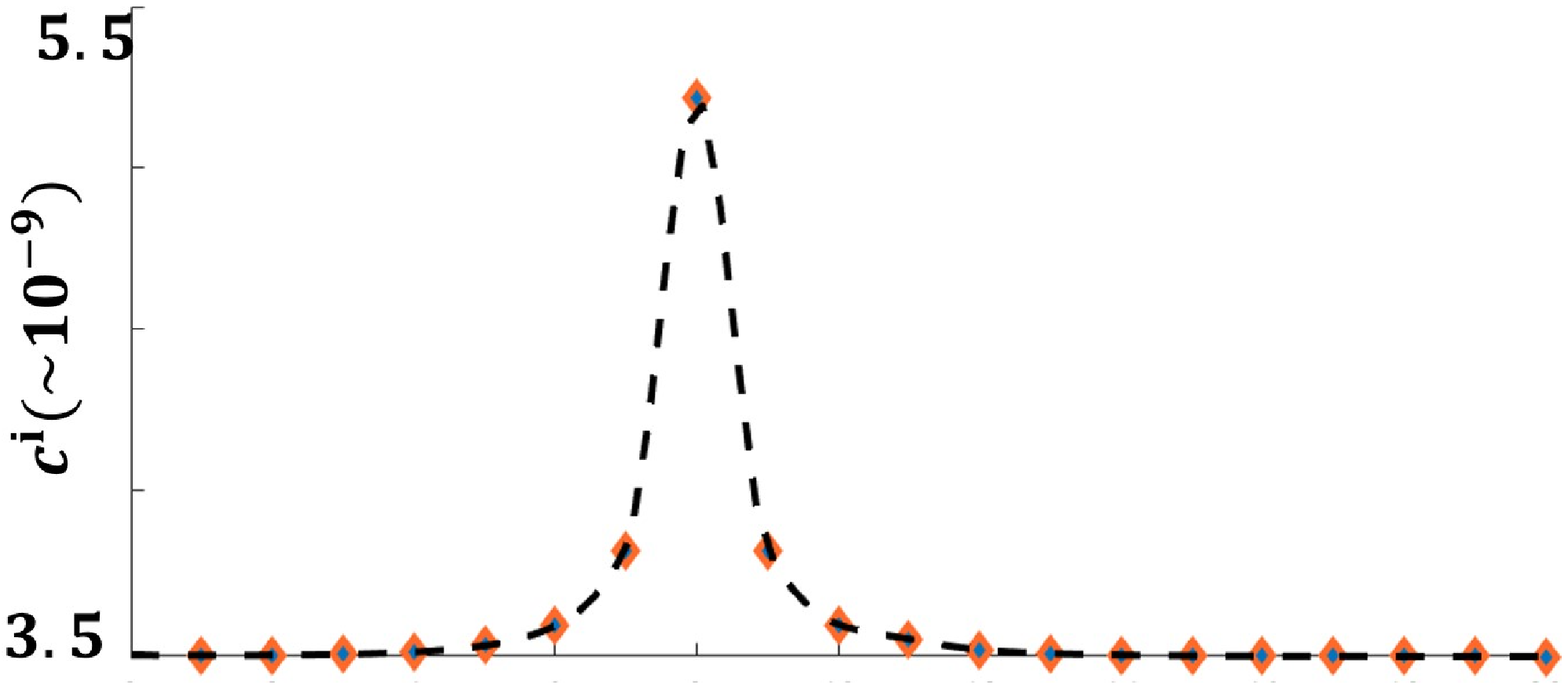}
}
\quad
\subfigure[NGAs]{
\includegraphics[width=0.45\textwidth]{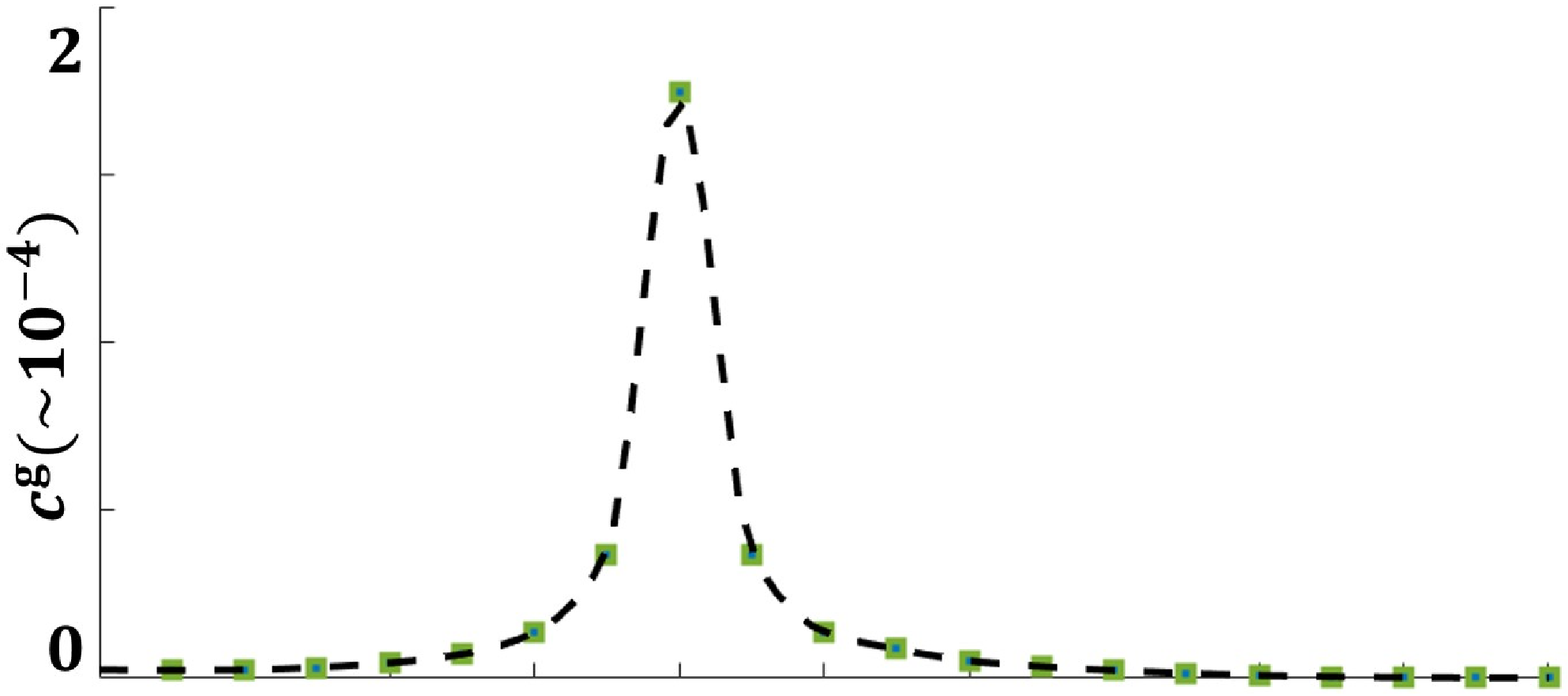}
}
\caption{PD fractional distribution along the long axis of $\Omega$ at 70 d.p.a.: (a) identification of the positions of interest; (b) the distribution of vacancy concentration; (c) the distribution of interstitial concentration; (d) the distribution of NGA concentration. Three slices are sampled sequentially away from the source region for further investigations, as shown in figure (a), and slice C coincides with the interface.}\label{Results-concentrationline}
\end{figure}

The distribution of PD fractional concentration is observed first along a longitudal line penetrating the computational domain $\Omega$, as shown in Fig.~\ref{Results-concentrationline}(a).
The concentration distributions for $c_{\text{v}}$, $c_{\text{i}}$, $c_{\text{g}}$ (at approximately 70 d.p.a.) are shown in Fig.~\ref{Results-concentrationline}(b)-(d), respectively. It is natural to see that high concentration is attained near the source region, and the value drops as moving away from it. Note that although $K_{\text{v}}$ is roughly ten times greater than $K_{\text{g}}$, the concentration for the two species are similar, suggesting that a large portion of vacancies have been absorbed by bubbles or annihilated with self-interstitial atoms.

\begin{figure}[!ht]
\centering
\subfigure[Slice A]{
\begin{minipage}{0.3\linewidth}
\includegraphics[height=0.12\textheight]{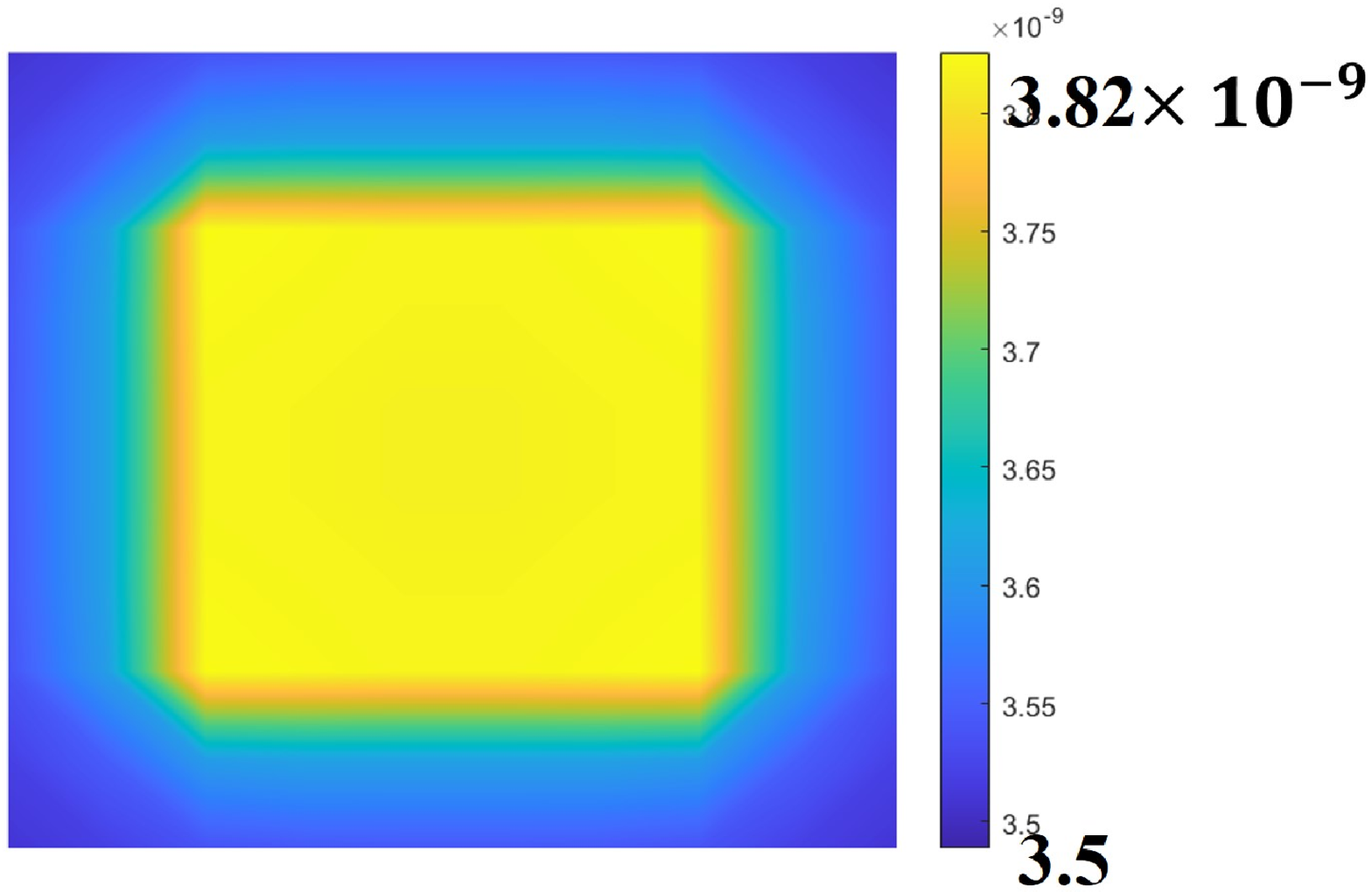}\\
\includegraphics[height=0.12\textheight]{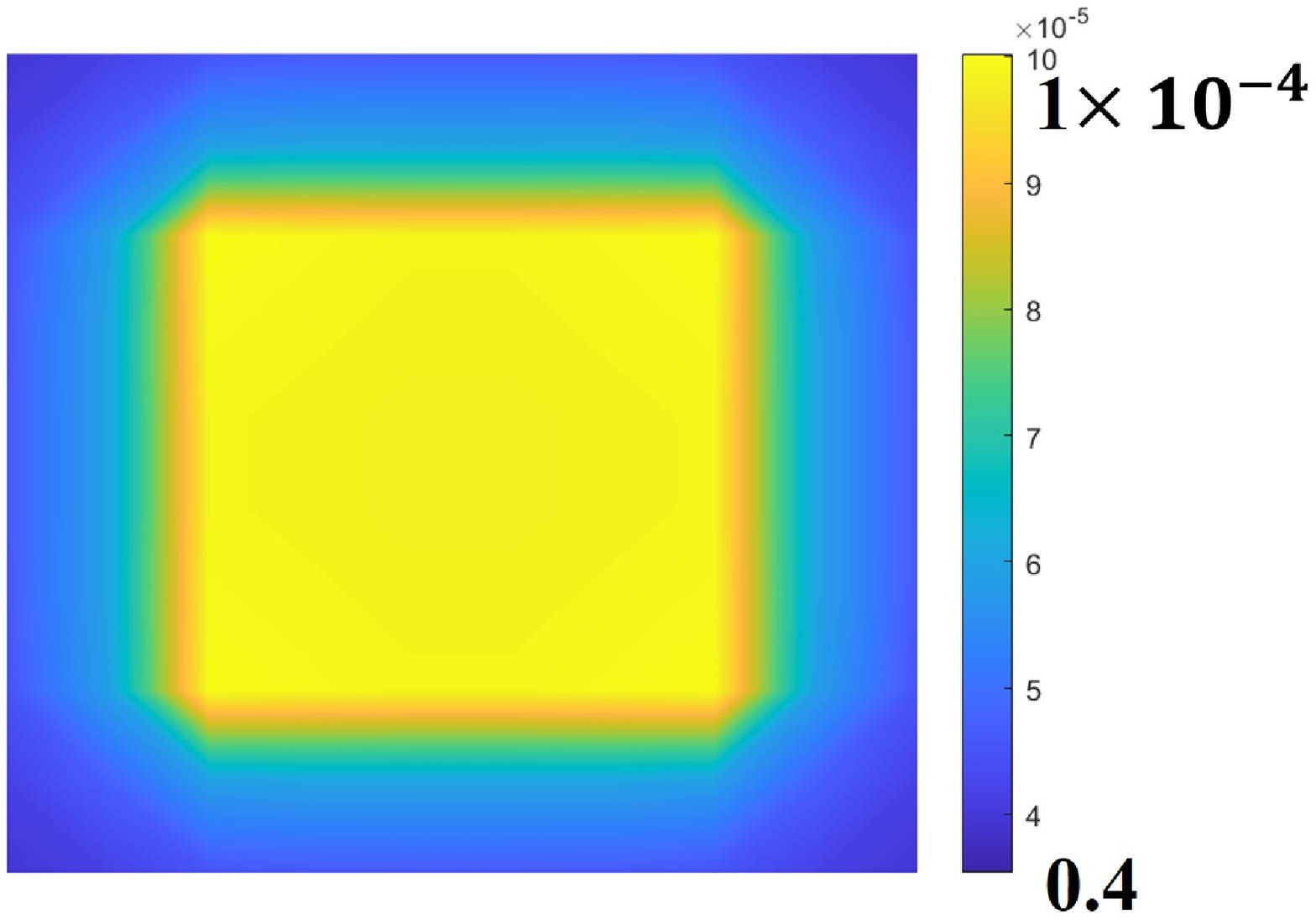}\\
\includegraphics[height=0.12\textheight]{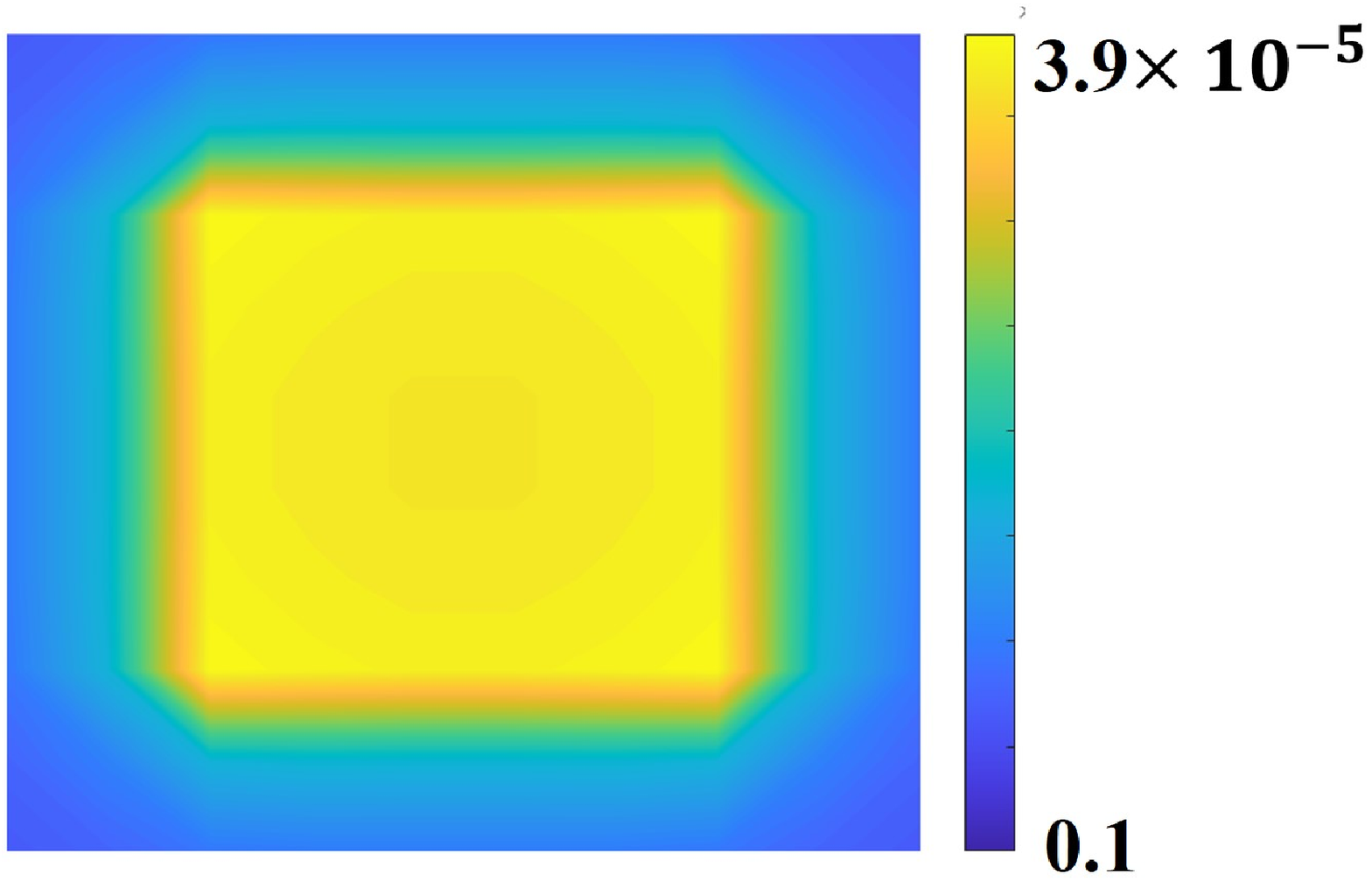}\\
\includegraphics[height=0.12\textheight]{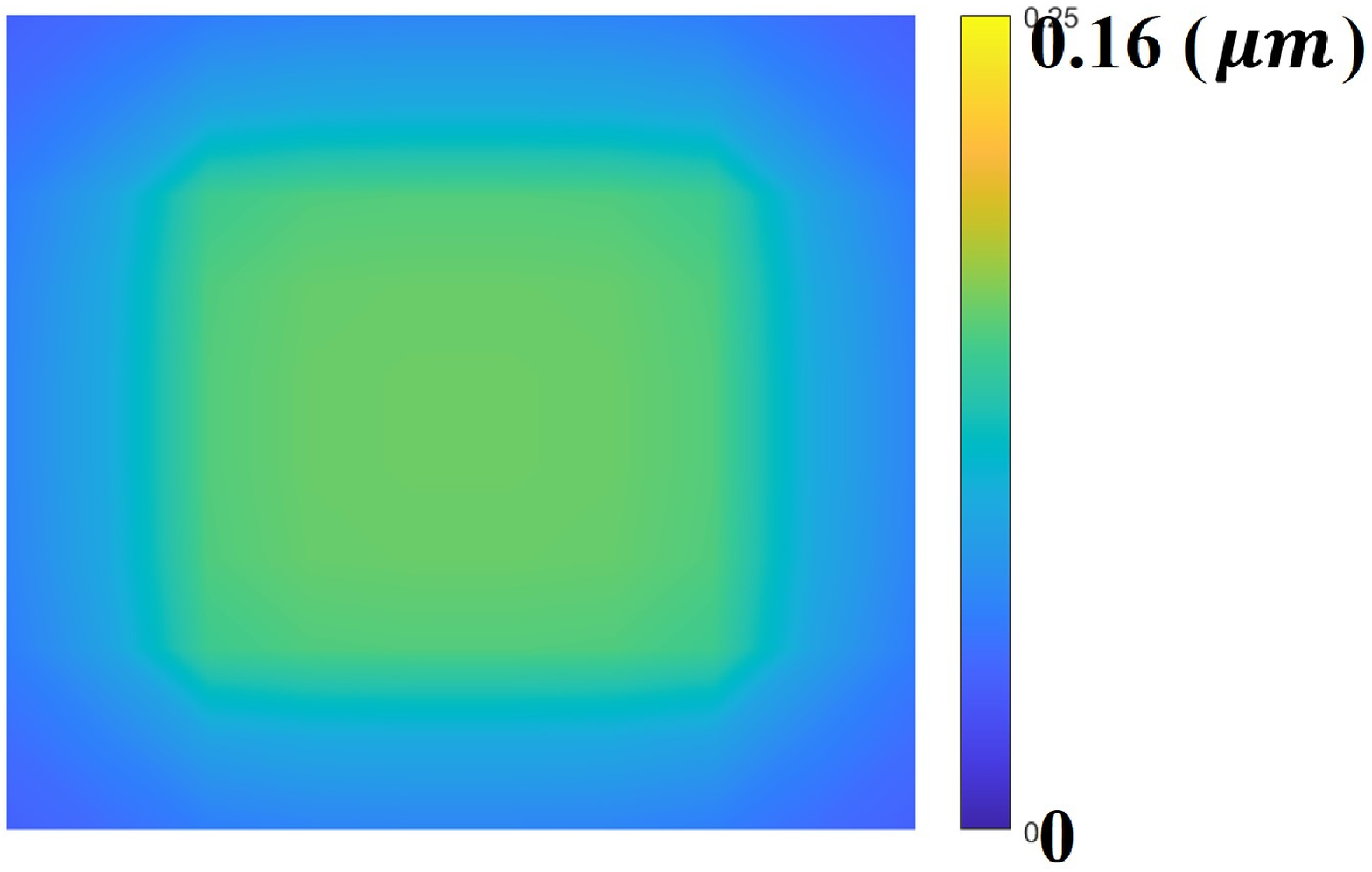}\\
\includegraphics[height=0.12\textheight]{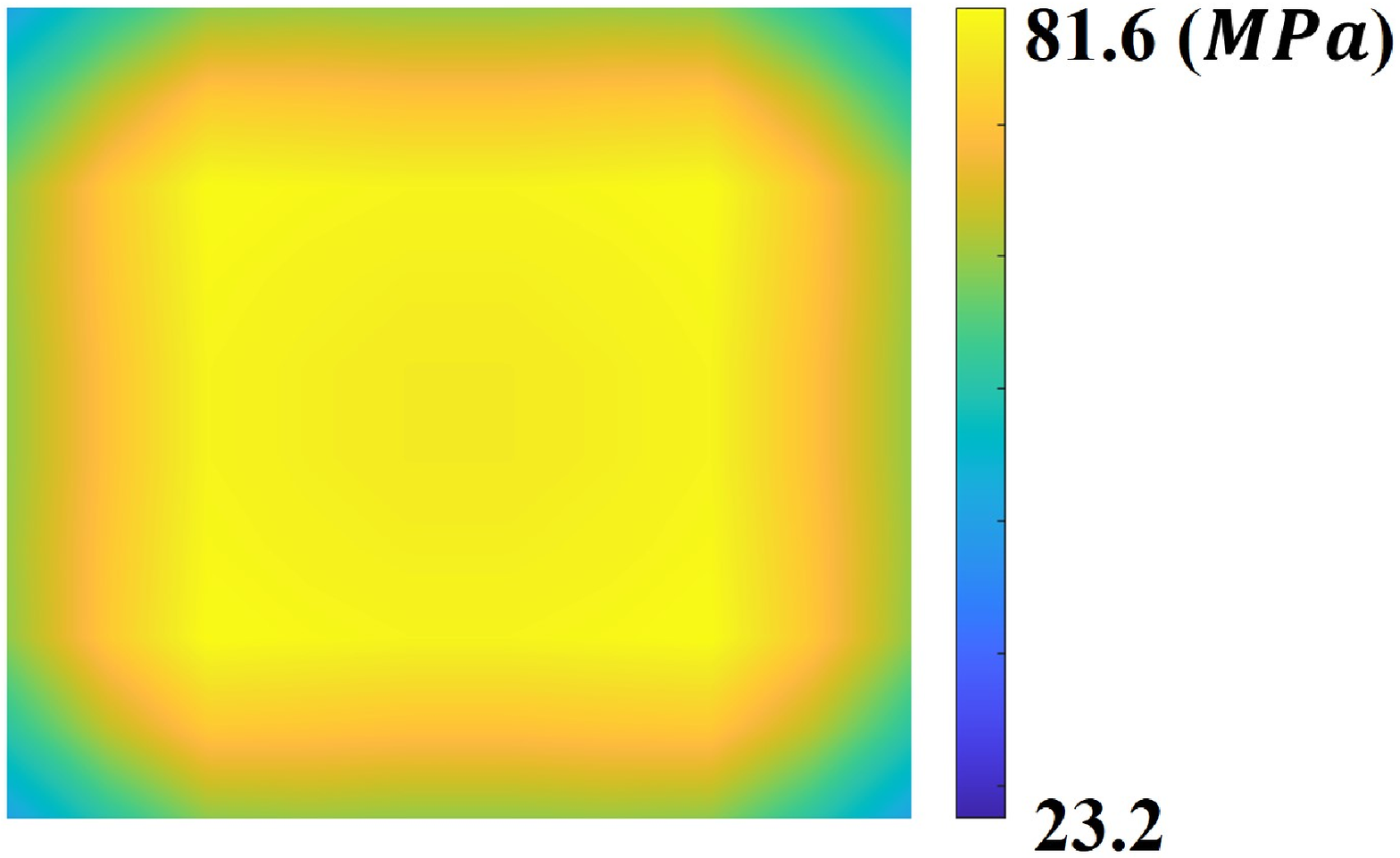}\\
\end{minipage}}
\subfigure[Slice B]{
\begin{minipage}{0.3\linewidth}
\includegraphics[height=0.12\textheight]{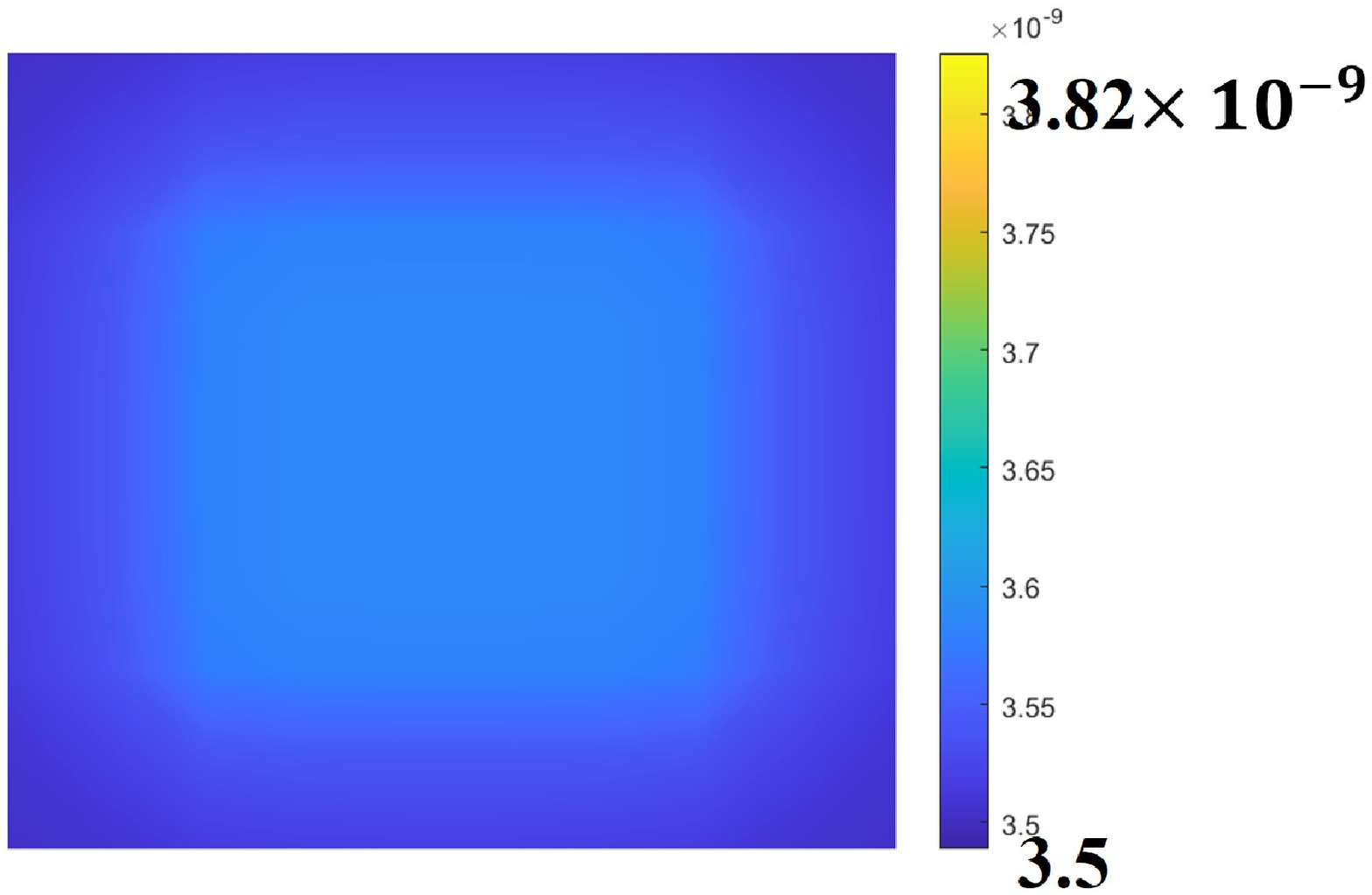}\\
\includegraphics[height=0.12\textheight]{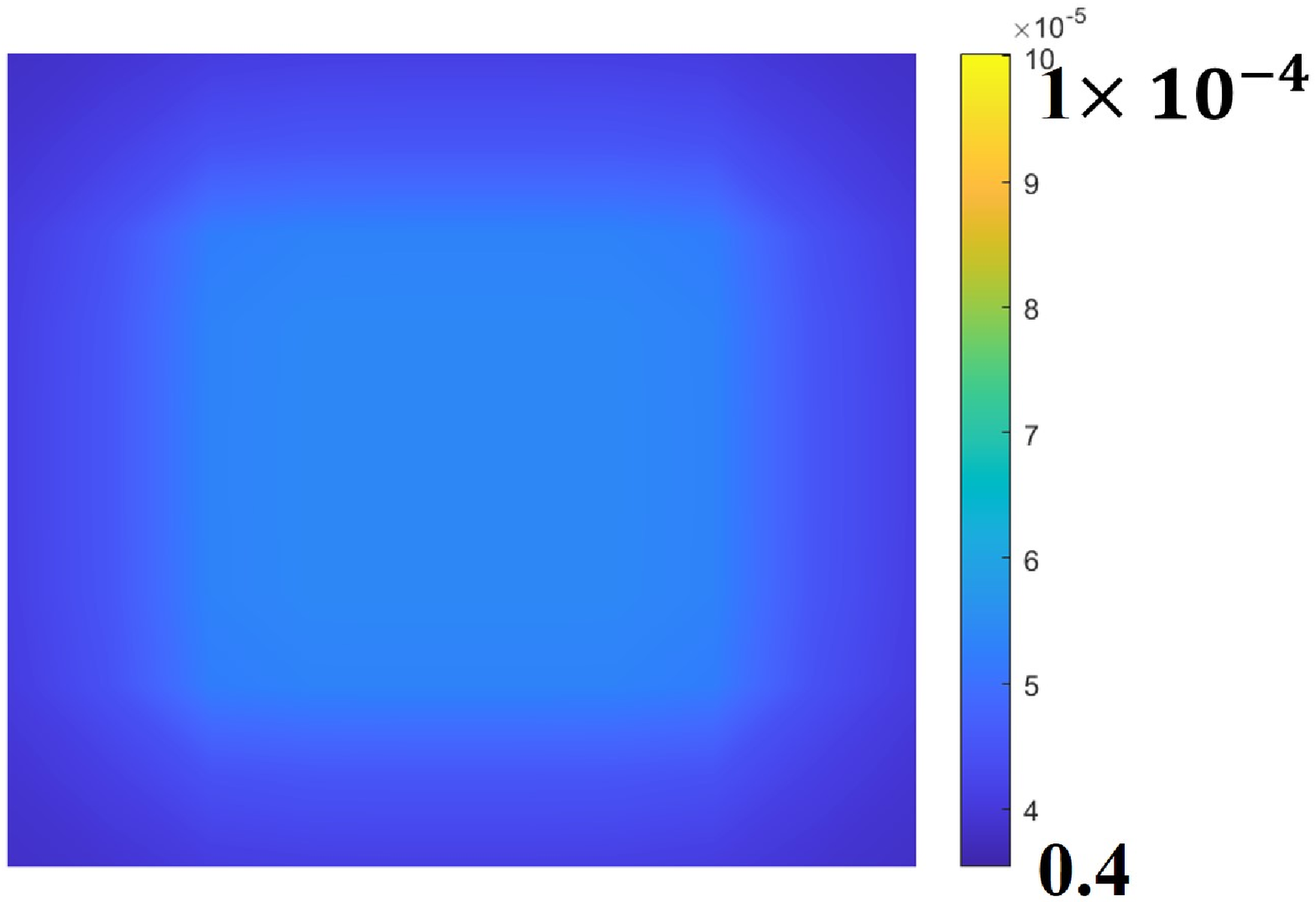}\\
\includegraphics[height=0.12\textheight]{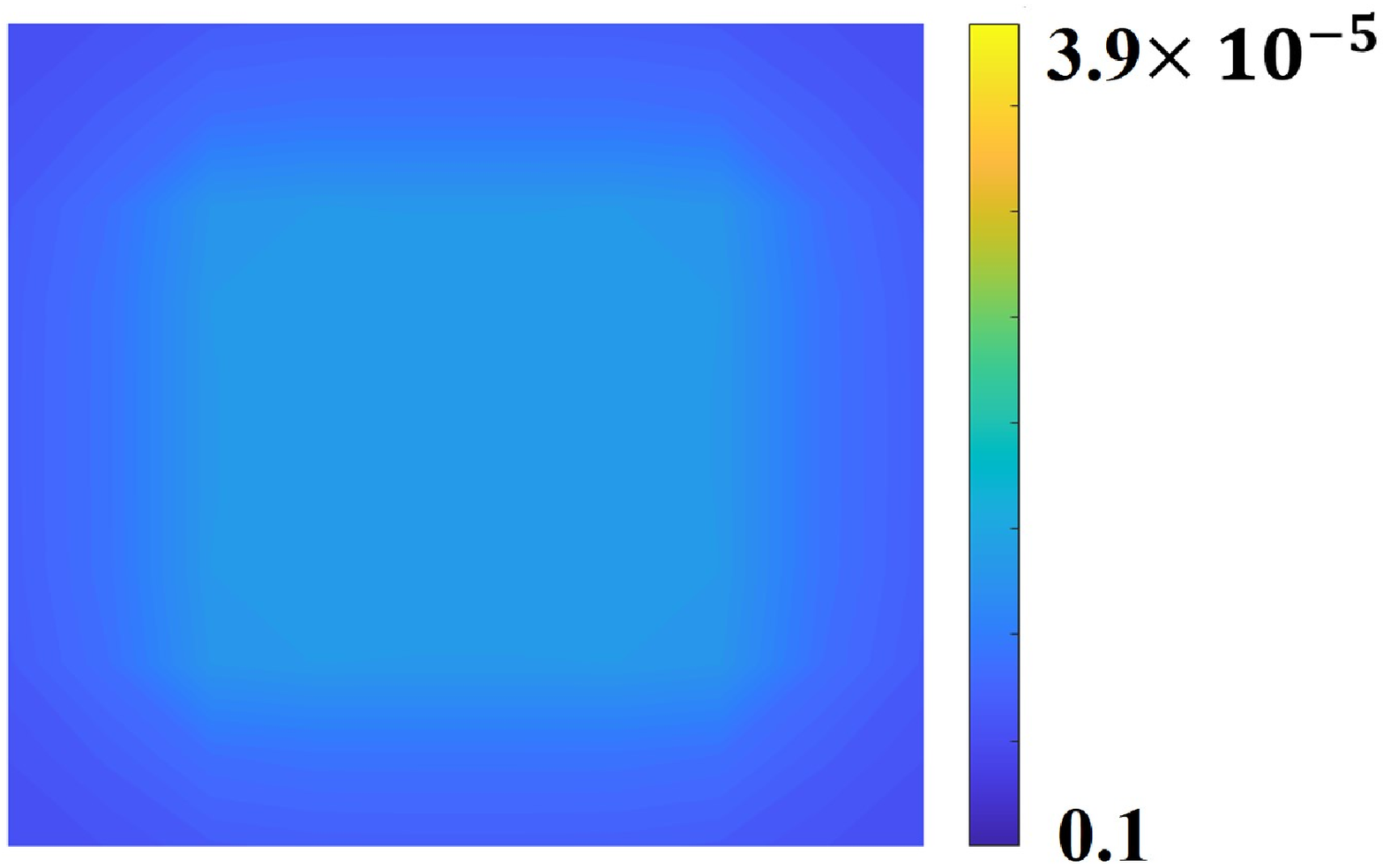}\\
\includegraphics[height=0.12\textheight]{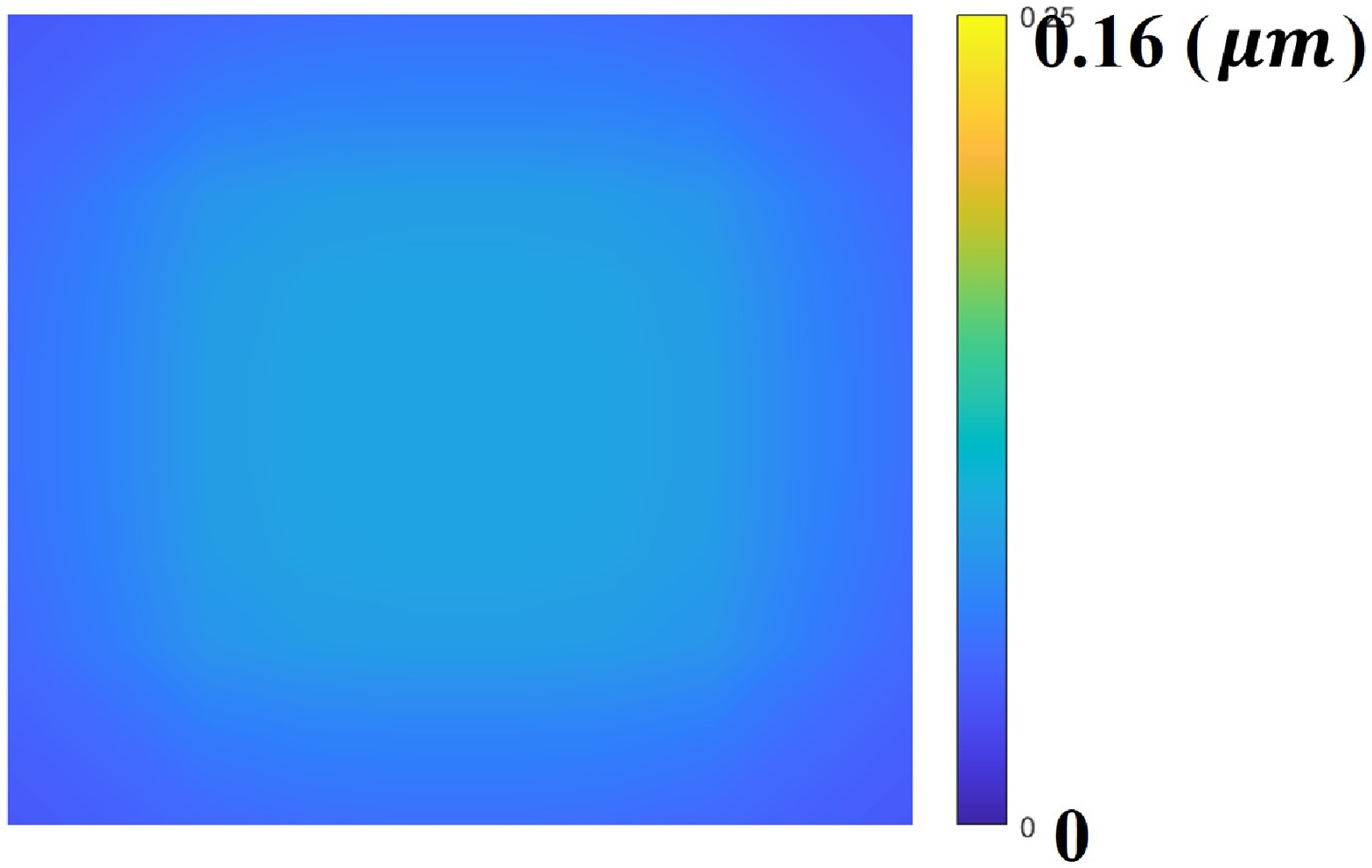}\\
\includegraphics[height=0.12\textheight]{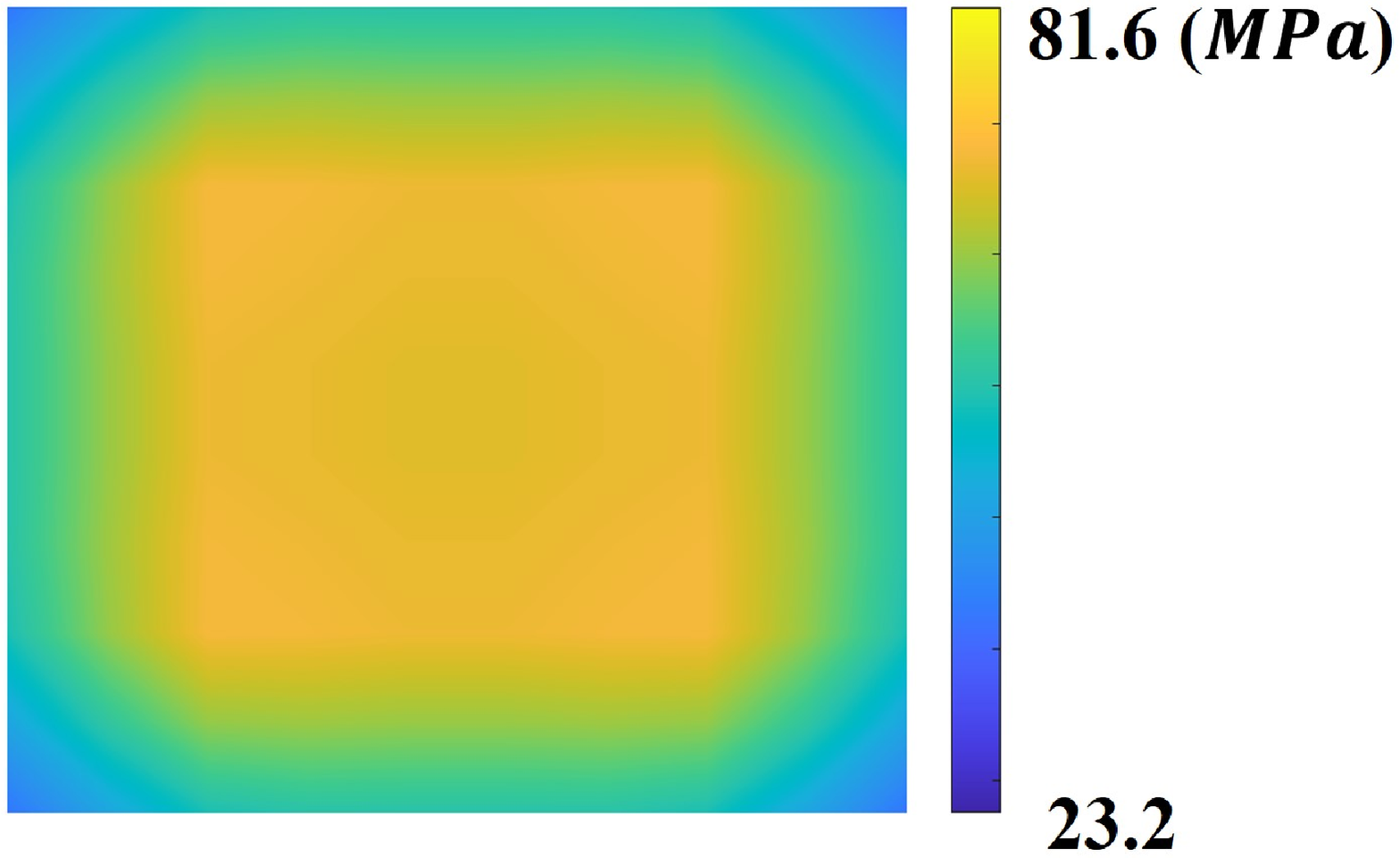}\\
\end{minipage}}
\subfigure[Slice C]{
\begin{minipage}{0.3\linewidth}
\includegraphics[height=0.12\textheight]{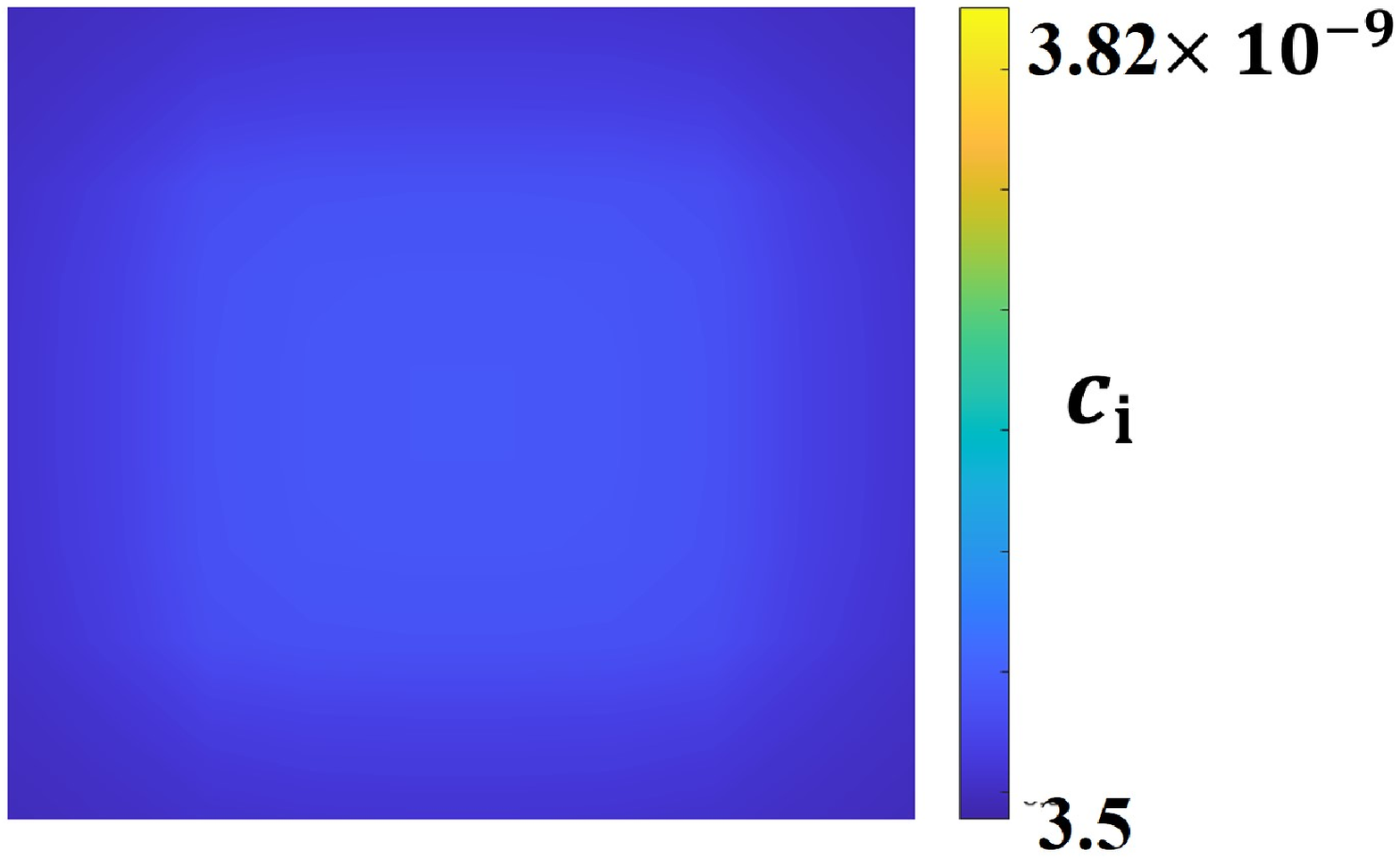}\\
\includegraphics[height=0.12\textheight]{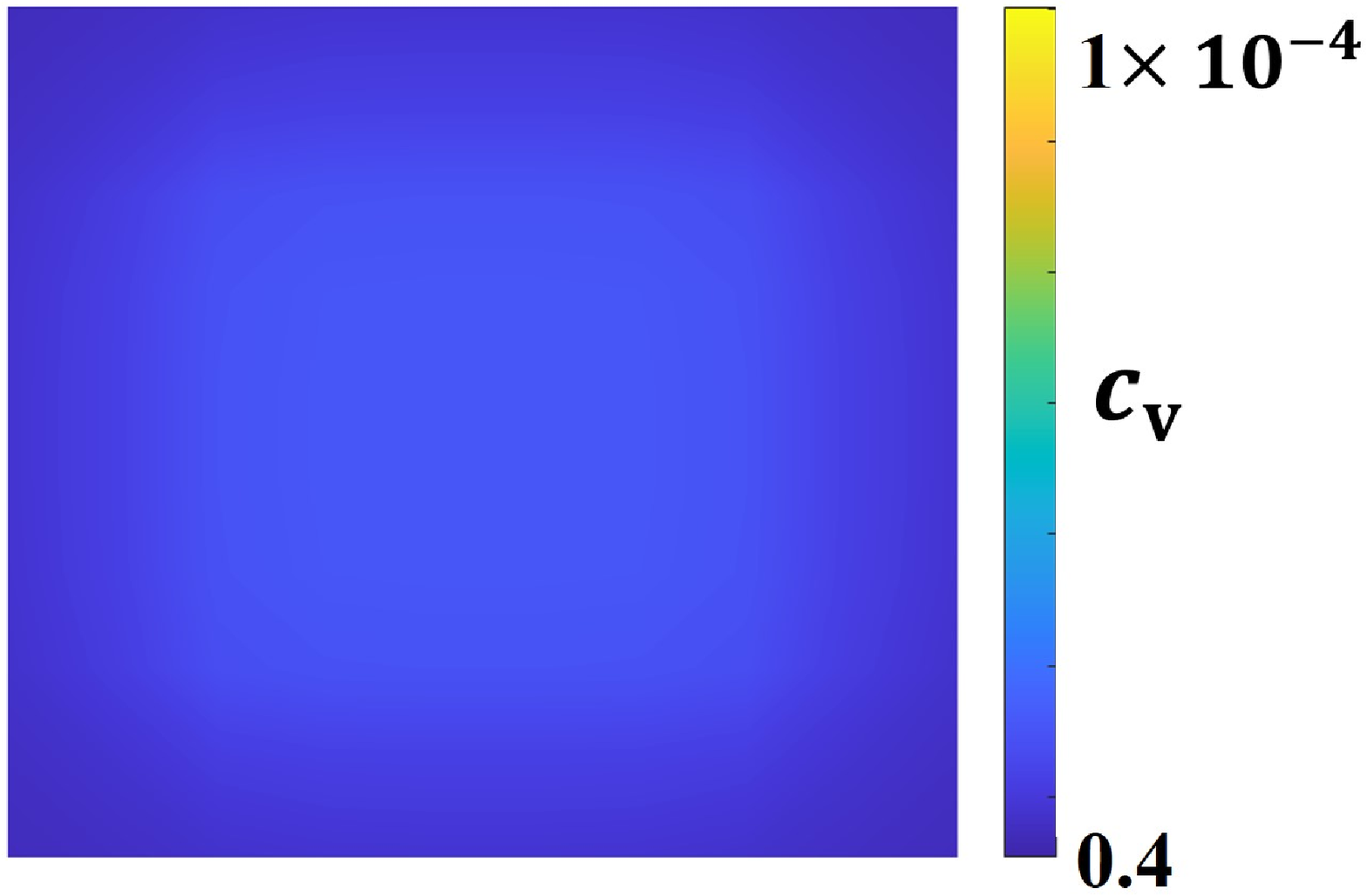}\\
\includegraphics[height=0.12\textheight]{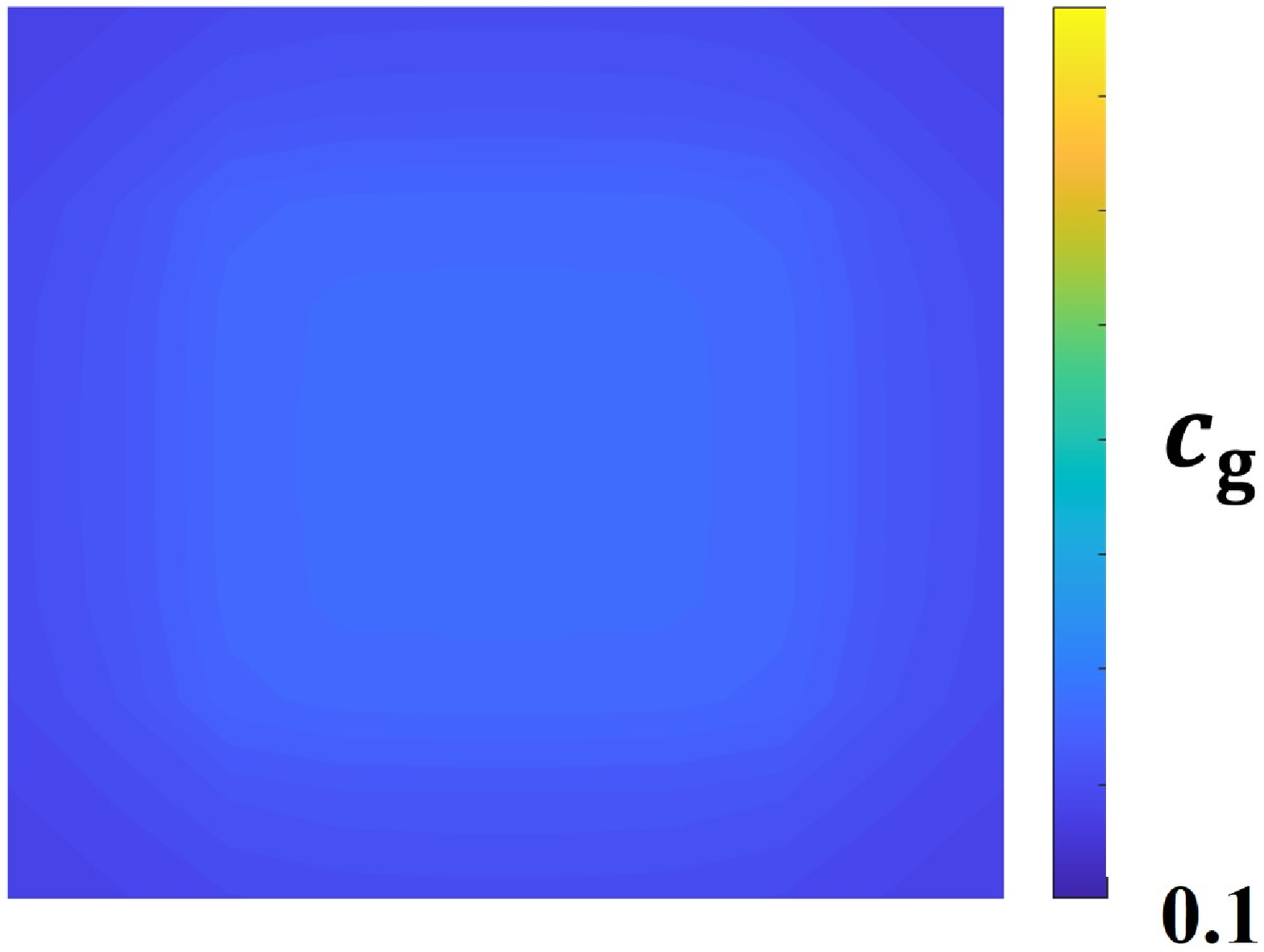}\\
\includegraphics[height=0.12\textheight]{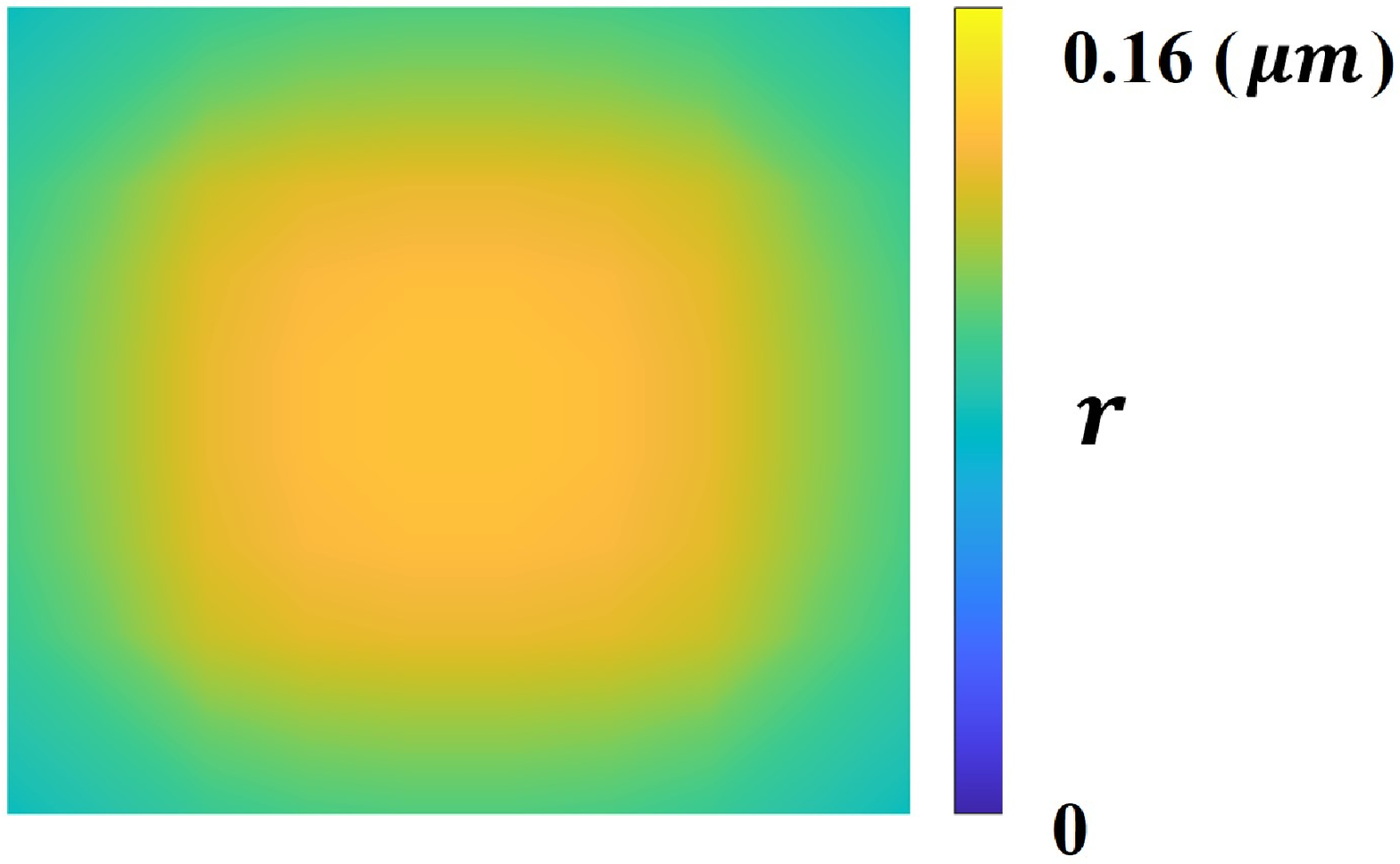}\\
\includegraphics[height=0.12\textheight]{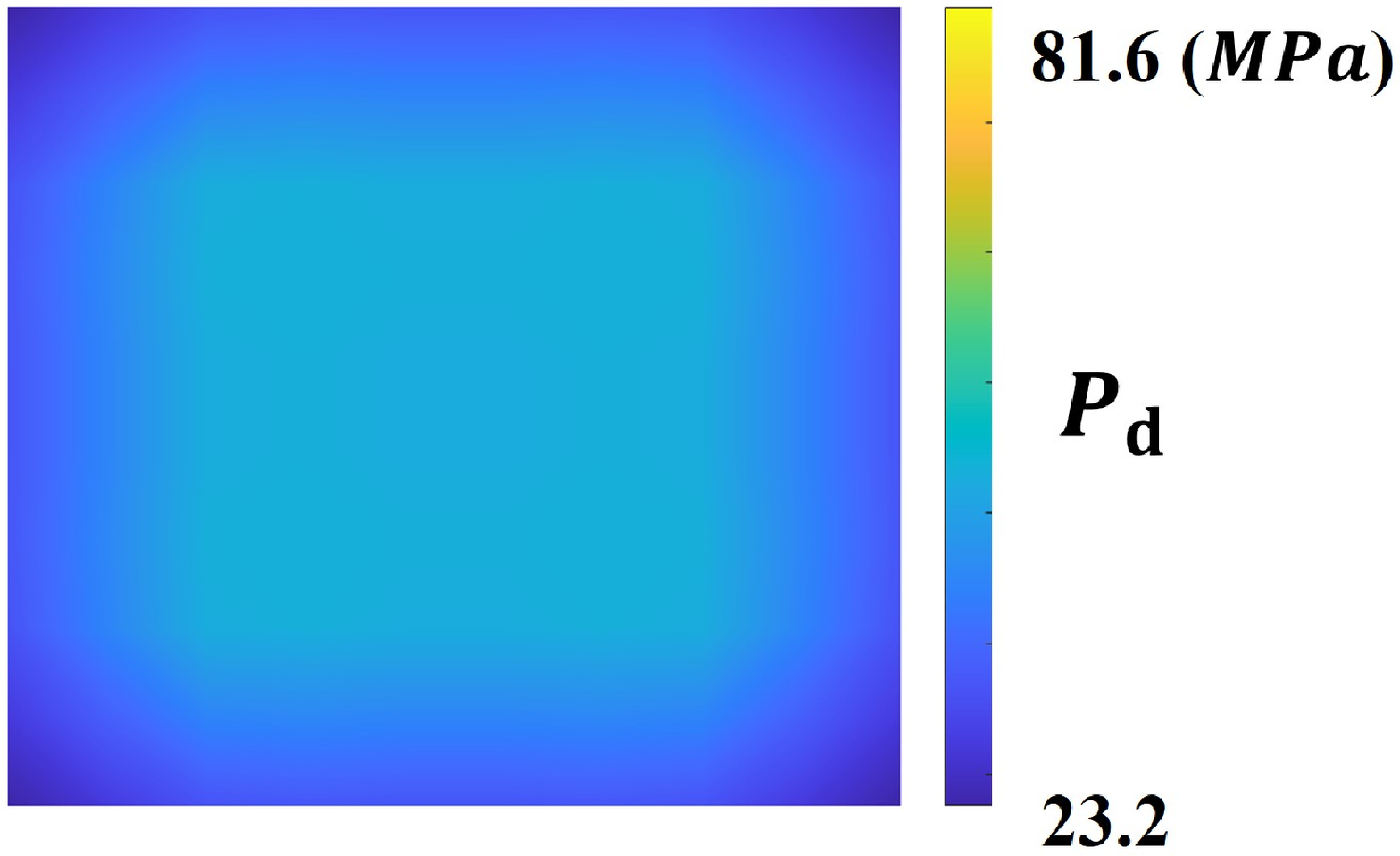}\\
\end{minipage}}
\caption{Properties on the three sampled slices shown in Fig.~\ref{Results-concentrationline}(a). Slice C coincides with the crystalline interface.} \label{Results-contour}
\end{figure}

Now the microstructural profiles are examined on three sampled slices, labelled by A, B and C in Fig.~\ref{Results-concentrationline}(a). These three slices lie sequentially away from the source region, and the slice C coincides with the crystalline interface. Then the
distributions of quantities of our interest on different slices are collectively shown in Fig.~\ref{Results-contour}.
By comparing the results on slices A and B, it is not surprising to find that the quantity values in the bulk drop as moving away from the high-concentration source region. But such a trend terminates on the crystalline interface (of slice C). For the bubble size, in particular, the values on the interface are higher than that on slice A, which is even closer to the source region. This is due to the much faster diffusion processes taking place on the interface. The distinctive role played by crystalline interfaces will be examined further in sec.~\ref{sec-role-interface}.

\begin{figure}[!ht]
\centering
\subfigure[]{
\includegraphics[width=0.45\textwidth]{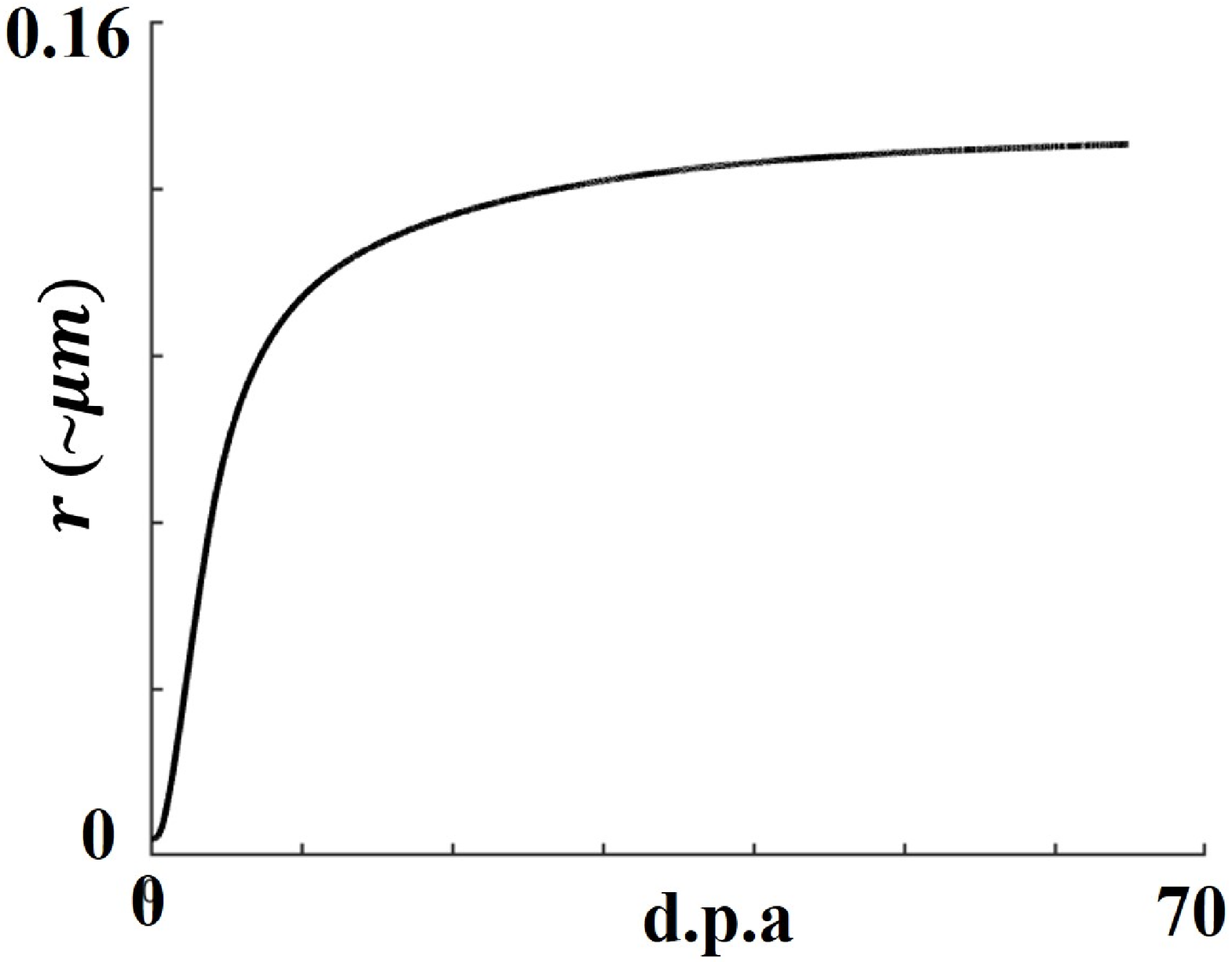}
}\quad
\subfigure[]{
\includegraphics[width=0.45\textwidth]{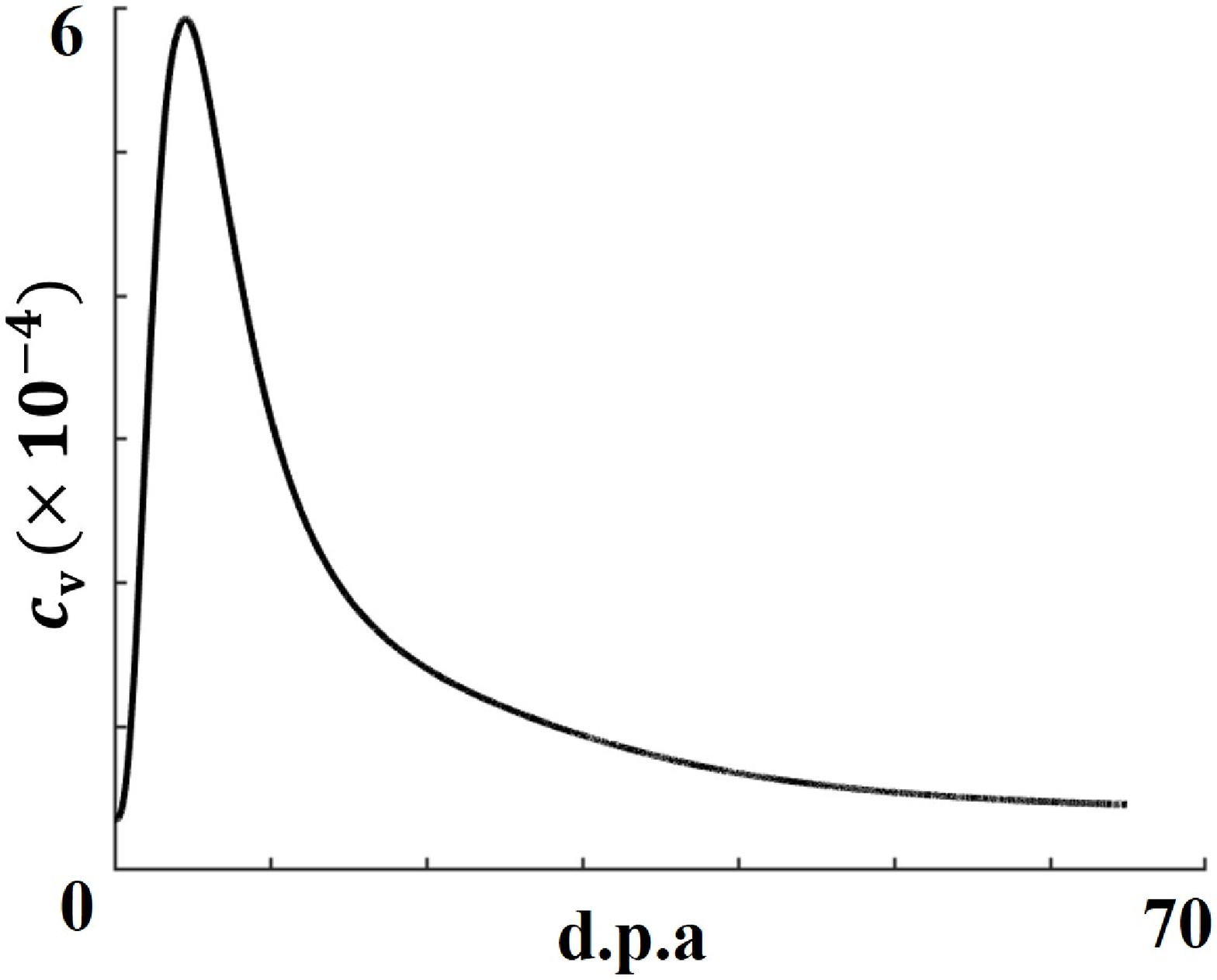}
}
\subfigure[]{
\includegraphics[width=0.45\textwidth]{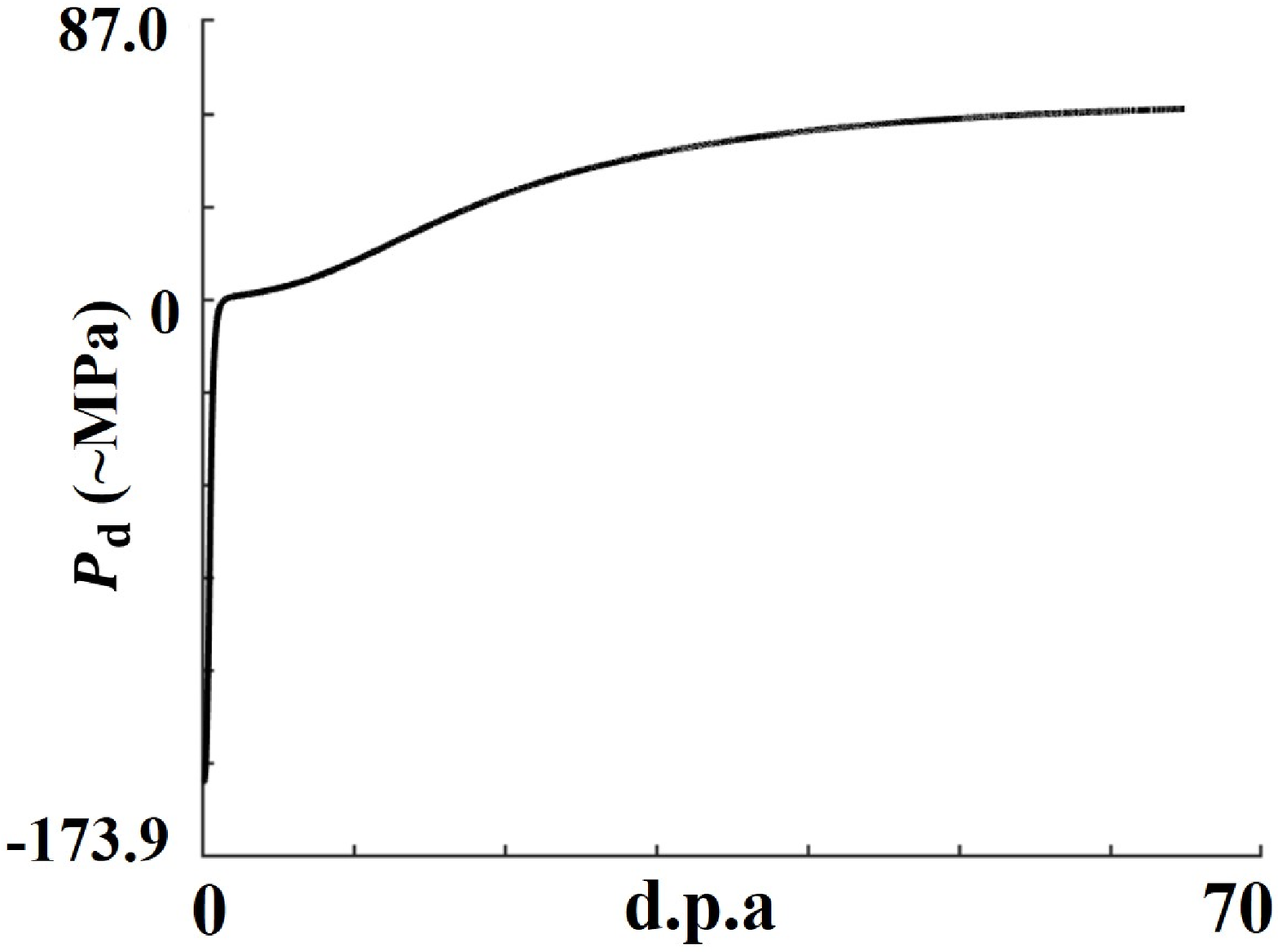}
}\quad
\subfigure[]{
\includegraphics[width=0.45\textwidth]{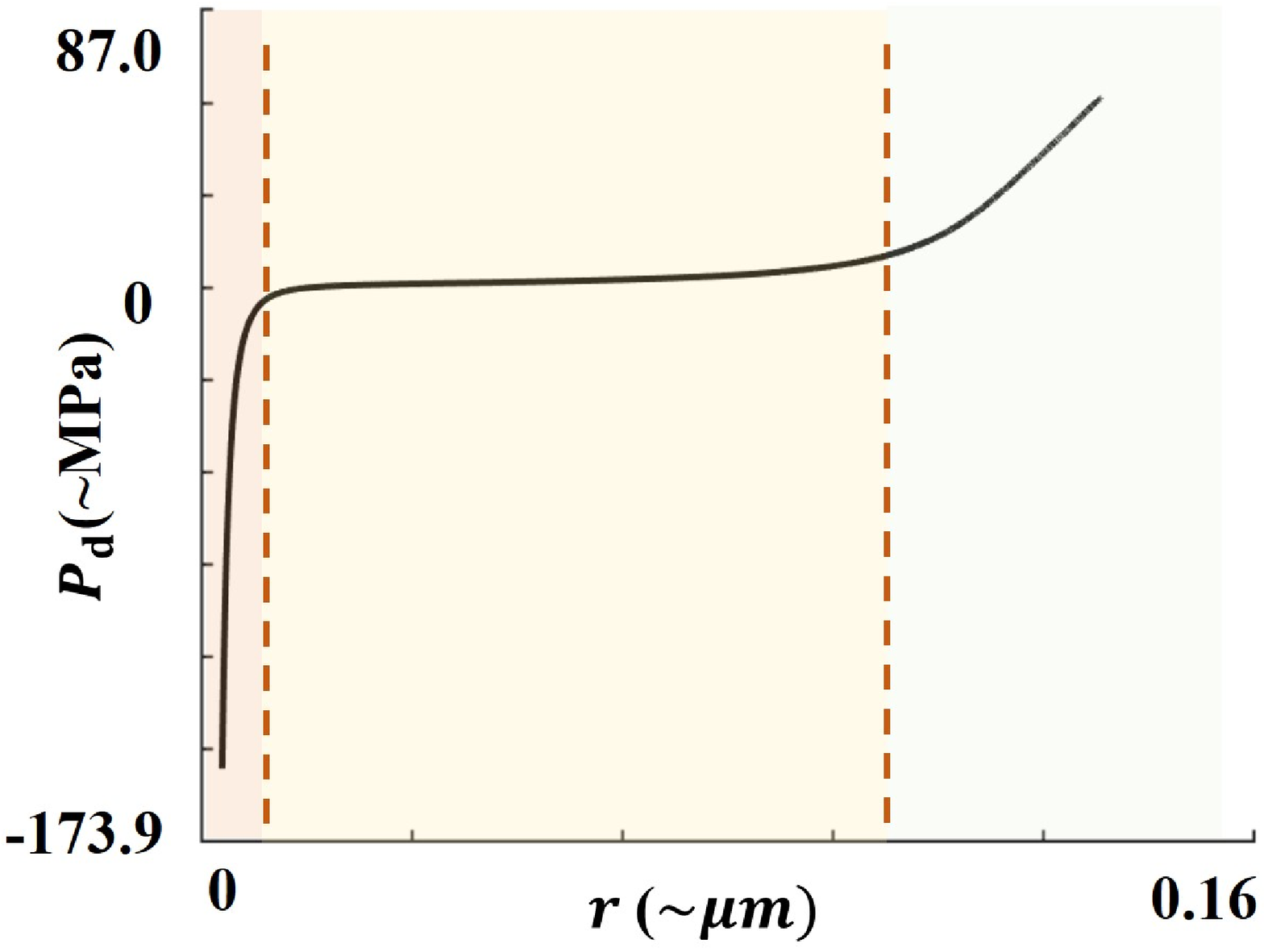}
}
 \caption{Evolution of bubble properties observed at the centre of the interface on slice C: (a) evolution of the (homogenised) bubble size $r $; (b) evolution of the neighbouring fractional vacancy concentration $c_{\text{v}}$; (c) evolution of actual pressure on the bubble surface $P_\text d$; (d) evolution of $P_{\text{d}}$ against $r $.}\label{Results_pressure_and_R_GB}
\end{figure}

Rate equations also enable us to monitor the bubble growth behaviour on crystalline interfaces as shown in fig.~\ref{Results_pressure_and_R_GB}, where the central point of the interface on slice C is selected for observation.
First, it is observed from Fig.~\ref{Results_pressure_and_R_GB}(a) that the bubble size on the interface generally grows at a decelerating rate. This is mainly caused by a drop in the concentration of surrounding vacancies, as revealed by Fig.~\ref{Results_pressure_and_R_GB}(b).

The evolution of the interior pressure per bubble on the interface is also tracked. Hence we introduce a quantity of ``effective pressure'' exerted on the bubble surface, denoted by $P_{\text{d}}$, which equals the gas pressure within the bubble less the surface tension, that is,
\begin{equation}\label{eqn-Pd}
    P_\text d=\frac{3n k_0 T}{4\pi r^3}-\frac{2\gamma}{r}.
\end{equation}
In Fig.~\ref{Results_pressure_and_R_GB}(c), the evolution of $P_{\text{d}}$ against irradiation time is shown. The initial value of $P_{\text{d}}$ is negatively large, because $n=0$ then and a diminishing $r$ induces a (negatively) large surface tensional force. After an initially short period, the value of $P_\text d$ surges to zero. This is because when $r$ is small, an increase in NGA number per bubble quickly escalates the first term in Eq.~\eqref{eqn-Pd}. But in a latter stage, the rising rate in $P_\text d$ drops, because $r$ in the denominator becomes large enough to cast its downside effect over the growth of $P_\text d$.

When the $P_\text d-r$ trend is examined as in Fig.~\ref{Results_pressure_and_R_GB}(d), a clear three-stage evolution is observed. Especially in stage 2, the value of $P_\text d$ roughly stagnates around zero, although $r$ keeps rising. Inserting $P_\text d=0$ into Eq.~\eqref{eqn-Pd}, we see that $n$ should roughly grow in proportional to $r^2$ in stage 2.

Such a three-stage evolution may shed lights on analysing the evolution of the mechanical behaviour of irradiated materials. During the early stage of bubble growth, the effective pressure $P_{\text{d}}$ barely grows, and cracks are unlikely to nucleate during this stage. But when this stage passes, the value of $P_{\text{d}}$ builds up more significantly, and channels linking neighbouring bubbles are likely to form then, so as to release the quickly built up interior pressure.

With the analytical results collected above, it is demonstrated that the present evolution model may be used for monitoring the bubble growth process carried out by a series of self-intertwined mechanisms taking place in irradiated crystalline materials.

\subsection{Examinations of various factors on bubble growth}
In this subsection, we will examine the combinative roles played by several key manipulatable factors that highly affect the radiation damage behaviour of crystalline materials. They are the presence of crystalline interfaces, the irradiation doses and the mechanical loads.

\subsubsection{Role of crystalline interfaces\label{sec-role-interface}}
The role played by crystalline interfaces as partial sinks is examined with the present model first. Note that for the case shown by Fig.~\ref{Results-model}, if the interface is treated as a perfect sink, as conventionally done, no point defects can be detected on its right side, and the bubble growth behaviour on the interface may be over-estimated.

\begin{figure}[!ht]
\centering
\subfigure[Bubble radius]{
\includegraphics[width=0.45\textwidth]{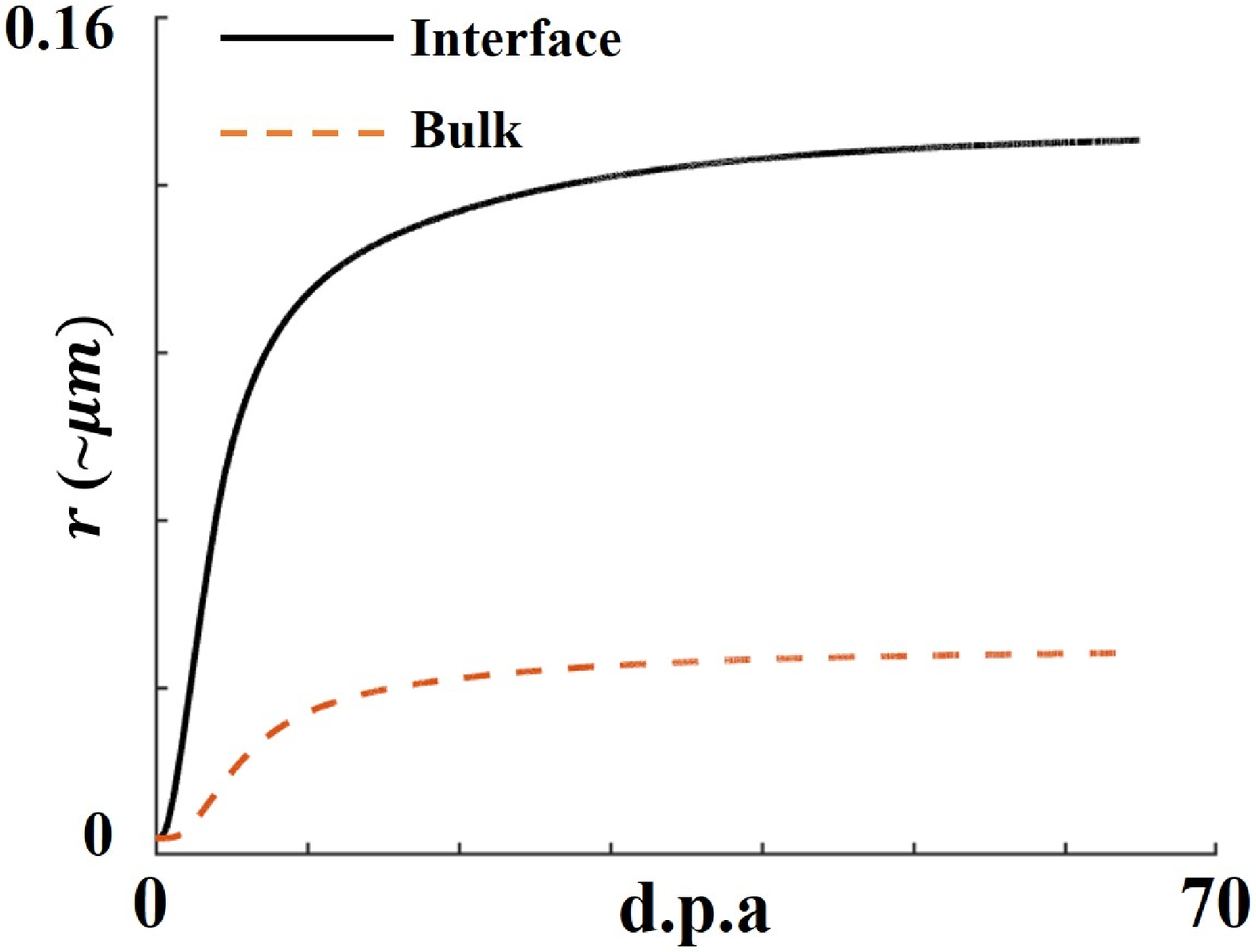}
}
\subfigure[Pressure on bubble surface]{
\includegraphics[width=0.45\textwidth]{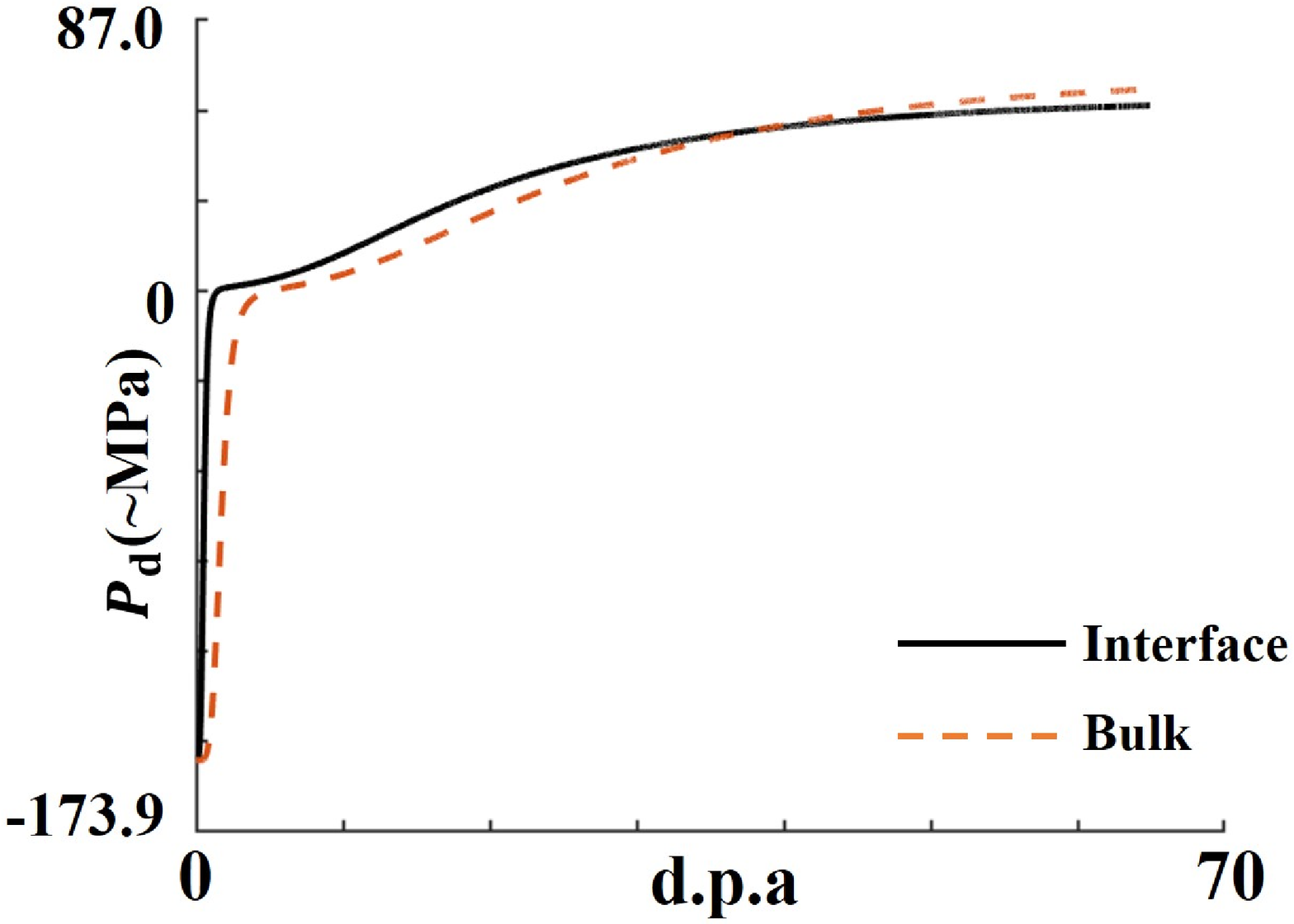}
}
  \caption{Comparison of microstructural behaviour in the bulk against that on the interface. The data are measured from points P$_1$ and P$_2$, as representatives on the interface and in the bulk, respectively. (a) Evolution of the bubble radius; (b) Evolution of the actual pressure on the bubble surface.}\label{Results-compare-evolution}
\end{figure}

To comparatively demonstrate the role of crystalline interfaces, two observation points P$_1$ and P$_2$ are identified as in Fig.~\ref{Results-model}. P$_1$ is located at the centre of the interface $\Gamma$, while P$_2$ is its mirrored point on the other side of the source region. Since the two points are of a same distance away from the source region, a comparison of the corresponding onsite values should help us capture the distinguished role played by the interface, and the result is summarised in Fig.~\ref{Results-compare-evolution}.
In Fig.~\ref{Results-compare-evolution}(a), the bubble size $r$ on the interface grows much faster on the crystalline interface than in the bulk. At roughly 70 d.p.a. (for this simplified case), the distance between adjacent bubbles on the interface is roughly four times shorter than that in the bulk. In contrast, the difference in effective surface pressure $P_\text d$ is not that obvious, as shown in Fig.~\ref{Results-compare-evolution}(b). In fact, one point on extending the present work is to further include the (multiscale) stress analysis in addition to the present rate equations, which may help quantitatively analyse the high probability of crack initiation observed on grain boundaries in fission fuels \cite{Gandhi_ActaMetal1979,Kapoor_JNucMater2007}.

\subsubsection{Role of irradiation dose\label{Sec_role_dose}}
\begin{figure}[!ht]
\centering
\subfigure[$r$ against d.p.a. dose]{
\includegraphics[width=0.45\textwidth]{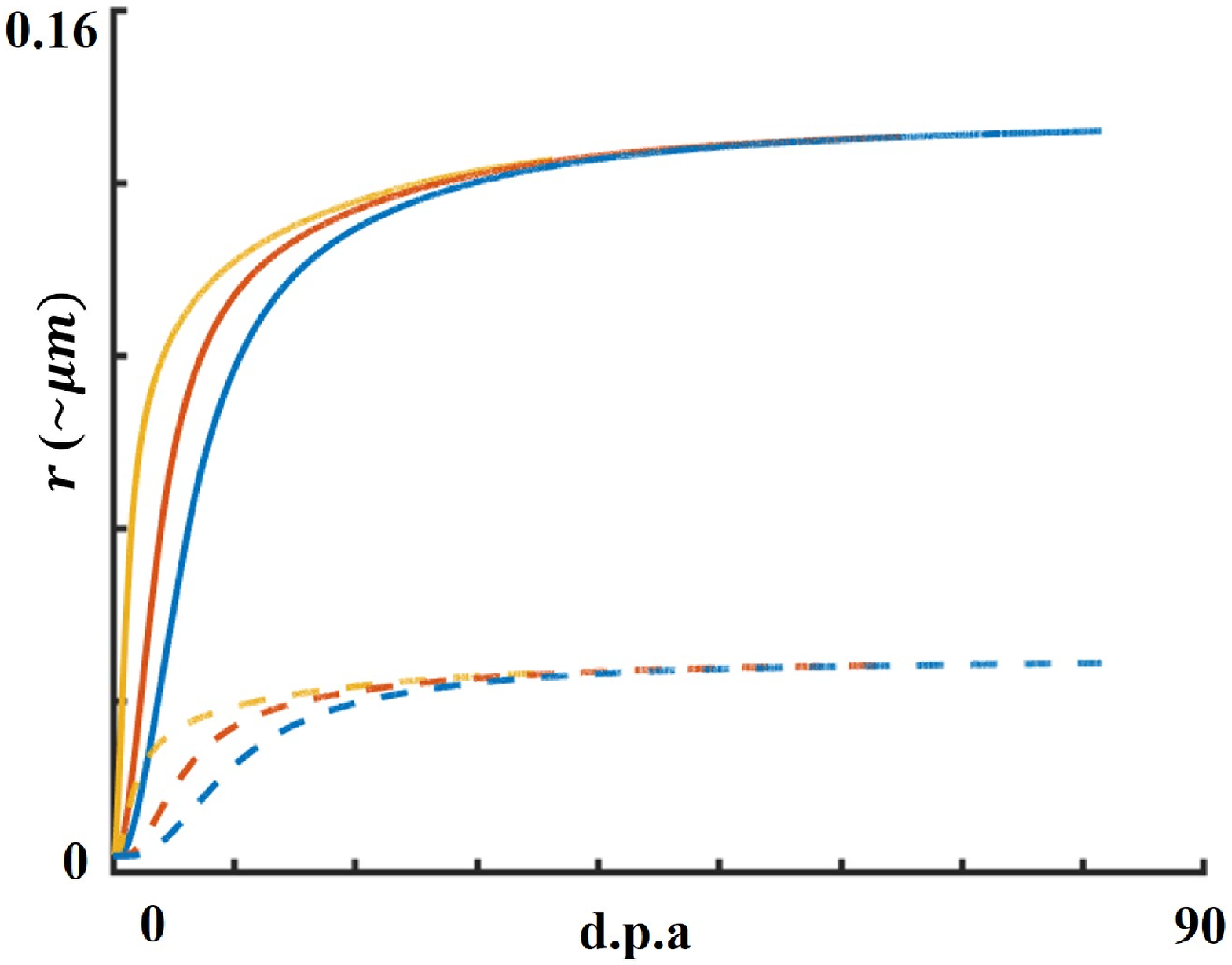}
}
\subfigure[$P_\text{d}$ against d.p.a. dose]{
\includegraphics[width=0.45\textwidth]{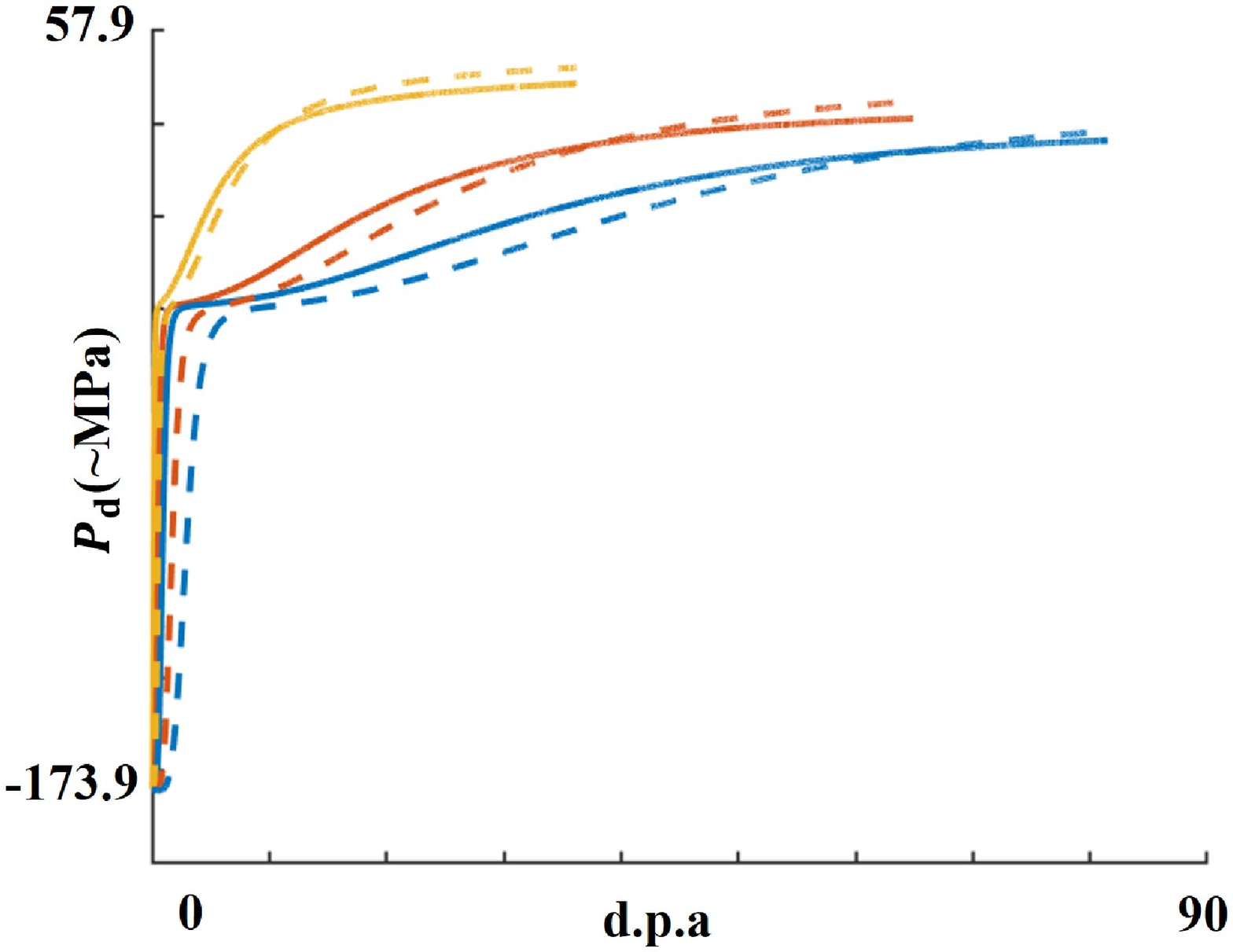}
}
\subfigure[$r$ against time]{
\includegraphics[width=0.45\textwidth]{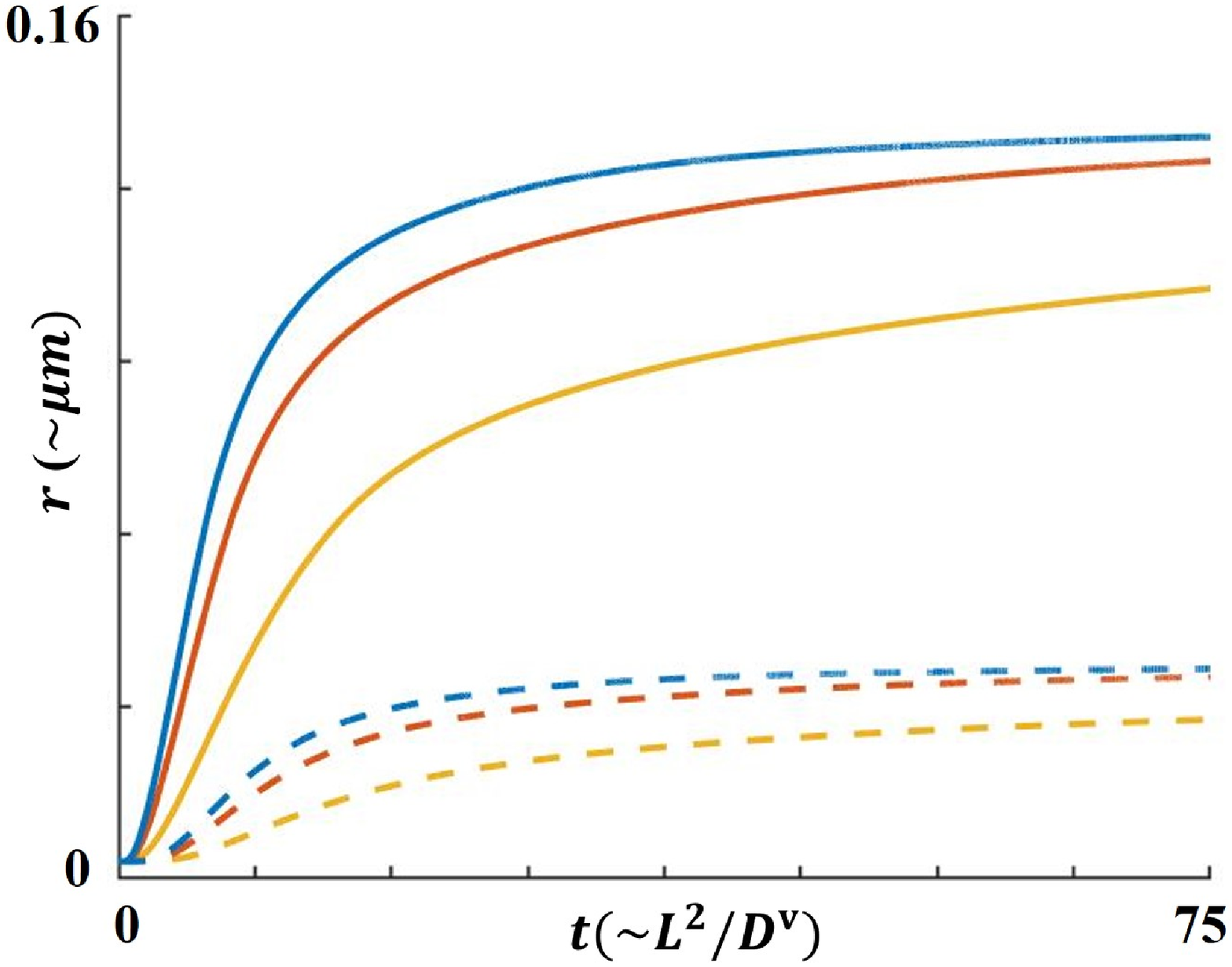}
}
\subfigure[$P_\text{d}$ against time]{
\includegraphics[width=0.45\textwidth]{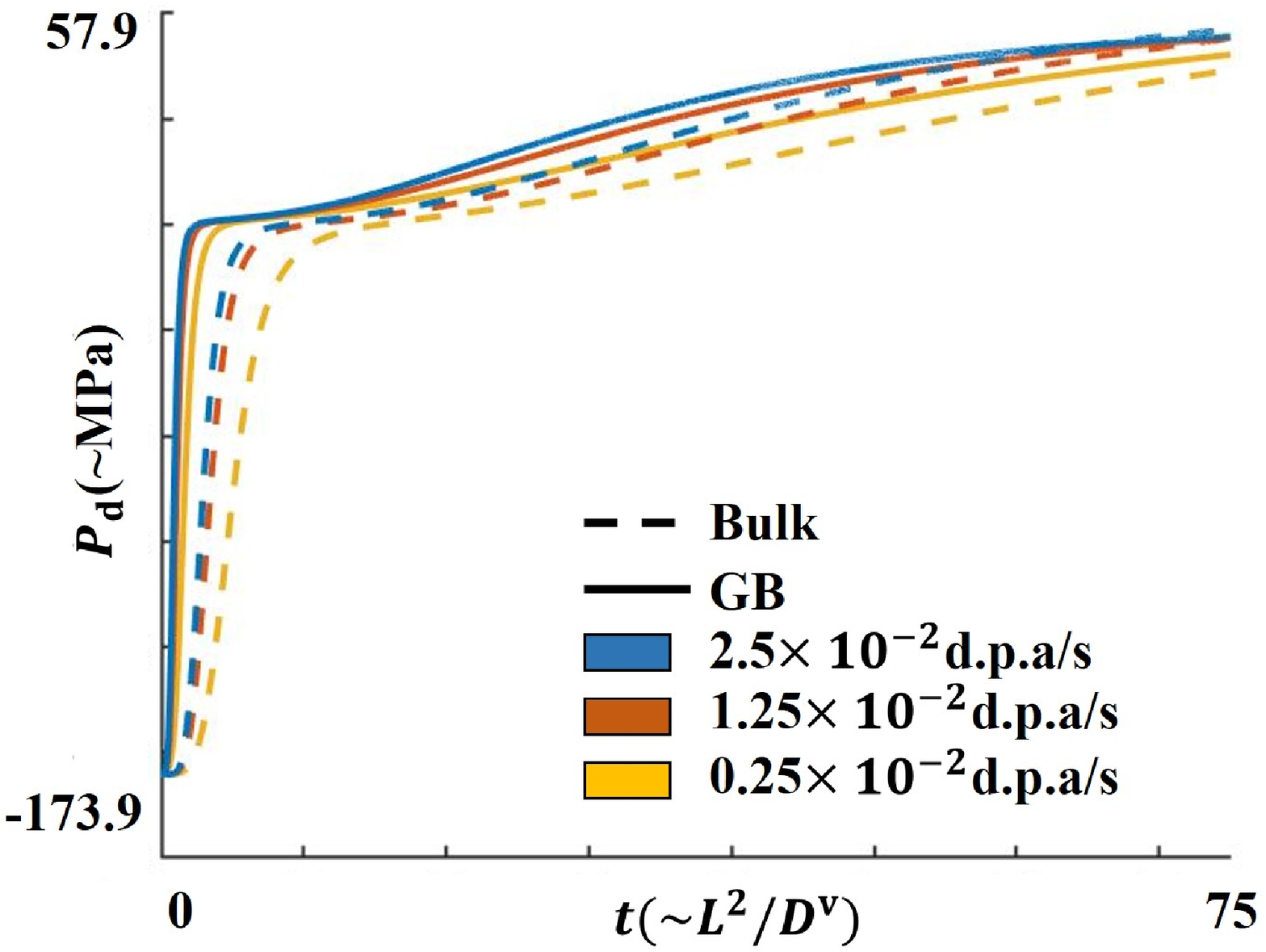}
}
  \caption{Microstructural evolution under different d.p.a. rates: (a) evolution of bubble radius $r $ against irradiation dose; (b) evolution of the actual pressure $P_{\text{d}}$ against irradiation dose; (c) evolution of bubble radius $r $ in time; (d) evolution of the actual pressure $P_{\text{d}}$ in time. The solid curves are generated by data measured at point P$_1$ on the interface as in Fig.~\ref{Results-model}, and the dashed curves are measured at P$_2$, the mirrored point of P$_1$ about the source region.}\label{Results-irrdiation-r}
\end{figure}

We also examined how the PD generation rate (empirically depending on the d.p.a. rate) affects the bubble growth behaviour. In Fig.~\ref{Results-irrdiation-r} (a)-(b) shows the evolution of bubble properties against irradiation dose at points P$_1$ and P$_{2}$ in Fig.~\ref{Results-model}.
It is found that at a same irradiation dose, a higher d.p.a. rate results in slightly lower values in microstructural properties. This is because PD concentration builds up more quickly then, leading to a stronger vacancy-interstitial recombination effect. As a result, (slightly) fewer vacancies are left available for bubble growth. Nonetheless, when measured in real time as shown in Fig.~\ref{Results-irrdiation-r}(c)-(d), the growth rate of $r $ is still faster at a higher d.p.a. rate.

Moreover, it is suggested from Fig.~\ref{Results-irrdiation-r}(a)-(b) that the discrepancies in microstructural properties (at a same irradiation dose) caused by different d.p.a. rates gradually vanish, implying that it seems eligible to assign a larger value to $\dot{s}_{\text{dpa}}$, so as to mimic the actual situations more efficiently with the present model (for the present case).

\subsubsection{Role of mechanical loads}
\begin{figure}[!ht]
\centering
\subfigure[Effect of applied load on bubble radius]{
\includegraphics[width=0.45\textwidth]{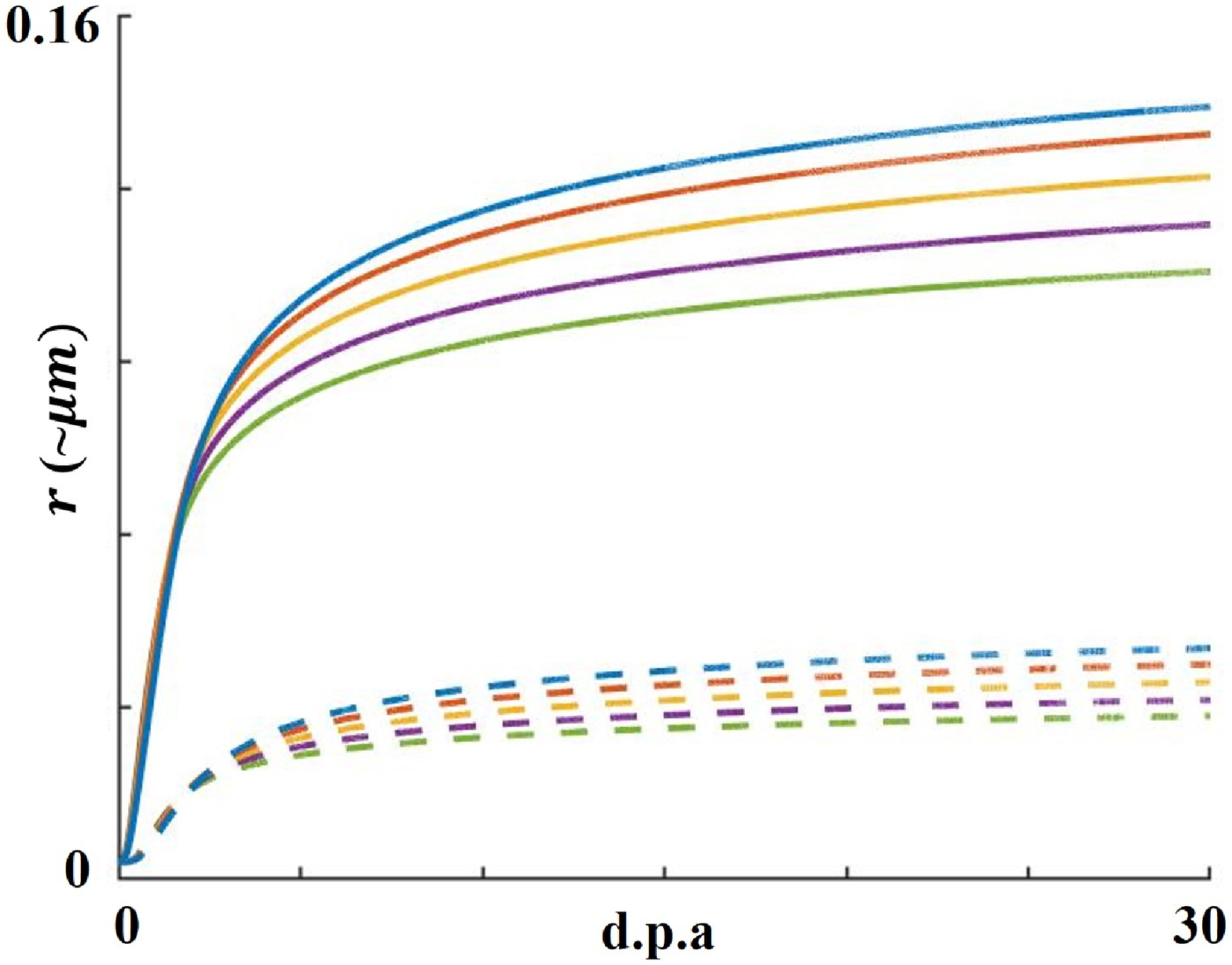}
}
\subfigure[Effect of applied load on effective pressure]{
\includegraphics[width=0.45\textwidth]{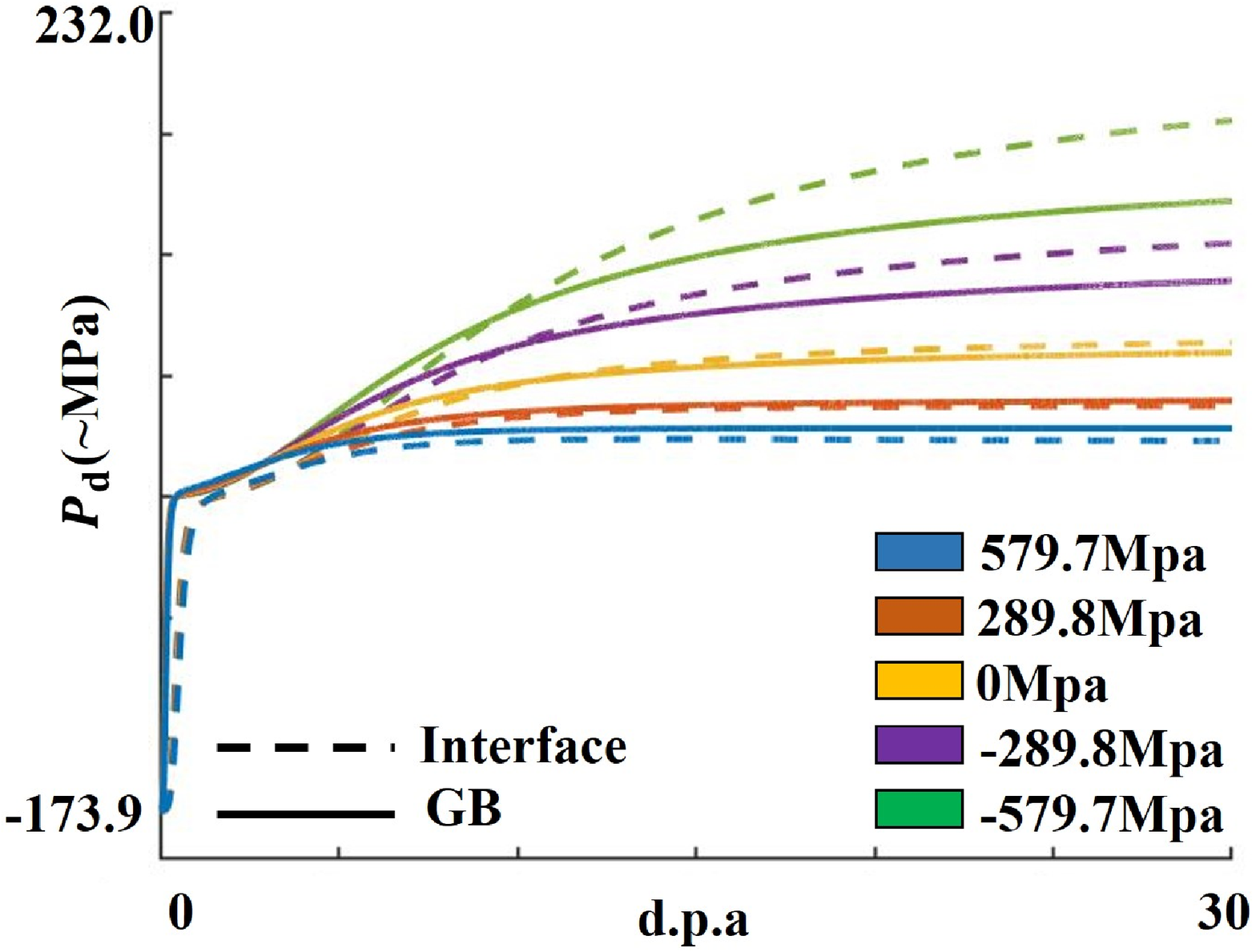}
}
  \caption{The role played by mechanical loads in radiation-induced microstructural evolution: (a) the effect of applied load on the growth of bubbles; (b) the effect of applied load on the inner pressure. The solid curves are cases on the interface, while the dashed curves are the cases in the bulk.}\label{Results-effect-load}
\end{figure}

The present evolution system also takes into consideration the effect due to the constant change in hydrostatic pressure gradient induced by microstructural evolution. Hence it can be adopted for investigating the (subtle) role of mechanical loads on bubble development. For this purpose, we let the value of the applied normal stress $\sigma_{\text{a}}$ vary from roughly -550MPa to 550MPa. Here the range of $\sigma_{\text{a}}$ is quite wide, so as to fully examine the effect due to the mechanical loads. The simulation results are summarised in Fig.~\ref{Results-effect-load} under five different values of $\sigma_\text a$.

In general, the differences caused by a change in $\sigma_{\text{a}}$ are not so obvious. But it can be read from Fig.~\ref{Results-effect-load} that bubble size tends to be larger under a loading environment of traction. This is because under a higher traction ($\sigma_{\text{a}}>0$), the local gradient in the hydrostatic pressure is sharper, as a bubble always tends to compress its surrounding region. In this scenario, the vacancy flux towards the bubble is more favoured, while the influx of interstitial atoms and NGAs is more abased. Consequently, the bubble size grows faster. As more NGAs tend to be absorbed by bubbles in a loading environment of compression, the effective pressure $P_{\text{d}}$ grows faster with a stronger compressive load, as suggested by Fig.~\ref{Results-effect-load}. Note that the local hydrostatic pressure gradient constantly develops in time, and this effect can be fully captured by the sink strengths expressed by the machine learning models here.

\begin{figure}[!ht]
  \centering
  \subfigure[Bubble radius]{\includegraphics[width=.43\textwidth]{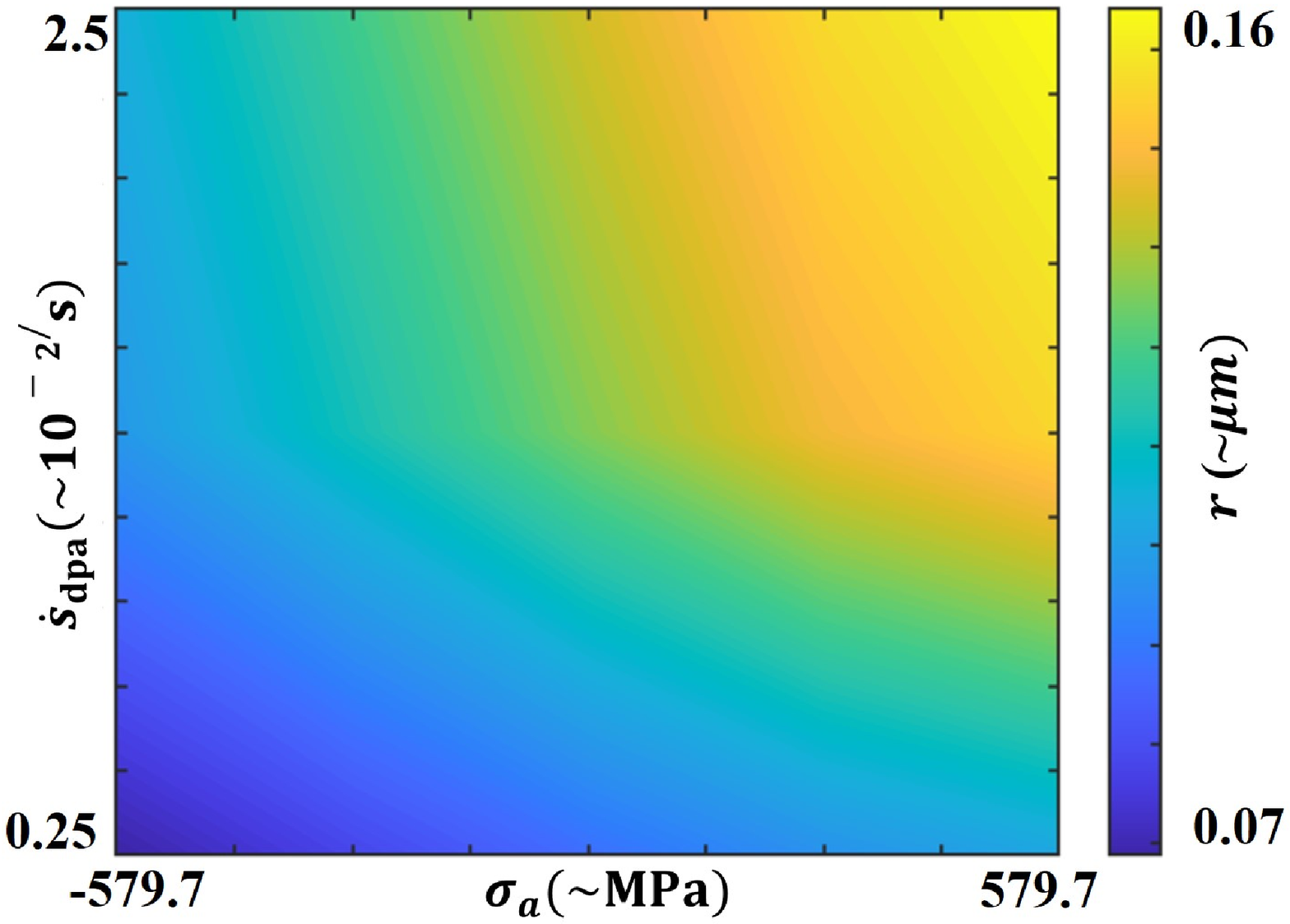}}  \subfigure[Effective pressure]{\includegraphics[width=.45\textwidth]{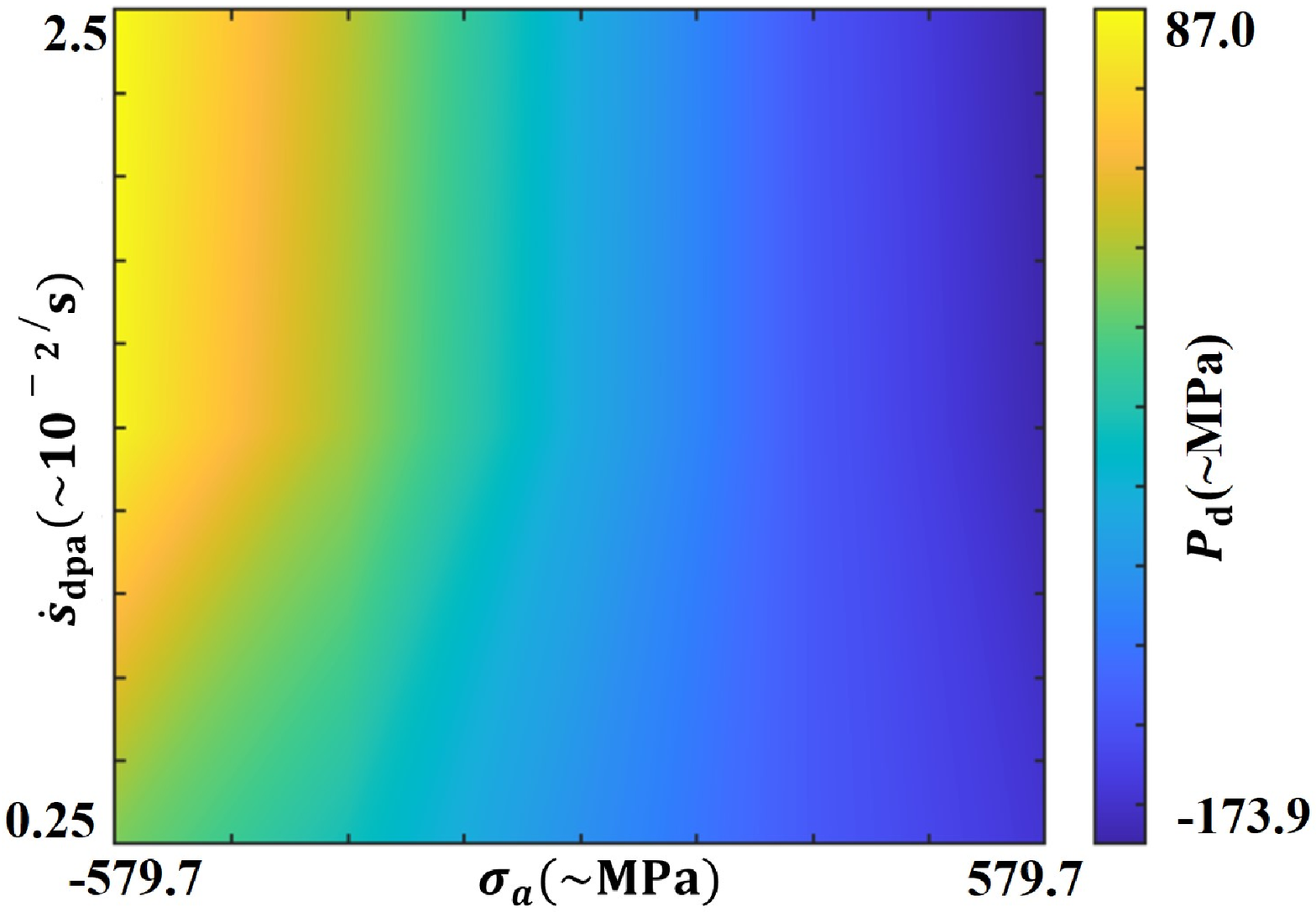}}
  \caption{Phase diagrams showing the combinative effects caused by the external loads and d.p.a. rates. The measurement is made at P$_1$, the centre of the crystalline interface.}\label{Results-phase-diagram}
\end{figure}

Finally, the present model of rate equations also enables us to examine the integrated effects caused by mechanical loads and d.p.a. rate, and two phase diagrams are produced as shown in Fig.~\ref{Results-phase-diagram}. The data for generating Fig.~\ref{Results-phase-diagram} are collected at P$_1$ indicated in Fig.~\ref{Results-model}, and at a same time slot of $t=0.7L^2/D_{\text{v}}$.

\section{Conclusion\label{Sec_conclusion}}
In this work, the point defect sink strengths due to bubbles in the bulk and on crystalline interfaces are formulated through machine learning models. In particular, crystalline interfaces are modelled as partial sinks to PDs, which are described by flux jumps across the interfaces. Instead of direct implementation of machine learning tools, scale analysis is performed against the evolution system of rate equations, so as to sort out, from the mean-field kinetics, a set of locally high-resolution cell problems that govern the underlying PD-bubble interactions. A further rescaling of the LHR cell problems is also suggested, so as to minimise the number of the input arguments for machine learning. The usage of machine learning effectively removes one barrier limiting the conventional treatments for sink strength evaluation, that is, the involved LHR cell problems should yield analytic solutions to a certain degree. Therefore, we are enabled in this work to take into account the subtle role played by the local oscillation in the hydrostatic pressure field, which results from the presence of evolving microstructural bubbles. With the derived evolution system, we examine the combinative effects on bubble development in irradiated materials, caused by a number of factors, such as crystalline interfaces (as partial sinks), mechanical loads, as well as the irradiation dose.

The present work is among early-stage efforts of expressing the sink strengths of rate equations through machine learning representation. It can be extended in a multitude of aspects. First, we need to include more microstructural mechanisms that may take place in irradiated materials. In theory, considering more types of microstructural behaviour should further complicate the outlook of the corresponding LHR problems. Thus extra efforts are needed for sorting out a curriculum to guide the corresponding machine learning treatments. Second, a void lattice configuration is used to define the effective medium for setting up the LHR problems. But the role played by the local spatial disorder in defect clusters is worth further investigations. Third, more accurate asymptotic analysis is needed for capturing the multiscale stress distribution in a continuum background of bubble distributions. This is especially helpful in predicating crack initiations observed in fission fuels. Finally, as crystalline interfaces are still treated as individual objects, the present work is mainly effective for simulations where several interfaces are at present. Hence the present formulation itself needs being further upscaled for analysing the behaviour of polycrystalline materials undergoing irradiation. Upon upscaling, crystalline interfaces are also treated as spatial distributions, and the present scale-transition strategy may shed lights on that piece of work.

\section*{Acknowledgement}
The financial supports from National Key Research and Development Plan (2016YFB0201601) from the Ministry of Science and Technology of the People's Republic of China, the National Natural Science Foundation of China (11772076, 11675161, 11732004, 11821202) are gratefully acknowledged.

\section*{Appendix}
\subsection*{Derivation of the functional relation of Eq.~\eqref{sink_strength}}
Now we rescale the locally high-resolution cell problems outlined in Sec.~\ref{Sec_homogenisation}, aiming to reduce the number of input arguments for machine learning. We first introduce a family of non-dimensional quantities of
\begin{equation} \label{rescaling_appendix}
    \hat{c}_{\alpha} = c_{\alpha}^{0}c^*, \; \hat{\boldsymbol{\sigma}} = \frac{k_0T\boldsymbol{\sigma}^*}{\lambda_{\alpha}\Delta v_{\alpha}}, \; \hat{\mathbf{u}} = \frac{L v_0 \mathbf{u}^*}{\lambda_{\alpha}\Delta v_{\alpha}}, \; \hat{p} = \frac{k_0T p^*}{\lambda_{\alpha}\Delta v_{\alpha}}.
\end{equation}
Note that the newly introduced quantities $c^*$, $\boldsymbol{\sigma}^*$, $\mathbf{u}^*$ and $p^*$ are effectively independent of the species type $\alpha$. Hence one only needs to solve a single set of cell problem, and uses the results to compute the sink strengths for all species involved.

Now we first consider rescaling the mechanical problem defined by Eqs.~\eqref{eqn_fb_sr} and \eqref{eqn_Hookean_sr} and the boundary conditions of Eq.~\eqref{BC_stress_inner_sr}. Incorporating Eq.~\eqref{rescaling_appendix} into them, we end up with a rescaled mechanical problem given by
\begin{equation}\label{problem_mechanical_rescale_appendix}
    \left\{\begin{aligned}
    & \nabla_{\bar{\mathbf{X}}} \cdot \boldsymbol{\sigma}^* = \boldsymbol{0}, \quad \text{in }\Upsilon;\\
    & \boldsymbol{\sigma}^* = \frac{\lambda v_0}{k_0T} \text{tr}(\nabla_{\bar{\mathbf{X}}}\mathbf{u}^*) + \frac{\mu v_0}{k_0T} \left(\nabla_{\bar{\mathbf{X}}}\mathbf{u}^* + (\nabla_{\bar{\mathbf{X}}}\mathbf{u}^*)^{\mathrm{T}}\right), \quad \text{in }\Upsilon;\\
    & \boldsymbol{\sigma}^* \cdot \mathbf{m} = -\lambda_{\alpha} \left(\frac{3n \Delta v_{\alpha}}{4\pi(r )^3} - \frac{2\gamma \Delta v_{\alpha}}{k_0T r}\right), \quad \text{on }\partial\mathcal{O}(\bar{r}^*);\\
    & \left.\boldsymbol{\sigma}^* \cdot \mathbf{m}\right|_{\bar{X}_i=\pm\frac1{2}} = \pm \sigma_i  \mathbf{e}^i, \quad \text{for }i=1, 2, 3,
    \end{aligned}\right.
\end{equation}
where $\{\mathbf{e}^i\}_{i=1}^3$ form an orthogonal triad; $\sigma_i$ are recalled to be the principle components of the onsite mean-field stress field. Problem~\eqref{problem_mechanical_rescale_appendix} is effectively problem~\eqref{problem_mechanical_rescale} in Sec.~\ref{Sec_reduction}, provided that Eqs.~\eqref{beta_appendix} hold.

From Eq.~\eqref{problem_mechanical_rescale_appendix}, one may express the the rescaled stress field $\boldsymbol{\sigma}^*$ by
\begin{equation} \label{stress_exp_appendix}
  \boldsymbol{\sigma}^*= \boldsymbol{\sigma}^*(\bar{\mathbf{X}};\boldsymbol{\beta}),
\end{equation}
where $\bar{\mathbf{X}}$ denotes its spatial variables and $\boldsymbol{\beta}$ given by Eq.~\eqref{beta_appendix} are parameters.

Then a Laplacian equation should be solved for the rescaled (non-dimensional) hydrostatic pressure field $p^*$, which satisfies
\begin{equation}\label{problem_pressure_rescale_appendix}
    \left\{\begin{aligned}
    & \nabla_{\bar{\mathbf{X}}}^2 p^* = 0, \quad \text{in }\Upsilon;\\
    & \left.p^*\right|_{\partial\mathcal{O}(\beta_1)} = - \left. \frac{\text{tr}(\boldsymbol{\sigma}^*)}{3}\right|_{\partial\mathcal{O}(\beta_1)};\\
    & \left.p^*\right|_{\bar{X}_i = \pm \frac1{2}} = - \left. \frac{\text{tr}(\boldsymbol{\sigma}^*)}{3}\right|_{\bar{X}_i = \pm \frac1{2}},
    \end{aligned}\right.
\end{equation}
for $i=1$, $2$ and $3$. Eq.~\eqref{problem_pressure_rescale_appendix} corresponds to Eq.~\eqref{problem_pressure_rescale} in the main text. Since $p^*$ is purely determined by $\boldsymbol{\sigma}^*$, we should have
\begin{equation} \label{p_exp_appendix}
  p^* = p^*(\bar{\mathbf{X}};\boldsymbol{\beta}).
\end{equation}

With $p^*$ obtained, we then incorporate Eqs.~\eqref{rescaling_appendix} into Eq.~\eqref{eqn_flux_sr} (and its associated boundary conditions) to reach a problem for the rescaled LHR fractional concentration $c^*$ given by
\begin{equation}\label{problem_c_rescale_appendix}
    \left\{\begin{aligned}
    & \nabla_{\bar{\mathbf{X}}} \cdot \left(\nabla_{\bar{\mathbf{X}}} c^* - c^* \nabla_{\bar{\mathbf{X}}} p^*\right) = 0, \quad \text{in }\Upsilon;\\
    & c^* = 0, \quad \text{on }\partial\mathcal{O}(\beta_1);\\
    & \left.c^*\right|_{\bar{X}_i = \pm \frac1{2}} = 1.
    \end{aligned}\right.
\end{equation}
Eq.~\eqref{problem_c_rescale_appendix} corresponds to Eq.~\eqref{problem_c_rescale} in the main text.

Moreover, one also needs to relate $c_{\alpha}^0$ in Eq.~\eqref{rescaling_appendix} to the mean-field fractional concentration $c_{\alpha}$. This can be done by inserting Eq.~\eqref{rescaling_appendix} into \eqref{c_macro_to_micro}, and we obtain
\begin{equation} \label{c0_by_c_appendix}
   c_{\alpha}^0 = \frac{c_{\alpha}}{\int_{\Upsilon} c^* \,\mathrm{d} \bar{\mathbf{X}}}\cdot \left(1-\frac{4\pi \beta_1^3}{3}\right).
\end{equation}
Inserting Eq.~\eqref{c0_by_c_appendix} back to Eq.~\eqref{rescaling_appendix} relates the locally high-resolution fractional concentration $\hat{c}_{\alpha}$ to $c^*$ by
\begin{equation} \label{c_sr_exp_appendix}
   \hat{c}_{\alpha} = \frac{c_{\alpha}}{\int_{\Upsilon} c^* \,\mathrm{d} \bar{\mathbf{X}}} \left(1-\frac{4\pi \beta_1^3}{3}\right) \cdot c^*(\bar{\mathbf{X}};\boldsymbol{\beta}).
\end{equation}

Incorporating Eq.~\eqref{c_sr_exp_appendix} into \eqref{A_macro_to_micro} gives
\begin{equation}\label{A_rescale_appendix0}
  k_{\alpha\text{B}}^2 = - \frac{\varrho^{\frac2{3}}}{\int_{\Upsilon} c^* \,\mathrm{d} \bar{\mathbf{X}}} \left(1-\frac{4\pi \beta_1^3}{3}\right) \int_{\partial \mathcal{O}(\beta_1)} \mathbf{m} \cdot \left(\nabla_{\bar{\mathbf{X}}}c^* - c^* \nabla_{\bar{\mathbf{X}}}p^*\right) \, \mathrm{d} S_{\bar{\mathbf{X}}}.
\end{equation}
Note that $\left.c^*\right|_{\partial \mathcal{O}(\beta_1)} = 0$ as indicated by the inner boundary condition in Eq.~\eqref{problem_c_rescale_appendix}. Eq.~\eqref{A_rescale_appendix0} thus becomes
\begin{equation} \label{A_rescale_appendix}
    \begin{aligned}
    \frac{k_{\alpha\text{B}}^2}{\varrho^{\frac2{3}}\left(1-\frac{4\pi \beta_1^3}{3}\right)} =  - \frac1{\int_{\Upsilon} c^* \,\mathrm{d} \bar{\mathbf{X}}} \int_{\partial \mathcal{O}(\beta_1)} \mathbf{m} \cdot \nabla_{\bar{\mathbf{X}}}c^* \, \mathrm{d} S_{\bar{\mathbf{X}}}.
    \end{aligned}
\end{equation}

Note that the right side of Eq.~\eqref{A_rescale_appendix} is now fully determined by the vector $\boldsymbol{\beta}$ given by Eq.~\eqref{beta_appendix}, we can introduce an implicit functional relation $F$ to represent it, i.e.,
\begin{equation} \label{F_star_appendix}
    F(\boldsymbol{\beta}) = - \frac1{\int_{\Upsilon} c^* \,\mathrm{d} \bar{\mathbf{X}}} \int_{\partial \mathcal{O}(\bar{r}^*)} \mathbf{m} \cdot \nabla_{\bar{\mathbf{X}}}c^* \, \mathrm{d} S_{\bar{\mathbf{X}}},
\end{equation}
which corresponds to Eq.~\eqref{F_star} in the main text. Inserting Eq.~\eqref{F_star_appendix} into \eqref{A_rescale_appendix} finally gives the expression for $k_{\alpha\text{B}}^2$ as formulated by Eq.~\eqref{sink_strength} in Sec.~\ref{Sec_reduction}.

\section*{References}
\bibliography{reference2}

\end{document}